\documentclass[acmlarge]{acmart}
\usepackage{balance}
\usepackage{graphicx,graphics}
\usepackage{amsfonts}
\usepackage{amsmath,units} 
\usepackage{cases}
\usepackage{mathrsfs}
\usepackage{multirow}
\usepackage{caption}
\usepackage{color}
\usepackage{flushend}
\usepackage{float}
\usepackage[ruled,vlined,linesnumbered]{algorithm2e}
\usepackage{algorithmic}
\usepackage{subfigure}
\usepackage{extarrows}
\usepackage{graphicx,graphics} 
\usepackage{array}
\usepackage{tabularx}
\usepackage{stmaryrd}
\usepackage{diagbox}
\usepackage{bm}
\usepackage{enumitem}

\usepackage{amssymb}
\usepackage{pifont}
\usepackage{makecell}
\usepackage{soul}
\usepackage{adjustbox}
\usepackage[figuresright]{rotating}
\usepackage{pdfpages}

\usepackage{titletoc} 

\titlecontents{section}          
  [0em]                          
  {\normalfont\large\bfseries}                  
  {\contentslabel{2em}}          
  {}                             
  {\titlerule*[.5pc]{.}\contentspage}  

\titlecontents{subsection}       
  [2em]                          
  {\normalfont}                  
  {\contentslabel{2em}}          
  {}                             
  {\titlerule*[.5pc]{.}\contentspage}  

\titlecontents{subsubsection}    
  [5em]                          
  {\normalfont}                  
  {\contentslabel{3em}}          
  {}                             
  {\titlerule*[.5pc]{.}\contentspage}  
  
\newcommand{\partitle}[1]{\medskip \noindent \textbf{#1.}}

\newcommand{\nop}[1]{}
\definecolor{mygray}{gray}{0.8}
\newcommand{\cmark}{\ding{51}}%
\newcommand{\xmark}{\color{mygray}\ding{55}}%

\AtBeginDocument{%
  \providecommand\BibTeX{{%
    \normalfont B\kern-0.5em{\scshape i\kern-0.25em b}\kern-0.8em\TeX}}}

\setcopyright{acmcopyright}
\copyrightyear{2024}
\acmYear{2024}

\acmJournal{CSUR}
\acmVolume{XX}
\acmNumber{X}
\acmArticle{XXX}
\acmMonth{11}

\begin{document}

\title{A Survey on Data Markets}

\author{Jiayao Zhang}
\email{jiayaozhang@zju.edu.cn}
\author{Yuran Bi}
\email{stellabyr@zju.edu.cn}
\author{Mengye Cheng}
\email{mengyecheng@zju.edu.cn}
\author{Jinfei Liu}
\email{jinfeiliu@zju.edu.cn}
\authornote{Corresponding author.}
\author{Kui Ren}
\email{kuiren@zju.edu.cn}
\author{Qiheng Sun}
\email{qiheng_sun@zju.edu.cn}
\author{Yihang Wu}
\email{wuyihang@zju.edu.cn}
\affiliation{%
  \institution{Zhejiang University}
}

\author{Yang Cao}
\affiliation{%
  \institution{Tokyo Institute of Technology}}
\email{cao@c.titech.ac.jp}

\author{Raul Castro Fernandez}
\email{ruoxijia@vt.edu}
\author{Haifeng Xu}
\email{ruoxijia@vt.edu}
\affiliation{%
  \institution{The University of Chicago}}

\author{Ruoxi Jia}
\affiliation{%
  \institution{Virginia Tech}}
\email{ruoxijia@vt.edu}

\author{Yongchan Kwon}
\affiliation{%
  \institution{Columbia University}}
\email{yk3012@columbia.edu}

\author{Jian Pei}
\affiliation{%
  \institution{Duke University}}
\email{j.pei@duke.edu}

\author{Jiachen T. Wang}
\affiliation{%
  \institution{Princeton University}}
\email{tianhaowang@princeton.edu}

\author{Haocheng Xia}
\affiliation{%
  \institution{University of Illinois Urbana-Champaign}}
\email{hxia7@illinois.edu}

\author{Li Xiong}
\affiliation{%
  \institution{Emory University}}
\email{lxiong@emory.edu}

\author{Xiaohui Yu}
\affiliation{%
  \institution{York University}}
\email{xhyu@yorku.ca}

\author{James Zou}
\affiliation{%
  \institution{Stanford University}}
\email{jamesz@stanford.edu}

\renewcommand{\shortauthors}{XXX, et al.}

\begin{abstract}
Data is the new oil of the 21st century. The growing trend of trading data for greater welfare has led to the emergence of data markets. 
A data market is any mechanism whereby the exchange of data products including datasets and data derivatives takes place as a result of data buyers and data sellers being in contact with one another, either directly or through mediating agents. It serves as a coordinating mechanism by which several functions, including the pricing and the distribution of data as the most important ones, interact to make the value of data fully exploited and enhanced.
In this article, we present a comprehensive survey of this important and emerging direction from the aspects of data search, data productization, data transaction, data pricing, revenue allocation as well as privacy, security, and trust issues. We also investigate the government policies and industry status of data markets across different countries and different domains. Finally, we identify the unresolved challenges and discuss possible future directions for the development of data markets.
\end{abstract}

\keywords{Data Market, Transaction, Pricing, Auction, Privacy, Security, Trust, Policy}

\maketitle

\newpage
\tableofcontents

\newpage
\section{Introduction}

Data is considered an invaluable resource in the digital economy. The last decades have witnessed the explosive growth of data. As raw material for acquiring knowledge and developing products, data generates value in an indirect way. After remodeling the commercial perspective of data, data is directly monetized like other material commodities nowadays. Individuals and organizations extensively trade datasets and derived data products. In this new vision, data is no longer the enabler of products, but also the product itself. Governments around the world are seizing this new opportunity. For example, the Chinese government unveiled a guideline to improve the market-based allocation of data factors, which is the first to list data as a production factor following land, labor, capital, and entrepreneurship \cite{productionfactors}. The United States created the Federal Data Strategy Action Plan aimed at leveraging data as a strategic asset \cite{federal-data-strategy}.

Driven by the tides of data monetization, data markets have emerged. Data markets, a nascent interdiscipline of computer science and economics, are growing rapidly and evolving in myriad research directions. The history of data markets can be traced back to 1986. A seminal work by \citet{admati1986monopolistic} studies a market where traders purchase information from a monopolistic seller. The information they trade is the data endowed or produced by individual agents. To the best of our knowledge, the term ``data market'' is put forward by \citet{DBLP:reference/gis/Keenan08} in 2008 for the first time in the literature. They propose to exchange spatial data collected by geographic information systems in the market. In 2011, \citet{DBLP:journals/pvldb/BalazinskaHS11} present a vision of a more general data market where commodities are derivative data products. They outline key challenges in a relational cloud data market for the database research community. Since then, data markets have experienced rapid development. \citet{koutris2012query} design the first query-based data market; \citet{DBLP:conf/sigmod/DeepK17} propose a scalable and flexible pricing framework for relational queries; \citet{DBLP:conf/ec/AgarwalDS19} design the first two-sided marketplace for trading training data directly; \citet{DBLP:conf/sigmod/ChenK019} introduce the first model-based data market; and more recently \citet{DBLP:journals/pvldb/LiuLL0PS21} propose the first end-to-end (model-based) data market involving the interactions among sellers, brokers, and buyers. 

With the growing demand for data transactions, many data marketplaces have sprung up, such as AWS Data Exchange \cite{AWS}, Dawex \cite{dawex}, BDEX \cite{BDEX}, Factual \cite{Factual}, and Snowflake \cite{Snowflake}. Data marketplaces are online transaction locations or exchanges that facilitate the buying and selling of data products. They are authorized to host data products and conduct data transactions for the benefit of stakeholders. 

We propose a definition of the data market in this survey as follows.

\begin{quote}
\textit{
A \textbf{data market} is any mechanism whereby the exchange of data products including datasets and data derivatives (such as query results and trained models) takes place as a result of data buyers and data sellers being in contact with one another, either directly or through mediating agents.
}
\end{quote}
The data market serves as a coordinating mechanism by which several functions, including the pricing and the distribution of data as the most important ones, interact to make the value of data fully exploited and enhanced. In data markets, the life chain of data covers the process of data search, productization, monetization in pricing and transaction, and finally destruction. Trading data products naturally raises privacy, security, and trust concerns, and faces regulatory barriers to achieving compliance and traceability. Whether in academia or industry, there are rich explorations on designing data markets, where different data markets vary from each other in terms of data products, underlying functions, and market mechanisms.

In this article, we present a comprehensive survey of this important and emerging direction from the aspects of data search, data productization, data transaction, data pricing, revenue allocation as well as privacy, security, and trust issues. We also investigate the government policies and industry status of data markets across different countries and different domains. Finally, we identify the unresolved challenges and discuss possible future directions for the development of data markets.

\subsection{Related Surveys}




The existing surveys on data markets can be generally categorized based on the scope: (1) surveys on academic research \cite{thomas2016big,DBLP:journals/access/DriessenMH22,DBLP:journals/access/LiangYAYFZ18,DBLP:journals/jtaer/AbbasAVZR21}, (2) surveys on industry status \cite{DBLP:journals/sigmod/SchommSV13,DBLP:journals/cacm/LiQW18,kennedyrevisiting,DBLP:journals/corr/abs-2201-04561}, and (3) surveys on data pricing \cite{DBLP:conf/birte/MuschalleSLV12,DBLP:conf/icsob/FrickerM17,zhang2020survey,DBLP:journals/tkde/Pei22,DBLP:journals/kais/CongLPZZ22,DBLP:journals/tbd/ZhangBL23,DBLP:journals/corr/abs-2306-04945,DBLP:conf/caibda/ChiZFZJY23}.

\partitle{Surveys on academic research}
Efforts \cite{thomas2016big, DBLP:journals/jtaer/AbbasAVZR21, DBLP:journals/access/DriessenMH22, DBLP:journals/access/LiangYAYFZ18} have been made to survey academic research for data markets within the whole lifecycle. \citet{thomas2016big} provide managers with a literature review on the commercialization of big data. From a managerial and commercial perspective, they introduce six business models in the data ecosystem, led by data suppliers, data managers, data custodians, application developers, service providers, and data aggregators. Based on this taxonomy, they discuss the characteristics of the data ecosystem and conclude with the challenges faced by managers and corresponding guidelines in trading big data including pricing and privacy concerns which we reinforce in this paper. \citet{DBLP:journals/jtaer/AbbasAVZR21} examine 133 academic articles using a Service-Technology-Organization-Finance (STOF) model. They find that the existing literature on data marketplaces is primarily dominated by technology studies. \citet{DBLP:journals/access/DriessenMH22} present a statistical analysis of works related to data markets up until 2021, discuss practical application areas for data markets, categorize the problems of designing data markets, and find corresponding solutions in the literature. 
\citet{DBLP:journals/access/LiangYAYFZ18} use 4V (Volume, Velocity, Variety, and Value) to define big data and survey the lifecycle of trading big data, including data pricing, data trading, and data protection, for each of which they review corresponding issues and models. The above works \cite{thomas2016big, DBLP:journals/jtaer/AbbasAVZR21, DBLP:journals/access/DriessenMH22, DBLP:journals/access/LiangYAYFZ18} mainly focus on techniques for trading data, while our survey covers state-of-the-art literature for trading general data products, including raw data and its derivatives such as queries, statistical inferences, and machine learning models. Moreover, our survey comprehensively covers key issues in main procedures in data markets from data search to data destruction.

\partitle{Surveys on industry status}
There are four works \cite{DBLP:journals/sigmod/SchommSV13,DBLP:journals/cacm/LiQW18,kennedyrevisiting,DBLP:journals/corr/abs-2201-04561} conducting industry surveys on data marketplaces. \citet{DBLP:journals/sigmod/SchommSV13} present an initial survey of data marketplaces and data vendors by investigating 46 data suppliers from twelve dimensions (type, time frame, domain, data origin, pricing model, data access, data output, language, target audience, trustworthiness, size of vendor, and maturity) up until Summer 2012. \citet{DBLP:journals/cacm/LiQW18} introduce policies of China for developing data markets and discuss concerns and research opportunities including preprocessing, pricing, security, privacy, and verifiability. \citet{DBLP:journals/corr/abs-2201-04561} investigate 180 entities which trade data on the Internet, summarize different business models, and discuss open challenges. \citet{kennedyrevisiting} introduce different types of data marketplaces and describe data transaction lifecycle from the perspective of buyers and sellers. They also interview buyers and sellers to understand the current status and challenges of online data marketplaces in 2022.  The above works \cite{DBLP:journals/sigmod/SchommSV13,DBLP:journals/cacm/LiQW18,kennedyrevisiting,DBLP:journals/corr/abs-2201-04561} provides an understanding of data marketplaces through practical investigations of entities and marketplaces. In contrast, our survey not only examines mainstream data marketplaces worldwide but also provides a list of government policies.


\partitle{Surveys on data pricing}
There have been several surveys \cite{DBLP:conf/birte/MuschalleSLV12, DBLP:conf/icsob/FrickerM17, zhang2020survey, DBLP:journals/tkde/Pei22, DBLP:journals/kais/CongLPZZ22,DBLP:journals/tbd/ZhangBL23,DBLP:journals/corr/abs-2306-04945} specializing in data pricing, a subtopic that receives the most attention in data markets. \citet{DBLP:conf/birte/MuschalleSLV12} investigate seven established vendors for their potential market situations, pricing approaches, and trends. \citet{DBLP:conf/icsob/FrickerM17} report a literature survey of 18 papers regarding several research questions, including the maturity and targets of pricing models, types of data products, and pricing mechanisms. \citet{zhang2020survey} review novel data pricing studies and categorize data pricing methods based on data granularity and privacy. \citet{DBLP:journals/tkde/Pei22} starts with the economics of data pricing and reviews pricing models based on a set of fundamental principles. He also discusses the differences between digital products and data products, and the corresponding pricing methods. Very recently, \citet{DBLP:journals/kais/CongLPZZ22} survey data pricing methods in machine learning pipelines, including pricing raw data sets, pricing data labels, and pricing in collaborative machine learning models. 
\citet{DBLP:journals/tbd/ZhangBL23} categorize and review pricing methods for queries from the aspects of market structure, privacy notion, query type, and pricing method. \citet{DBLP:journals/corr/abs-2306-04945} classify data pricing techniques into three strategies and analyze thirteen pricing models. \citet{DBLP:conf/caibda/ChiZFZJY23} outline the fundamental concepts of data pricing, categorize data pricing strategies into query-based and privacy-based approaches, and offer an overview of data pricing from a data science standpoint.
In addition to covering other data market functions, our survey includes a comprehensive analysis of data pricing that examines both revenue allocation for allocating compensations to data sellers and data product pricing for pricing data products to data buyers and their interactions. Furthermore, it systematically reviews emerging game-theoretic approaches to data pricing for the first time.


In summary, while existing surveys approach data markets from either academic or industry perspective, our survey provides a comprehensive and general review of data markets covering both academic research and industry status including government policies across representative countries and domains. We also discuss the differences between data and other production factors and the corresponding impact on the design of data markets. While existing surveys investigate a few significant challenges in data markets,  we study the interaction between key entities, summarize important desiderata for designing a well-functioning data market, and review techniques regarding data search, productization approaches, pricing mechanisms, data transactions, privacy concerns, etc, based on a formal framework as in Figure \ref{fig:framework}. 

\partitle{Contributions}
We present a comprehensive survey of data markets in both academia and industry. The purpose of this survey is to delve into subtopics of data markets in terms of computer science while covering mechanisms, regulations, and challenges in economics, law, and governance. The main contributions of this survey are summarized as follows.

\begin{itemize}
    \item Identify the unique properties of data and discuss the difference between data markets and other markets for the four production factors (land, labor, capital, and entrepreneurship).

    \item Introduce the framework of data markets, formalize the abilities and restrictions of key roles, and illustrate the main procedures in the operations of data markets.
        
    \item Present important desiderata for well-functioning data markets.

    \item Summarize various methods of data search for various purposes, including crowdsourced dataset collection, dataset discovery in databases, data discovery in machine learning, and general dataset search.
    
    \item Introduce various approaches to data productization based on versioning and data market categories.

    \item Outline advertising strategies for data sellers and data purchase methods for data buyers in data transactions.

    \item Review different approaches for revenue allocation and data product pricing, along with game-theoretic pricing methods.
    
    \item Describe possible attacks on privacy preservation, fairness, profitability, and traceability from dishonest entities and corresponding solutions.
    
    \item Present guidelines and regulations and investigate actual data marketplaces in representative countries and domains.
    
    \item Discuss various open challenges and emerging directions for future research.
\end{itemize}

\begin{figure}[htb]
 \centering
 \includegraphics[width=1\textwidth]{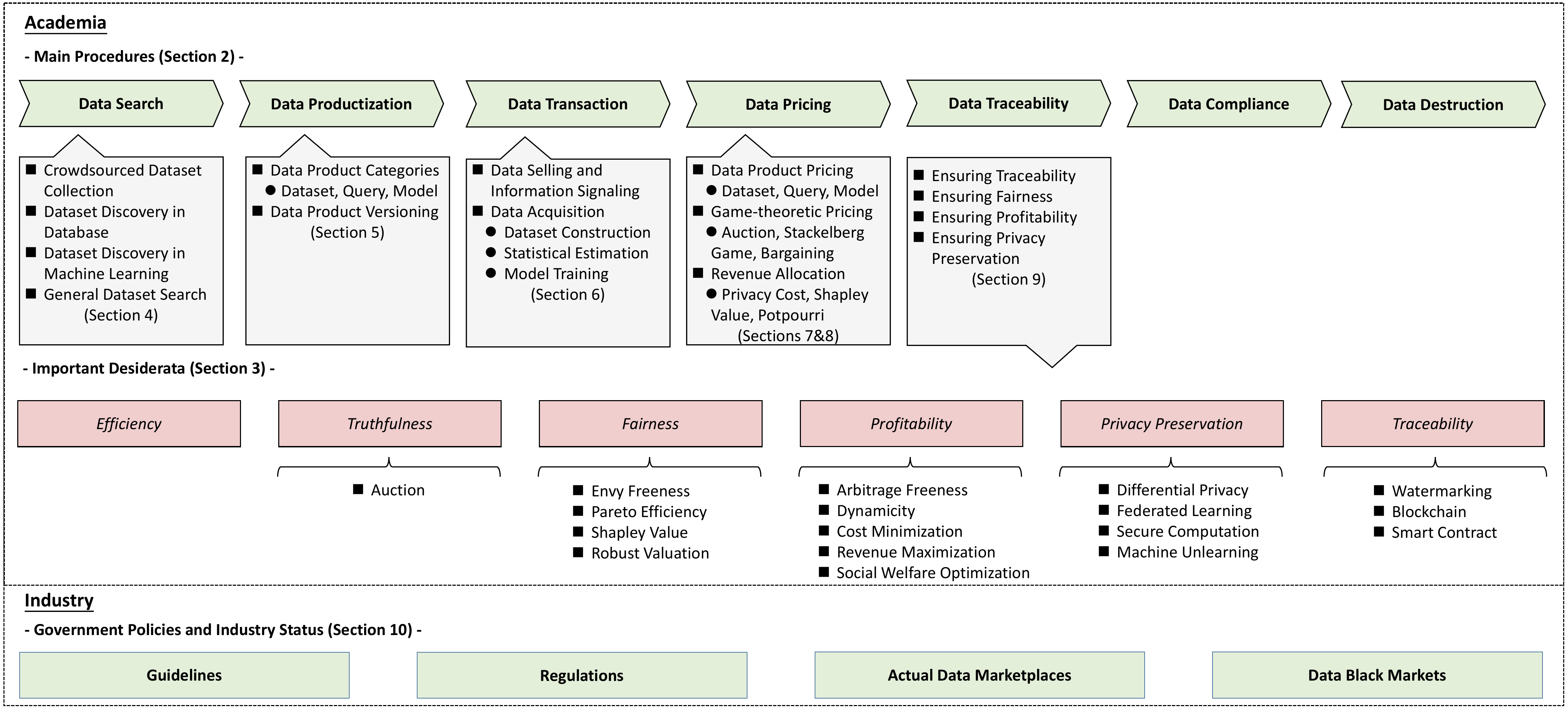}
 \caption{The structure of the survey.}
 \label{fig:structure}
\end{figure}

\subsection{Structure of The Survey}

Figure \ref{fig:structure} shows the structure of the survey. Section \ref{sec:architecture} first presents a diagram that shows different problems affecting data markets and their relationship. Section \ref{sec:desiderata} summarizes important desiderata for building well-functioning data markets. Section \ref{sec:search} reviews methods of data search including crowdsourced dataset collection, dataset discovery in databases, dataset discovery in machine learning, and general dataset search. Section \ref{sec:productization} describes techniques for data productization. Section \ref{sec:transaction} overviews strategies of both data buyers and data sellers in data transactions. Section \ref{sec:pricing} reviews studies investigating data product pricing and game-theoretic pricing. Section \ref{sec:revenueallocation} reviews studies investigating revenue allocation. Section \ref{sec:untrusted} deals with issues related to dishonest participants in untrusted data markets. Section \ref{sec:status} provides government policies and industry status in representative countries and regions. Section \ref{sec:conclusion} draws a conclusion and discusses open challenges and opportunities for future work. 

\section{Framework of Data Markets}\label{sec:architecture}
In this section, we first show the differences between data and other production factors in Section \ref{subsec:difference}. We then introduce key entities in data markets and their interactions in Section \ref{subsec:key-entities}. Finally, we describe seven main procedures to illustrate how data markets operate in Section \ref{subsec:mainprocedures}.

\subsection{The Differences Between Data and Other Production Factors}\label{subsec:difference}

Production factors, also known as factors of production, are the essential inputs required to produce goods and services. These factors play a key role in the production process by contributing to the creation of value. The classic economic classification of production factors includes land, labor, capital, and entrepreneurship \cite{smith1937wealth, schumpeter2021theory}. Differences in various properties between data and other production factors (land, labor, capital, and entrepreneurship) are listed in Table \ref{tab:difference}, which provides insights into the burgeoning data markets. Land refers to the natural resources used to create a good or service, e.g., forests and oil. Labor is the work done by the people in a workforce. Capital refers to capital goods or man-made resources such as tools and infrastructure while entrepreneurship represents innovation and technology used in the production of a good or service. The characteristics of data make data markets more complicated than other markets. 

 \begin{table}[h]
    \centering
    \caption{Difference between data and other production factors.}\label{tab:difference}
    \resizebox{\linewidth}{!}{
    \begin{tabular}{llccccc}
        \hline
            \multicolumn{1}{c}{Property} &
            \multicolumn{1}{c}{Description} &
            \multicolumn{1}{c}{Land} & \multicolumn{1}{c}{Labor} & \multicolumn{1}{c}{Capital}& \multicolumn{1}{c}{Entrepreneurship}& \multicolumn{1}{c}{Data}\\
        \hline
            Replicability     & can be replicated (almost) without cost & 
            \xmark & \xmark & \xmark & \cmark & \cmark \\
            Non-rivalry     & can be used by multiple parties simultaneously & 
            \xmark & \xmark & \xmark & \cmark & \cmark \\
            Externality       & indirect cost or benefit arising from others' activity & 
            \cmark & \cmark & \cmark  & \cmark & \cmark \\
            Composability     & can be integrated to achieve higher value & 
            \cmark & \cmark & \cmark & \cmark & \cmark \\
            Divisibility  & can be feasibly divided & 
            \cmark & \xmark & \xmark & \xmark & \cmark \\
            Heterogeneity & unit element has different value & 
            \cmark & \cmark & \cmark & \cmark & \cmark \\
            Persistence & is reusable and not expendable & 
            \cmark & \xmark & \xmark & \cmark & \cmark \\
            Natural Growth & can grow without extra artificial effort & 
            \xmark & \xmark & \xmark & \xmark & \cmark \\
            Physical Object  & is a physical object & 
            \cmark & \xmark & \cmark & \xmark & \xmark \\
            Tradability       & can be traded       & 
            \cmark & \cmark & \cmark & \cmark & \cmark \\
            High Liquidity  & can be easily converted into cash in a mature market & 
            \cmark & \cmark & \cmark & \cmark & \xmark \\
        \hline
    \end{tabular}}
 \end{table}
\partitle{Concerning replication} 
One of the most significant properties of data is the \emph{replicability}. Data, once obtained by someone, can be easily replicated almost without cost, which is the same as entrepreneurship, while land, labor, and capital cannot be reproduced unlimitedly. Research on the challenges brought by the replicability of data are summarized in Section \ref{subsec:fairness}, along with the corresponding techniques including data replication attacks and robust-to-replication data pricing. The replicability of data leads to another property, \emph{non-rivalry}, which means that multiple parties can share the usage of the same data at the same time. The replicability of data makes the externality especially prominent in data markets \cite{DBLP:journals/corr/abs-2302-08012} while externality exists in other factors as well. The \emph{externality} of data suggests an indirect impact on the data value for a data user when other users have the same data. The externality can be either positive or negative. A positive externality arises in cooperation because data users specializing in distinctive fields can benefit from each other due to the insights from data, e.g., co-workers in the same product line. A negative externality arises in competition because the use of data confers on a given user a competitive advantage that hinders the performance of its competitors. The externality of data has been taken into account in data pricing as shown in Section \ref{sec:pricing}.


\partitle{Concerning utilization} 
In terms of utilization, several similarities and differences appear between data and other production factors. The \emph{composability} implies data can be combined for different uses which is in line with other factors, while the \emph{divisibility} differs data from labor, capital, and entrepreneurship, indicating the data can be easily divided for independent uses. In terms of value, the value of data extraordinarily varies in different tasks, making the \emph{heterogeneity} much more prominent for data than other factors. Data is also \emph{persistent}, which indicates that data can be reused and cannot be consumed, while labor and capital can be considered expendables, i.e., once used, never recycled. Besides persistence, data can even grow without artificial effort, e.g., weather data can be gathered naturally as time goes by, while entrepreneurship, e.g., technology and management expertise, must be created by human beings with mental labor. In other words, data can be constantly updated, which raises concerns about how to measure, acquire, and value data updates. This \emph{natural growth} property of data brings a greater challenge for data pricing and data acquisition, which will be further discussed in Sections \ref{sec:transaction} and \ref{sec:pricing}. While the properties of \emph{physical object} and \emph{tradability} can be easily understood, the lack of \emph{high liquidity} of data implies that currently data can hardly be turned into cash easily since there is no mature market system as shown in the review of industrial development of data markets in Section \ref{sec:status},  giving data low liquidity. 

\subsection{Key Entities}\label{subsec:key-entities}

\begin{figure}[htb]
 \centering
 \includegraphics[width=1\textwidth]{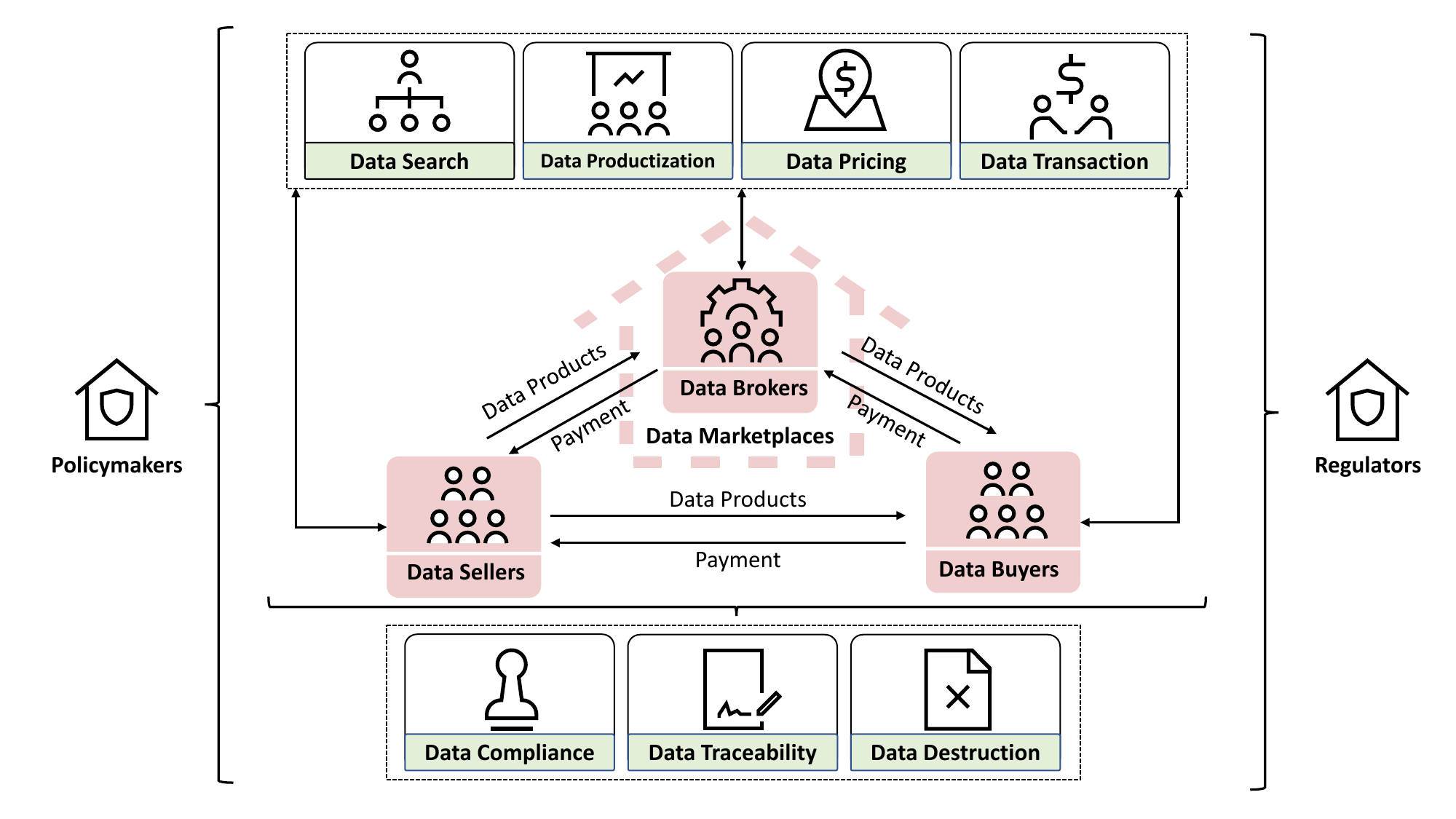}
 \caption{A framework of data markets.}
 \label{fig:framework}
\end{figure}

In this section, we introduce key entities in data markets and how they interact with others. Well-functioning data markets must bring together key entities with clear benefits for all.
The framework is illustrated in Figure \ref{fig:framework}.

An entity (an individual or organization) can play multiple roles—data owner, data buyer, data seller, and data broker—depending on the specific context and transaction they are involved in.

\partitle{Data owners} 
Data owners are individuals or organizations that have the ownership of data. They may sell their data, making them data sellers, or buy data from others, making them data buyers.

\partitle{Data sellers}
Data sellers are individuals or organizations that sell data products in a data transaction.

\partitle{Data buyers}
Data buyers are individuals or organizations that purchase data products in a data transaction.

\partitle{Data brokers}
Data brokers are individuals or organizations that arrange data transactions between data sellers and data buyers. Data brokers act as data buyers when interacting with data sellers and as data sellers when interacting with data buyers. It cannot be ignored that data brokers serve as agents, managing and negotiating data transactions on behalf of both data sellers and data buyers. A data broker is not necessary for a data transaction, as data sellers and buyers trade data products directly.

\partitle{Data marketplaces}
Data marketplaces are platforms that connect data buyers with data sellers and provide free or paid services for data transactions, such as data product searches and negotiation areas. In addition to providing fast and centralized market information, data marketplaces usually need to operate as a data management platform to ensure that the data trading process is trustworthy, fair, profitable, secure, traceable, and efficient. Data brokers are practitioners in data marketplaces and may retain a portion of transaction proceeds as profits.

\partitle{Policymakers and regulators}
Policymakers propose and establish guidelines and regulations for data transactions, while regulators oversee and enforce them. The two authorities play effective roles in ensuring that entities comply with norms in data markets.

\subsection{Main Procedures}\label{subsec:mainprocedures}
In this section, we introduce seven main data market procedures, which generally follow a sequence: data search, data transaction, data compliance, data pricing, data traceability, and data destruction. It is important to note that not all of these procedures will occur, nor do they necessarily follow a fixed order, and they may overlap or interact with each other. The data market begins with data search, where data sellers gather valuable data from multiple sources. The collected data is further productized to generate various data products. Data products that are identified as compliant with data laws will be priced and traded in marketplaces. Reaching from data search to data transaction, the source and transformation of data will be tracked and recorded for traceability. Finally, when the ownership of data is revoked, meaning the data seller chooses to withdraw the control over the data from the data buyer, the data buyer is required to destroy the corresponding data.

\paragraph*{Data Search}
Data search is the initial phase of data markets where data sellers search and gather data from a wide range of sources, integrate them, and ensure they are valuable and accurate to meet the needs of data buyers. Given the data proliferation and the heterogeneous value of data as shown in Section \ref{subsec:difference}, how to identify valuable data from plentiful data sources remains an important and challenging problem. \citet{DBLP:journals/vldb/ChapmanSKKIKG20} survey techniques and implementations of dataset search, including information retrieval, databases, entity-centric, and tabular search. \citet{DBLP:journals/tkde/RohHW21} survey subproblems of data aggregation for machine learning, including data acquisition, data labeling, and improving existing data. \citet{DBLP:journals/sj/VimercatiFLS21} discuss the issues, possible existing techniques, and challenges of enabling individuals to trade data using a data marketplace platform while maintaining control over it. While various types of data vendors offer integrated data sources \cite{DBLP:journals/sigmod/SchommSV13,DBLP:conf/sigmod/RekatsinasDDGS16}, inaccuracies in data attributes sold by data brokers have been identified. A study by \citet{DBLP:conf/www/VenkatadriSRMGM19} investigating Facebook advertising system and its partnership with six data brokers shows that more than $40\%$ of attributes sold by data brokers are not accurate. Data search methods are summarized from the data usage perspective in Section \ref{sec:search}.

\paragraph*{Data Productization}
Data productization is the process by which data sellers analyze the possible needs of data buyers, and then create sellable, standardized, repeatable, and comprehendible data products \cite{harkonen2015productisation}. It can be sold independently or integrated to achieve higher value (property ``composability'' in Table \ref{tab:difference}). Raw data can be sold directly or taken as materials to create new data products with more value. Thus, data productization offers an extensive variety of commodities derived from raw data such as web interfaces, query results, and machine learning models. In general, data sellers develop data products by actualizing a product differentiation strategy. They conduct market research and release different versions of data products tailored to current and potential data buyers with different needs and sold at different prices (versioning) \cite{shapiro6versioning}. Yet it presents technical challenges to versioning strategy and data pricing, such as maximizing revenue \cite{DBLP:journals/pvldb/LiuLL0PS21,DBLP:conf/soda/GuruswamiHKKKM05}, no-arbitrage pricing \cite{koutris2012query,DBLP:journals/jacm/KoutrisUBHS15}, and fine-grained pricing \cite{DBLP:journals/dss/LiR14}. We provide a detailed discussion of data productization and versioning strategies in Section \ref{sec:productization}.

\paragraph*{Data Transaction}
A data transaction is an agreement between a data seller and a data buyer to exchange data products in return for payment. The interactions between data sellers and buyers are central to achieving a data transaction. Since data buyers do not have exact prior value of data, data sellers can take strategic actions to shape the beliefs of data buyers for greater sales. Data sellers send signals of data products to promote the sale, which can include descriptions, schema, and pricing details. Data buyers choose cost-efficient data products according to their needs. \citet{DBLP:journals/pvldb/FernandezSF20} propose a research agenda of data market platforms to tackle the problems of data sharing, discovery, and integration in the data transaction. The challenges of data transactions lie in marketing, selling, and purchasing data products, as well as addressing issues related to discovery and integration, and the tradeoffs between monetary costs and quality. In contrast, the challenges of data search lie in integrating raw data from diverse sources with data quality assurance. While both processes involve discovery and integration, the fundamental difference lies in the nature of the data being processed, i.e., data products versus the raw data. Detailed analysis of problems in the data transaction is provided in Section \ref{sec:transaction}.


\paragraph*{Data Pricing}
Data pricing is to set prices for data products, which are influenced by interactions among all entities. The price and cost of a product are critical factors in calculating profitability \cite{harkonen2015productisation}. A well-designed data pricing system can widely facilitate data transactions. In markets selling tangible products, one can determine the minimum price once the costs are understood, where the costs can be easily divided into material costs, processing costs, etc. Nevertheless, the costs of data products are hard to decompose and clearly estimate. The manufacturing cost of a data product includes human expertise, computing resources, and other factors, which are challenging to evaluate in practice according to any theoretical model, while its replication cost quickly comes closer to zero. In other words, data products may have high fixed costs with low marginal costs. The traditional cost-based pricing and competition methods tend to drive prices of data products to the level of marginal costs, leaving data sellers unable to recover their up-front investments \cite{shapiro6versioning}. Moreover, data products are task-oriented, making their prices application-specific. Customer-perceived value differs for the same data product. To tackle these problems, many researchers have proposed data pricing strategies that are more value-based and personalized by incorporating economics or game theory, such as Shapley value \cite{shapley1953value,DBLP:conf/ec/AgarwalDS19, DBLP:journals/pvldb/JiaDWHGLZSS19, DBLP:journals/pvldb/LiuLL0PS21, DBLP:journals/pvldb/LiuLZR0L0PS21,DBLP:journals/tmc/ZhengPWTC20} and auctions \cite{roughgarden2010algorithmic,DBLP:journals/mor/Myerson81,10.5555/1109557.1109677,chen2018blockchain,bernhardt1994note,friedman2018double}. See details in Section \ref{sec:pricing}.

\paragraph*{Data Traceability} 
Data traceability describes where data comes from and how it is transformed, which is a way to guarantee the truthfulness of data search, data productization, data pricing, and data transactions. Data buyers face a pressing problem of how to verify whether the data sellers have truthfully provided data products. An opportunistic way for data sellers to reduce the expenditure for data acquisition is to mingle some falsified data into the raw dataset without notifying data buyers \cite{DBLP:conf/icde/NiuZWGC17}. In addition, malicious entities are likely to copy and transmit data products at almost zero cost, when the data market lacks a sound notion of authorship \cite{DBLP:journals/corr/abs-2201-04561}. Therefore, it is necessary to ensure data traceability, which can also mitigate the data quality problem and the revenue allocation problem. Detailed analysis of data traceability is provided in Section \ref{subsec:traceability}.

\paragraph*{Data Compliance}
Data compliance refers to the standards and regulations that all entities must adhere to in order to prevent sensitive data leakage, misuse, destruction, etc. Different types of data security laws have been laid out around the world, such as HIPAA \cite{HIPAA}, GDPR \cite{GDPR}, CPPA \cite{CCPA}, PCD-DSS \cite{pcidss}, and SOX \cite{sarbanes2002sarbanes}. The actors in data markets have to comply with these laws or face steep fines. See Sections \ref{subsec:guides-policies} and \ref{subsec:laws-regulations} for more guidelines and regulations. A rich body of studies delve into techniques for using data in compliance with data security laws and avoiding privacy violations, such as data desensitization \cite{DBLP:conf/icde/CastellanosZJRDDJ10}, differential privacy \cite{DBLP:conf/tcc/DworkMNS06,DBLP:conf/tamc/Dwork08,DBLP:journals/fttcs/DworkR14}, multi-party secure computation \cite{DBLP:conf/focs/Yao86,DBLP:conf/eurocrypt/FurukawaLNW17}, federated learning \cite{DBLP:journals/ftml/KairouzMABBBBCC21}, and machine unlearning \cite{DBLP:conf/sp/CaoY15}. See details in Section \ref{subsec:privacy}.

\paragraph*{Data Destruction}
Data destruction requires data buyers to destroy data as soon as it is no longer used or the data seller is no longer willing to share the data. It mainly includes the physical destruction of storage media or overwriting of storage contents. Recently, regulated by data privacy regulations (e.g., GDPR \cite{GDPR}, CCPA \cite{CCPA}, and PIPL \cite{PIPL}), the right to be forgotten becomes a part of the personal data protection standard. A data subject has the right to have their personal data erased from the controller under several specific circumstances. Data itself can exist persistently as pointed out in Section \ref{subsec:difference}, but data usage should be restricted. A new research direction coined “machine unlearning” focuses on making machine learning models comply with the right to be forgotten.

\section{Important Desiderata in Data Markets}\label{sec:desiderata}
In this section, we show the important desiderata that well-functioning data markets should enjoy. These include truthfulness, fairness, privacy preservation, and efficiency in all procedures, as well as profitability for data pricing, which should be free from arbitrage, dynamic, revenue maximization and cost minimization, and emphasizing social welfare optimization. Additionally, it is important to have traceability in data transactions to ensure transparency and accountability.

\subsection{Truthfulness}
The most important concern in data markets is truthfulness, which is also referred to as incentive compatibility (IC) \cite{roughgarden2010algorithmic} in economic theory and mechanism design. A data market is considered to be truthful when entities act truthfully and reveal their real thoughts to gain their maximum utility while selfish entities cannot benefit from lying. The desired property of truthfulness is supposed to hold in the interactions among all entities of data markets. It can convince data sellers that data brokers have truthfully used their data and convince data buyers that data brokers have truthfully produced and managed data products. Specifically in data pricing, truthfulness indicates that entities truthfully reveal their data valuations and bids. Researchers seek to design mechanisms to motivate entity behaviors to be truthful. Examples are shown as follows.

\begin{itemize}

\item 
Truthful report: \citet{DBLP:conf/nips/ChenSZ20} propose a pricing mechanism based on peer prediction where reporting true data is the only optimal way for the data sellers to maximize the expected payment. \citet{zheng2024truthful} use pointwise mutual information to guarantee that data providers always maximize their expected score by truthfully reporting their observed data.

\item
Truthful bidding: The Vickrey–Clarke–Groves (VCG) auction \cite{roughgarden2010algorithmic} incentivize bidders to bid their true valuations for multiple items by ensuring that truthfulness is an optimal strategy for each bidder, regardless of what other bidders do (a strong truthfulness guarantee known as  \emph{dominant strategy incentive compatibility}). The VCG auction has been extensively applied to data markets \cite{DBLP:conf/www/GrubenmannBMS18, DBLP:conf/sigecom/GhoshR11, wang2016strategy}. See details in Section \ref{subsec:gametheorypricing}.

\item
Truthful storage: A blockchain \cite{DBLP:journals/comsur/SalmanZEJS19} is a distributed data structure used to store immutable records of information, featuring decentralization, verifiability, and integrity. It has been applied to data markets \cite{chen2018blockchain,DBLP:conf/blockchain2/BajoudahDM19,DBLP:conf/coinco/ColmanCC19} to prevent fraudulent behaviors and enhance truthfulness.

\end{itemize}

\subsection{Fairness}
Fairness is one of the fundamental requirements for dividing resources or allocating revenue among participants. The universal notion of fairness can be tricky to formalize since it is natural that different phenomena call for different notions of fairness. A typical data market requires some of them. In this section, we review different definitions of fairness in two areas: resource division in Section \ref{subsubsec:resource-division} and revenue allocation in Section \ref{subsubsec:revenue-allocation}.

\subsubsection{Resource Division}\label{subsubsec:resource-division}
Fair resource division distributes data products to data buyers in a fair manner, typically in auction-style data transactions. There is a rich body of studies in economics, sociology, and mathematics on the problem of fair division, which could date back to \citet{fairdivision} in 1948. We review several multifaceted notions of fairness in the fair division. Given a set of entities $N = \{1,\ldots,n\}$ and a set of resources $O = \{o_1,\ldots,o_m\}$, an allocation $\Pi = (\pi_1,\ldots,\pi_n)$ is a $n$-partition of the set of resources $O$, where $\pi_i \subseteq O$ is a bundle of resources allocated to entity $i$. Utility measures the satisfaction or value that an entity derives from consuming or owning the resource. Let $u_{i}(\pi_j) \in \mathbb{R}_{\ge 0}$ be the utility of entity $i$ toward $\pi_j$ for the allocation $\Pi = (\pi_1,\ldots,\pi_n)$.
\begin{itemize}
    \item Envy freeness \cite{foley1966resource}: An allocation $\Pi$ is envy-free iff $\forall i,j \in N$, $u_{i}(\pi_i)\ge u_{i}(\pi_j)$. That is, each entity prefers its own allocation over that of any other entity.
    
    \item Pareto efficiency \cite{pareto1896new}: An allocation $\Pi$ is Pareto-efficient iff there is no allocation $\Pi' = (\pi'_1,\ldots,\pi'_n)$ such that $\forall i \in N$, $u_{i}(\pi'_i)\ge u_{i}(\pi_i)$ and $\exists j \in N$, $u_{j}(\pi'_j) > u_{j}(\pi_j)$. That is, no entity can be better off without making any other entity worse off or without any loss. An envy-free allocation is proven to be Pareto efficiency when the number of entities equals the number of indivisible resources and the utilities are quasi-linear \cite{svensson1983large}.

    \item Max-min fairness (also known as the Santa Claus problem) \cite{shoham2008multiagent}: An allocation $\Pi$ is max-min fair iff $\Pi = max_{\Pi}\ min_i u_i(\pi_i)$. Since it is somehow difficult to provide an exact notion of fairness or to find a mechanism that satisfies all desirable notions of fairness, max-min fairness is proposed in this case, which is defined as making the least-happy entity as happy as possible \cite{shoham2008multiagent}.

    \item Utilitarian social welfare \cite{gul1999walrasian}: An allocation $\Pi$ is utilitarian social welfare iff $\Pi = max_{\Pi} \sum_{i\in N} u_i(\pi_i)$. That is, utilitarian social welfare is to maximize the total sum of the utilities.

    \item Nash social welfare \cite{kaneko1979nash}: Given weight $w_i > 0$ of entity $i$, an allocation $\Pi$ is Nash social welfare iff $\Pi = max_{\Pi} \left(\prod_{i \in N} u_{i}\left(\pi_{i}\right)^{w_{i}}\right)^{\frac{1}{\sum_{i \in N} w_{i}}}$. \citet{DBLP:journals/teco/CaragiannisKMPS19} argue that Nash social welfare guarantees ``unreasonably'' fairness for allocating divisible resources. Nash social welfare is considered a trade-off between max-min fairness and utilitarian social welfare \cite{DBLP:conf/stoc/GargHV21}. A characteristic of Nash social welfare is its invariance under the scaling of utility functions.  Different from utilitarian social welfare and max-min fairness, the allocation of Nash social welfare remains unchanged when scaling the utility functions of the entities by arbitrary positive constants.
\end{itemize}


\subsubsection{Revenue Allocation}\label{subsubsec:revenue-allocation}
Fair revenue allocation distributes the jointly-produced revenue to the data sellers in the process of data pricing in a fair manner. Consider a set of $n$ players $\mathcal{N}=\{\bm{z}_1,\ldots,\bm{z}_n\}$. A \emph{coalition} is a subset of players $\mathcal{S} \subseteq \mathcal{N}$ that cooperate to complete a task. A utility function $\mathcal{U}(\mathcal{S})$ $(\mathcal{S} \subseteq \mathcal{N})$ evaluates the utility of a coalition $\mathcal{S}$ for a task. 
The \emph{marginal contribution} of $\bm{z}_i$ with respect to a coalition $\mathcal{S}$ is $\mathcal{U}(\mathcal{S}\cup \{\bm{z}_i\})-\mathcal{U}(\mathcal{S})$. The Shapley value measures the expectation of marginal contributions by $\bm{z}_i$ in all possible coalitions. That is,
\begin{equation}\nonumber\label{equ:svp}
  \mathcal{SV}_i=\frac{1}{n} \sum_{\mathcal{S}\subseteq \mathcal{N} \setminus \{\bm{z}_i\}}\frac{\mathcal{U}(\mathcal{S}\cup \{\bm{z}_i\})-\mathcal{U}(\mathcal{S})}{\binom{n-1}{|\mathcal{S}|}}.
\end{equation}
The Shapley value is a fair revenue distribution in the sense that it is the only measure satisfying all four fundamental requirements of balance, symmetry, additivity, and zero element.
\begin{itemize}
    \item {Balance} requires that the total revenue should be fully allocated to all players.
    \item {Symmetry} specifies that two players should receive the same revenue if they have the same marginal contributions.
    \item {Additivity} indicates that the sum of revenue on two individual tasks should be the revenue on their combined task.
    \item {Zero element} specifies that a player should not be rewarded anything if the player does not make any marginal contribution, i.e., no contribution, no payment.
\end{itemize}
More recently, the Shapley value has been used to estimate data value and allocate revenue among data sellers \cite{DBLP:journals/pvldb/JiaDWHGLZSS19,DBLP:journals/pvldb/LiuLL0PS21,DBLP:conf/ec/AgarwalDS19}. Several other concepts in cooperative game theory also consider how to allocate revenue generated by the grand coalition to its players. 
Among them, the Banzhaf value and the core have been studied for data valuation \cite{DBLP:conf/aaai/YanP21,DBLP:conf/aistats/WangJ23}.
\begin{itemize}
    \item The Banzhaf value \cite{banzhaf1964weighted}: 
     The Banzhaf value of player $\bm{z}_i$ is defined as 
     $\mathcal{BV}_i=\frac{1}{2^{n-1}} \sum_{\mathcal{S}\subseteq \mathcal{N} \setminus \{\bm{z}_i\}}[\mathcal{U}(\mathcal{S}\cup \{\bm{z}_i\})-\mathcal{U}(\mathcal{S})]$.
    The Banzhaf value, as well as the Shapley value, are two well-known instances of generalized semivalues.
    Unlike the Shapley value, the Banzhaf value satisfies symmetry, zero element, and additivity, except for balance. The Banzhaf value measures the influence or power of each player in achieving a certain revenue, regardless of the coalition size.
    
    \item The core \cite{edgeworth1881mathematical}: Let an allocation $x: N \rightarrow R_{+}$ represent the payoff distributed to the players in $\mathcal{N}$ with $x(\mathcal{N})=\mathcal{U}(\mathcal{N})$. The excess is defined by $e(\mathcal{S},x)=x(\mathcal{S})-\mathcal{U}(\mathcal{S})$, where $x(\mathcal{S})=\sum_{\bm{z}_i\in\mathcal{S}}x_i$. The core is the set of all allocations whose excesses of all coalitions are non-negative.
    
    \item Nucleolus \cite{schmeidler1969nucleolus}: The allocation $x$ is called an imputation if $x_i\ge \mathcal{U}(\{\bm{z}_i\})$ holds for all $\bm{z}_i \in \mathcal{N}$. For an imputation, sort excesses of all coalitions in non-decreasing order in a vector to obtain an excess vector. 
    The nucleolus is the imputation that has the excess vector of maximum lexicographical order. The nucleolus always exists and is a member of the core when the core is non-empty. 
\end{itemize}

\subsection{Profitability}
Profitability refers to the ability to generate profit. It is challenging to make accurate and profitable market decisions. Below we analyze and list several properties that a profitable strategy might have. 
Arbitrage freeness, dynamicity, revenue maximization, and cost minimization are desired for creating and pricing data products from the perspective of data sellers while social welfare optimization is preferred from the perspective of the whole society.

\subsubsection{Arbitrage Freeness}
Arbitrage is the activity of buyers taking advantage of price differences to buy something else at a lower price and using this to infer something that they would have to pay a higher price for.
For example, suppose that the price of query ``$age > 50$'' is set to $\$100$, and the price of query ``$age > 30$'' is only set to $\$80$. Sophisticated buyers can purchase query ``$age > 30$'' and then filter out data that satisfies ``$50 \ge age > 30$'', and finally get the result of query ``$age > 50$'' at the price of $\$80$ to achieve arbitrage. 

Data sellers provide multiple data products with different characteristics to satisfy the various needs of data buyers. They should carefully set the price for each data product to anticipate and block buyers from engaging in arbitrage. Speculators would usually act quickly if an arbitrage opportunity exists, thereby costing sellers a loss of revenue. In the absence of arbitrage, data buyers can have confidence that prices are fair and reflect their actual value, which can make the decision-making process easier and faster. Thus, providing arbitrage-free pricing is highly desirable in data markets. Naturally, setting a uniform price for all data products makes no arbitrage in some cases but comes with limitations in increasing revenue. A differential pricing model is preferred to achieve revenue maximization but comes with a much higher computational cost.

\subsubsection{Dynamicity}
Dynamicity comes from the inherent dynamism of data products as data is updated over time. Data sellers may face the issue of adjusting the price of data products over time in response to various factors. This includes the enhanced or decreased value of data as it gets updated or shifts occur in data buyer demand, and charges for overlapping information according to the transaction history of data buyers.

\subsubsection{Revenue Maximization}
Data sellers make market decisions from the perspective of maximizing revenue.  Versioning provides a practical way to price nuanced data products differently without incurring high costs or offending customers. With versioning, data sellers aim to maximize overall revenue by getting each customer to pay the highest possible price for its favorite product. Therefore, for revenue maximization, sellers face a fundamental problem of how to create the right number of versions and set the right prices for them \cite{shapiro6versioning,DBLP:journals/pvldb/LiuLL0PS21}.

\subsubsection{Cost Minimization}
Achieving cost minimization is a proper choice of competitive strategy for data sellers. A cost leadership strategy is the act of establishing a competitive advantage by producing goods with customer-acceptable characteristics at the lowest cost relative to competitors \cite{stahl1997strategic}. The data seller collects data and offers products with the goal of minimizing total payouts while achieving a target (such as model accuracy and information entropy) or maximizing a target with a fixed total budget, which can always be formulated as an optimization problem \cite{DBLP:journals/pvldb/LiuLL0PS21}.

\subsubsection{Social Welfare Optimization}
As mentioned before, (utilitarian) social welfare refers to the total sum of the utilities of all agents. Taking production cost into account, social welfare, also known as social surplus, can be considered as the ``profit'' of the whole society, the difference between total consumption utility and production cost. Specifically, the social welfare in data markets can be evaluated by adding up the profit of every participant including both data sellers and data buyers, or by measuring the increase in the value of data with its ownership transferred from data sellers to data buyers. Social welfare optimization, which suggests that the value of data can be fully exploited, is desired in data markets and can be enforced through data pricing and allocation mechanisms. For example, besides truthfulness, another advantage of VCG mechanism \cite{roughgarden2010algorithmic} is the guarantee of social welfare maximization, which will be introduced in Section \ref{sec:pricing}.

\subsection{Privacy Preservation}
Privacy preservation involves protecting the privacy of all participants and potentially ensuring that data owners are properly compensated as an exchange for their private data. Privacy is defined as ``the claim of individuals, groups, or institutions to determine for themselves when, how, and to what extent information about them is communicated'' by \citet{westin1968privacy}. Usually, users are reluctant to reveal sensitive information, such as their real identities, browsing history, and trading records. From an economic perspective, privacy protects users from price discrimination \cite{DBLP:conf/webist/MakrisPS16}. From a computer science perspective, privacy preservation is to protect the data from being inferred and abused by potential attackers and malicious parties.

While data owners value their privacy, they may be open to data exchanges for economic benefits or other indirect benefits, e.g., better services (contributing data for training recommendation model \cite{DBLP:conf/icml/SundararajanK23}) or social good (data for training better medical diagnosis models). Privacy can be considered a commodity and traded naturally as personal data is traded. Data owners have the right to determine how much information is divulged in a regulated data marketplace \cite{DBLP:journals/cacm/Laudon96}. The question for data buyers is how to properly measure the level of privacy loss for data owners and compensate them. For example, \citet{DBLP:journals/pvldb/LiuLL0PS21} propose a pricing approach that incorporates not only the quality of data but also the privacy breaches measured by differential privacy. Differential privacy introduced by \citet{DBLP:conf/tcc/DworkMNS06} provides a rigorous, quantitative way to reason about the privacy costs \cite{DBLP:journals/sigecom/PaiR13}.

\begin{definition}(Differential Privacy \cite{DBLP:conf/tcc/DworkMNS06})\label{def:DP}
A randomized algorithm $\mathcal{A}$ is $(\epsilon,\delta)$-differential private if for any pair of datasets, $\mathcal{S}$ and $\mathcal{S}'$ that differs in one data sample, and for all possible output $\mathcal{OUT}$ of $\mathcal{A}$, the following holds,
\begin{equation}\nonumber
    \mathbb{P}[\mathcal{A}(\mathcal{S}) \in \mathcal{OUT}] \leq e^{\epsilon} \mathbb{P}\left[\mathcal{A}\left(\mathcal{S}^{\prime}\right) \in \mathcal{OUT}\right]+\delta,
\end{equation}
where the probability is taken over the randomness of $\mathcal{A}$. $\epsilon$ captures the degree of indistinguishability, which intuitively can be seen as the upper bound of the privacy loss for each data sample. The smaller $\epsilon$ is, the more indistinguishable the outputs are, and the less the privacy loss is. The other parameter $\delta$ captures the probability that the privacy loss is out of the upper bound. Given a specified $\epsilon$, the closer $\delta$ is to 0, the less the probability of the privacy breach is. A key desideratum is making data products derived from the raw data to ensure differential privacy and properly compensating data owners according to the level of differential privacy.
\end{definition}

\subsection{Traceability}
Traceability implies the capability to trace the entire lifecycle of data products from productization, pricing, and transaction to destruction. Traceability describes where data comes from and how it is transformed, which is a way to guarantee truthfulness. In data markets, traceability can be used for two purposes. First, it would enable different entities to access and identify the source inputs and processing schemes of a data product, and audit its completeness and compliance using the recorded identification. Second, it would enable different entities to track data ownership and verify the trade legitimacy. Multiple techniques can be useful to provide traceability, such as blockchain and watermarks. See details in Section \ref{subsec:traceability}.

\subsection{Efficiency}
Efficiency requires that all mechanisms in data markets are completed within tractable computation time and communication costs. The difficulty for researchers in designing new mechanisms is whether there is an inherent conflict or interplay between efficiency and other desiderata. For example, the Shapley value is widely adopted to allocate revenue thanks to its fairness property. However, computing the exact Shapley value in general cases is an \#P-hard problem \cite{DBLP:journals/mor/DengP94}. There exists a clash between the two desiderata of fairness and efficiency when employing the Shapley value, thus fairness comes at the expense of reduced efficiency. See details in Section \ref{subsubsec:Shapley-Value-Based-Revenue-Allocation}.

\section{Data Search}\label{sec:search}

Data search for sellers refers to discovering and gathering data from personal or institutional databases and external sources such as open data repositories. Several approaches have been developed to explore the vast number of data sources. These approaches can be generally categorized based on the purposes: (1) crowdsourced dataset collection for generating new datasets by crowdsourcing in Section \ref{subsec:crowdsourced}; (2) dataset discovery for database to find datasets with either joinable attributes to increase the number of attributes or unionable attributes to increase the number of records in Section \ref{subsec:join}; (3) dataset discovery in machine learning for enriching training datasets by attribute expansion or record augmentation to improve the model performance in Section \ref{subsec:modeldata}; and (4) general dataset search for finding relevant datasets based on querying in Section \ref{subsec:search}.

\subsection{Crowdsourced Dataset Collection}\label{subsec:crowdsourced}
Crowdsourced dataset collection publishes data collection tasks on the platform and collects answers from the crowd to generate new datasets \cite{DBLP:conf/sigmod/LiZFWC17,DBLP:journals/tkde/FanWZYD19,DBLP:journals/tkde/ShanM0C0Z20,DBLP:conf/sigmod/ParkW14,DBLP:conf/sigmod/ChaiLLDF16}. \citet{DBLP:conf/sigmod/ParkW14} present a crowdsourced data collection system, CrowdFill, where multiple workers cooperate to fill in an evolving table. Workers can also upvote and downvote data entered by others, and these votes are aggregated to determine whether the row is acceptable. All operations by workers are immediately sent to a central server which resolves conflicts arising from concurrent operations and updates the table displayed to all workers. Specifically, CrowdFill has a component called Central Client (CC) that adds rows to the table, ensuring that the final table satisfy constraints. Once workers execute operations, CC computes a maximum bipartite matching for a bipartite graph that represents the relation between template rows and probable rows, and inserts probable rows into the candidate table. \citet{DBLP:journals/tkde/ShanM0C0Z20} use crowdsourcing to fill missing values in a given tabular relation. In particular, they consider heterogeneity and dependencies in the truth inference process by maximizing the likelihood of worker answers.

The above works \cite{DBLP:conf/sigmod/ParkW14,DBLP:journals/tkde/ShanM0C0Z20} do not specify any requirements for the distribution of datasets. To address the gap, \citet{DBLP:journals/tkde/FanWZYD19} present a method for distribution-aware crowdsourced entity collection. The main objective of this task is to minimize the difference (Kullback-Leibler divergence) between the distribution of collected entities and a given distribution on an attribute when using crowdsourcing to collect entities. To achieve this,  their proposed method involves an iterative worker selection process and entity distribution estimation. The underlying entity distribution of workers is dynamically estimated based on their history of collected entities, and then the optimal set of workers is adaptively selected.

\subsection{Dataset Discovery in Database}\label{subsec:join}
Dataset discovery in database receives a dataset as input, uses signatures to summarize data sources in a concise manner, measures and indexes their similarities, and finds datasets that can be joined or unioned with the input dataset \cite{DBLP:conf/sigmod/FernandezAMS16}. 
Approaches can be generally categorized into matching-based \cite{DBLP:conf/sigmod/ZhuDNM19,DBLP:conf/icde/DongT0O21,DBLP:conf/icde/KoutrasSIPBFLBK21,DBLP:conf/sigmod/SarmaFGHLWXY12}, hashing–based \cite{DBLP:journals/pvldb/Miller18,DBLP:journals/pvldb/ZhuPNM17,DBLP:journals/pvldb/ZhuNPM16,DBLP:journals/pvldb/NargesianZPM18,DBLP:conf/icde/FernandezMNM19,DBLP:journals/pvldb/CasteloRSBCF21,DBLP:conf/icde/SantosBMF22}, and learning-based \cite{DBLP:journals/pvldb/FanWLZM23,DBLP:journals/pvldb/Dong0NEO23, DBLP:journals/corr/abs-2403-07653}.

\partitle{Matching-based dataset discovery}
\citet{DBLP:conf/sigmod/ZhuDNM19} find joinable tables in data lakes by solving a top-$k$ overlap set similarity search problem, where the joinability between columns is measured by the number of overlapping (matched) records. However, this method shows limitations in handling more complex scenarios such as tables with heterogeneous data. To address the issue, \citet{DBLP:conf/icde/DongT0O21} embed columns into high-dimensional vectors by a pre-trained model and utilize pivot-based filtering to enhance the identification of joinable tables. Schema matching, a core component of dataset discovery \cite{DBLP:conf/sigmod/SarmaFGHLWXY12}, has seen significant evaluations through a unified, scalable, and open-source experiment by \citet{DBLP:conf/icde/KoutrasSIPBFLBK21}. They assess various existing schema matching algorithms \cite{DBLP:conf/vldb/MadhavanBR01, DBLP:conf/icde/MelnikGR02,DBLP:conf/vldb/DoR02,DBLP:conf/sigmod/ZhangHOPS11,DBLP:conf/icde/FernandezMQEIMO18,DBLP:conf/sigmod/CappuzzoPT20}, providing a comprehensive overview. Besides schema matching, entity matching \cite{DBLP:journals/vldb/HuangHBCQ23, DBLP:journals/pvldb/PaganelliBBG22, DBLP:journals/pvldb/WangZ0YMZG22,DBLP:conf/ijcai/WangLFH0XCLZ22,DBLP:conf/sigir/GeZCG22,DBLP:journals/pvldb/JainSS21,DBLP:journals/pvldb/PeetersB21,DBLP:journals/pvldb/Thirumuruganathan21,DBLP:journals/pvldb/WuSLCH21,DBLP:conf/icde/WangZWP21,DBLP:conf/www/ChenSZ21,DBLP:conf/sigmod/Miao0021,DBLP:conf/www/MaoWWL21} and functional dependency analysis \cite{DBLP:conf/cikm/AbedjanSN14dfd, DBLP:journals/cj/HuhtalaKPT99tane, DBLP:conf/icdt/NovelliC01fun, DBLP:conf/icdm/YaoHB02fdmine,DBLP:conf/cikm/BleifussBFRW0PN16aidfd, DBLP:journals/tcs/KivinenM95,DBLP:journals/pvldb/0001N18pyro,DBLP:conf/sigmod/PapenbrockN16hyfd,DBLP:conf/edbt/LopesPL00depminer,DBLP:conf/dawak/WyssGR01fastfds,DBLP:conf/icde/LinGSL00WPWL23} can help find datasets that can be joined or unioned. By identifying common entities across datasets or analyzing functional dependencies between attributes, one can determine how to join or merge datasets into a larger, more comprehensive dataset. \citet{DBLP:conf/icde/BogatuFP020} model the relationship between datasets as a weighted graph and discover the join path by traversing the graph in depth-first order. They aggregate five types of criteria to measure the similarities between datasets when computing the weight of edges. Additionally, \citet{DBLP:conf/icde/GongZGF23} focus on the problem of view discovery in large data repositories without any join information. They describe five challenges: noisy queries, noisy join paths, large number of join paths, noisy result views, and result view navigation. To address these challenges, they identify project-join views in collections of tables that do not possess any pre-defined pathways for joins. They further categorize views for filtering and adopt a bandit-based learning algorithm designed to adapt to user-specific preferences in view discovery, thereby aiding in the identification of optimal views among a multitude of options.


\partitle{Hashing–based dataset discovery}
Hashing has been widely used since it mitigates the problem of all-pairs similarity estimation between datasets \cite{DBLP:journals/pvldb/Miller18}. One of the well-known hashing techniques for indexing high-dimensional data is Locality Sensitive Hashing (LSH), which allows for fast approximate nearest neighbor search \cite{DBLP:conf/stoc/IndykM98}. \citet{DBLP:journals/pvldb/ZhuPNM17} propose an interactive navigation system for searching datasets that can join with the dataset provided by the user. The system uses LSH Ensemble proposed by \citet{DBLP:journals/pvldb/ZhuNPM16} to find data linkages between the datasets efficiently. LSH Ensemble \cite{DBLP:journals/pvldb/ZhuNPM16} partitions the domains based on their cardinality and constructs an LSH index for Jaccard similarity search for each partition. It supports fast domain search queries. Similarly, \citet{DBLP:journals/pvldb/NargesianZPM18} propose a set of probabilistic metrics to assess the unionability of attributes and use LSH to expedite the process of efficient attribute unionability search. Lazo proposed by \citet{DBLP:conf/icde/FernandezMNM19} has the advantage of being able to simultaneously estimate containment and similarity while standard LSH cannot. Lazo achieves this by redefining Jaccard similarity based on the cardinality of sets. \citet{DBLP:journals/pvldb/CasteloRSBCF21} demonstrate a specific search engine for datasets building on a general-purpose search engine Elasticsearch \cite{Elasticsearch} and Lazo \cite{DBLP:conf/icde/FernandezMNM19}. In addition, \citet{DBLP:conf/icde/SantosBMF22} propose a new hashing scheme based on Quadrant Count Ratio \cite{holmes2001correlation} and find joinable tables with the highest number of common hashed terms.

\partitle{Learning-based data discovery}
Learning approaches are introduced in data discovery. Traditional methods such as matching and hashing can be enhanced with machine learning techniques. To discover unionable tables in data lakes, a contrastive learning framework is developed by \citet{DBLP:journals/pvldb/FanWLZM23} to train column encoders from pre-trained language models (PLMs) without requiring labeled data. Column encoders produce column representations that facilitate the identification of datasets semantically similar to the target datasets. To discover joinable tables in data lakes, \citet{DBLP:journals/pvldb/Dong0NEO23} fine-tune PLMs (DistilBERT and MPNet) to generate column embeddings. Specifically, columns are converted into text sequences incorporating its metadata, which are input into PLMs to generate embeddings. Based on these embeddings, joinable columns close to a given query column are discovered through an approximate nearest neighbor search algorithm. Differently, \citet{DBLP:journals/corr/abs-2403-07653} employ Graph Neural Networks (GNNs) to predict joinablity among column pairs. They generate training data for GNNs by simulating joins and non-joins in a similarity graph where nodes represent columns, and edges denote similarities calculated by several similarity metrics.

\subsection{Dataset Discovery in Machine Learning}\label{subsec:modeldata}
Dataset discovery in machine learning focuses on finding datasets that can be utilized in learning pipelines to augment training data, ultimately aiming to enhance model performance \cite{DBLP:journals/pvldb/ChaiLTLL22,DBLP:conf/sigmod/KumarNPZ16,DBLP:journals/pvldb/ShahKZ17,DBLP:journals/pvldb/ChepurkoMZFKK20,DBLP:journals/pvldb/0001J23,wangoptimizing,DBLP:journals/corr/abs-2311-13712}. Approaches \cite{DBLP:conf/sigmod/KumarNPZ16,DBLP:journals/pvldb/ShahKZ17,DBLP:journals/pvldb/ChepurkoMZFKK20} join a base table with related tables to incorporate more features. However, such join operation can introduce redundancy and correlation between features and reduce model performance. Thus, \citet{DBLP:conf/sigmod/KumarNPZ16} and \citet{DBLP:journals/pvldb/ShahKZ17} explore the potential benefits of joining tables to improve the accuracy of the classifier in terms of schema and VC dimensions. They only exclude joins from a small subset of tables, which leads to the inclusion of irrelevant features and adds noise to the model. To address the issue, \citet{DBLP:journals/pvldb/ChepurkoMZFKK20} propose an automated system for relational data augmentation called ARDA, which can be used in automated machine learning pipelines. ARDA uses an external data discovery system \cite{DBLP:conf/icde/FernandezAKYMS18} to obtain candidate joinable tables, constructs a coreset by sampling and sketching to represent the original table, and joins the candidate tables with the coreset. To remove a large number of irrelevant features, ARDA combines feature rankings obtained from random forests and sparse regression using a reweighting strategy and employs a modified exponential search \cite{DBLP:journals/ipl/BentleyY76}.

Instead of incorporating more features by joining datasets \cite{DBLP:conf/sigmod/KumarNPZ16,DBLP:journals/pvldb/ShahKZ17,DBLP:journals/pvldb/ChepurkoMZFKK20,DBLP:conf/nips/KangJSJ23,DBLP:conf/nips/MahmoodLAFL22}, an alternative approach to enhance machine learning models is by merging more datasets to incorporate more data tuples. \citet{DBLP:journals/pvldb/ChaiLTLL22} demonstrate this method by merging data tuples from data sources in the wild. Datasets in the wild can be tables or images and may follow diverse distributions. They first use existing data discovery tools to discover relevant datasets from various sources to generate candidate datasets such as Google and Azure, then cluster all data points from candidate datasets using a multivariate Gaussian mixture model. Finally, they iteratively sample data points from a cluster picked by a multi-armed bandit-based solution and a Deep Q Networks-based reinforcement learning solution. The reward given to the picked cluster is set as the change in the model performance evaluated on a validation dataset after adding the sample to the current training dataset. Besides, a down-weighted reward will be given to neighboring clusters of the picked cluster. \citet{wangoptimizing} enhance the method \cite{DBLP:journals/pvldb/ChaiLTLL22} from three aspects: model training, cluster score estimation, and cluster sampling. Specifically, they employ online learning \cite{DBLP:conf/colt/McMahanS10} to reduce retraining overhead across multiple iterations. In each training iteration, they adaptively sample data from clusters based on the proportion of the cluster scores to the sum of all scores, instead of only sampling from one cluster. Model performance improvement is assigned to clusters in the form of Shapley values as rewards. These rewards are then used to compute adaptive scores for each cluster via adaptive mean estimation \cite{DBLP:journals/sac/BodenhamA17}. \citet{DBLP:conf/nips/MahmoodLAFL22} consider the costs of data collection and aim to train the model that achieves a performance target at minimal costs. Their proposed method uses bootstrap sampling to learn the data requirement distribution and then employs stochastic optimization to determine the amount of data to collect, factoring in the costs and penalties of not meeting the target. The process iterates through rounds of data collection and model evaluation until the target is reached or time runs out. In scenarios where data providers offer only a limited number of samples publicly, \citet{DBLP:conf/nips/KangJSJ23} introduce a data selection framework called projektor. Projektor first estimates model performance using the Optimal Transport distance between a validation dataset and a composition of the available data samples. It then extrapolates the performance estimation to larger data sizes by leveraging scaling laws. Based on the fitted model performance curve that incorporates data sizes and compositions, projektor calculates the optimal proportion of data selection among data sources by a gradient-based method.

There are also many data selection methods, such as active learning, coreset selection, and data augmentation, which offer useful tools for data markets. Active learning \cite{DBLP:journals/csur/LiuWRHZ22,DBLP:journals/csur/RenXCHLGCW22,DBLP:conf/uai/ZhengWLC18} involves iteratively selecting samples for annotation by an oracle, and using these samples to update the model. Alternatively, coreset selection \cite{DBLP:conf/iclr/ColemanYMMBLLZ20,DBLP:conf/icml/MirzasoleimanBL20,DBLP:journals/pvldb/WangC0L022} involves selecting a representative subset of the data that preserves the performance of the original dataset to speeding up training time. Active learning and coreset selection allow for more efficient use of data in machine learning. They can be used in conjunction with data augmentation \cite{DBLP:journals/pvldb/LiuZCL021, DBLP:conf/icde/LiuCLLFT22,DBLP:conf/sigmod/ZhaoF22,DBLP:conf/icde/GalhotraGF23,DBLP:conf/iclr/ZhouK23,DBLP:conf/kdd/ZhangZSKK22,DBLP:conf/icml/ShenBG22,DBLP:conf/iclr/Liu0SZ0SW23,DBLP:journals/pr/JiaZ20} to further improve model performance.


\subsection{General Dataset Search}\label{subsec:search}
General dataset search involves developing search engines to find datasets that match user-supplied keywords, datasets, or questions, and to rank and return them in order of relevance \cite{DBLP:conf/sigmod/HalevyKNOPRW16,DBLP:conf/www/ZhangB18,DBLP:conf/sigir/ChenTHX020,DBLP:conf/www/BrickleyBN19,DBLP:conf/dmah-ws/RezigVGPCS20,DBLP:journals/pvldb/RezigBFPVGS21,DBLP:journals/pvldb/YangWSP22,DBLP:conf/icde/FernandezAKYMS18,DBLP:journals/pvldb/EltabakhKEA23,DBLP:journals/pacmmod/WangF23,DBLP:conf/sigir/00600CPS21,  DBLP:conf/edbt/IonescuAIPPC0A023,DBLP:journals/tkde/WangCPKQ23,DBLP:journals/pvldb/OuelletteSNBZPM21,DBLP:conf/sigmod/NargesianPZBM20}. They do not make assumptions about the purpose of datasets and therefore can also handle aforementioned discovery tasks, including joinability discovery, unionability discovery, and machine learning augmentation. 

Several studies have developed systems that support keyword search \cite{DBLP:conf/sigmod/HalevyKNOPRW16,DBLP:conf/www/ZhangB18,DBLP:conf/sigir/ChenTHX020,DBLP:conf/www/BrickleyBN19}. \citet{DBLP:conf/www/ZhangB18} match queries to tables based on semantic similarities. They embed both queries and tables into vectors in multiple semantic spaces, and use early and late fusion strategies to measure the semantic similarity between these vectors. \citet{DBLP:conf/sigir/ChenTHX020} explore the application of BERT. Partial content of the dataset, query, and additional contextual information are encoded into the format of BERT. BERT then processes the encoded data to obtain a feature vector, which is fed into a regression layer to predict the relevance score between the query and the dataset. Google introduces Goods \cite{DBLP:conf/sigmod/HalevyKNOPRW16}, a system designed for managing datasets with a variety of forms from various departments within a company. Later, Google develops Google Dataset Search \cite{DBLP:conf/www/BrickleyBN19}, a dataset discovery engine that allows users to search datasets published on the web. It leverages metadata provided by dataset owners to facilitate search without semantic processing of the dataset content.

Several studies focus on searching datasets most relevant to an input dataset \cite{DBLP:conf/dmah-ws/RezigVGPCS20,DBLP:journals/pvldb/RezigBFPVGS21,DBLP:journals/pvldb/YangWSP22}. DICE proposed by \citet{DBLP:conf/dmah-ws/RezigVGPCS20,DBLP:journals/pvldb/RezigBFPVGS21} takes as input a few user-supplied example records, discovers candidate datasets by finding join paths in Primary Key-Foreign Key graphs of tables, and presents the user with a selection of records for feedback, repeating the process until the user is content with the results. \citet{DBLP:conf/sigmod/ZhangI20} propose a method for dataset search in Jupyter Notebook platform, which combines multiple table relationship measures, such as row and column overlap, new information, and provenance similarity, to identify and rank datasets that are most relevant to a target dataset. In search for spatial datasets, \citet{DBLP:journals/pvldb/YangWSP22} utilize the Earth Mover’s Distance as a similarity measure and return a ranked list of datasets that are most similar to the target dataset. 

Several studies have developed systems that support flexible querying \cite{DBLP:conf/icde/FernandezAKYMS18,DBLP:journals/pvldb/EltabakhKEA23,DBLP:journals/pacmmod/WangF23,DBLP:conf/sigir/00600CPS21}. \citet{DBLP:conf/icde/FernandezAKYMS18} capture relationships between datasets using a hypergraph called Enterprise Knowledge Graph (EKG). The EKG comprises nodes denoting columns of data sources and edges/hyperedges with weights that indicate the similarity relationships between these nodes. Hyperedges that link any number of nodes allow finding similar datasets at different granularity, e.g., columns and tables. A source retrieval query language and a graph index are designed to support and facilitate multiple complex discovery queries in the EKG. In addition,  \citet{DBLP:conf/sigir/00600CPS21} and \citet{DBLP:journals/pacmmod/WangF23} develop a self-supervised learned data discovery system that allows users to write natural language questions instead of standard query language. \citet{DBLP:conf/sigir/00600CPS21} convert a table into a multi-granular graph where each cell, row, and column is represented as a node.  They employ BERT to generate a query embedding, while a Graph Transformer is used to generate embeddings for nodes in the graph. A multi-layer perceptron then computes the relevance scores between the query embedding and the node embeddings for matching queries to tables. \citet{DBLP:journals/pacmmod/WangF23} collect training data by generating SQL queries that can be answered using tables and then converting those queries into natural language questions. The tables are decomposed down into triples, and the questions and triples are encoded using question and passage encoders, respectively. The relevance of tables with respect to the question is then evaluated using an OpenQA model. Considering that most data discovery efforts are limited to processing only the structured data, \citet{DBLP:journals/pvldb/EltabakhKEA23} propose a system CMDL that can discover both structured and unstructured data. Columns of tables and text documents are the basic units of discovery in CMDL. CMDL preprocesses documents and tables to get a unified column-style format. A profiler generates sketches and statistics for transformed documents and tables, which are then indexed to support efficient retrieval. As a key part of CMDL, an embedding-based joint representation is introduced to encode relationships between documents and columns. The joint representation model is trained on datasets generated by a weakly-supervised learning approach. After indexing and joint representation, the relationships among data elements are integrated into an EKG for querying.

\subsection{Discussion on Actual Data Marketplaces}
In actual data marketplaces, data sellers obtain data through a variety of methods. Some data sellers generate data directly through their operations by employing various methods such as using sensors to collect data or tracking user behavior. For instance, the Northern California Earthquake Data Center, part of the Amazon Sustainability Data Initiative, collects seismic data through a network of broadband, short period, strong motion seismic sensors, and GPS \cite{NorthernCaliforniaEarthquakeData}.
Other data sellers acquire data from external sources, including other companies, cooperative partners, and public datasets. One example is Rearc, which provides datasets including the Foreign Direct Investment (FDI) data sourced from World Bank Open Data \cite{Rearc}. Additionally, some data sellers utilize crowdsourced dataset collection methods, such as surveys and questionnaires, to gather data. For example, QUICK Corp. offers a product on AWS Data Exchange called the ``QUICK monthly market survey'' \cite{QUICKmonthlymarketsurvey}. This dataset is collected through surveys conducted among market professionals in Japan. A data seller on Datatang utilizes crowdsourcing to gather images about sign language gestures or 3D hand gestures \cite{datatang980}. The acquired data is then processed into products and offered to the marketplace.

\section{Data Productization}\label{sec:productization}
In this section, we categorize data products based on their types in Section \ref{subsec:datatranscate} and analyze methods for versioning data products in Section \ref{subsec:version}.

\nop{
\subsubsection{Data Categories}\label{subsec:datacate}
Raw data can be generally categorized as shown in Table \ref{tab:DataCategories}.

 \begin{table}[h]
    \centering
    \caption{Data Categories.}\label{tab:DataCategories}
    \begin{tabular}{|l|l|l|}
        \hline
        Property & Description & Details \\
        \hline
        Type & What kind of the data is? & \makecell[l]{Relational data;\\ Non-relational data (text, image, video); ...} \\ \hline
        Source & Where does the data come from? & \makecell[l]{Individuals; Enterprises; Businesses;\\ Public sources; Self-Generated; ...}  \\ \hline
        Ownership & What is the ownership of the data? & Personal data; Enterprise data; Public data\\ \hline
        Storage & How is the data stored? & \makecell[l]{Centralized/Decentralized public clouds;\\ Centralized /Decentralized private clouds;\\ Local device; ...} \\ \hline
        Security & How to ensure data security? & \makecell[l]{Secure Sockets Layer (SSL);\\ Distributed Ledger Technology (DLT);\\ User authentication; ...}  \\
        \hline
        Product & What products can be made from raw data? & Query; Model; ... \\ \hline
        Characteristic & The characteristics of data. & Dynamic; Incomplete; ... \\ \hline
    \end{tabular}
 \end{table}
}

\subsection{Data Product Categories}\label{subsec:datatranscate} 

Data products are derived from raw data in many forms and can be generally categorized based on their types: datasets, (database) queries, and (machine learning) models. Compared with queries and models, which are aggregated or transformed results derived from raw data or datasets, datasets carry a higher risk of leaking privacy. \textit{Dataset-based markets} \cite{DBLP:journals/dss/LiR14, DBLP:journals/tifs/XuJQZLR17, DBLP:conf/gis/KanzaS15, DBLP:conf/gis/NguyenKS20, DBLP:journals/isci/Parra-Arnau18, DBLP:journals/compsec/ShenGSDDZZJ22, heckman2015pricing, DBLP:conf/apweb/DingWZLG15, DBLP:journals/candie/YuZ17, DBLP:journals/toit/CaoPVTLD16, DBLP:journals/jsac/ZhengPWTC17, DBLP:conf/nips/ChenSZ20} offer fixed datasets for sale and allow buyers to access the dataset directly.
\emph{Query-based markets} \cite{koutris2012query,DBLP:journals/jacm/KoutrisUBHS15, koutris2013toward, DBLP:conf/webdb/LiM12, DBLP:conf/icdt/LiLMS13, DBLP:journals/tods/LiLMS14, DBLP:journals/cacm/LiLMS17, DBLP:journals/pvldb/LinK14, DBLP:journals/pvldb/UpadhyayaBS16, DBLP:conf/sigmod/DeepK17, DBLP:conf/kdd/NiuZWTGC18, DBLP:conf/soda/GuruswamiHKKKM05, DBLP:journals/pvldb/ChawlaDKT19, DBLP:conf/icde/NiuZWTC20, DBLP:conf/icml/ChenLX22, DBLP:journals/tkde/MiaoGCPYL22, DBLP:journals/tpds/CaiYYZLX22, DBLP:conf/icde/ChenYWWL22, DBLP:conf/mdm/ZhengCY20}
charge data buyers and compensate data sellers on a per-query basis. 
The marketplace makes decisions about data usage restrictions (e.g., return queries with privacy protection \cite{DBLP:journals/tods/LiLMS14}), revenue allocation, and query-based pricing. 
\emph{Model-based markets} \cite{DBLP:conf/ec/AgarwalDS19,DBLP:conf/sigmod/ChenK019,DBLP:journals/pvldb/JiaDWHGLZSS19,DBLP:journals/pvldb/LiuLL0PS21,han2023datacentric} have been recently proposed, where data brokers set model usage charge from model buyers and distribute revenue to data sellers.

\subsection{Data Product Versioning}\label{subsec:version}

Versioning data products means offering data products in different versions for different segments of data buyers. The versioning strategy has been investigated when designing data products in many works \cite{DBLP:conf/birthday/BalazinskaHKSU13,koutris2012query,DBLP:journals/jacm/KoutrisUBHS15,koutris2013toward,DBLP:conf/icdt/LiLMS13, DBLP:journals/tods/LiLMS14, DBLP:journals/cacm/LiLMS17,DBLP:journals/jsac/ZhengPWTC17,DBLP:journals/tmc/ZhengPWTC20,DBLP:conf/ec/AgarwalDS19,DBLP:conf/sigmod/ChenK019,DBLP:journals/pvldb/LiuLL0PS21,DBLP:conf/infocom/SunCLH22,DBLP:journals/isci/JiangXWYWL23}. These can be categorized into three dimensions. 
\begin{itemize}
    \item View/subset-based versioning \cite{DBLP:conf/birthday/BalazinskaHKSU13,koutris2012query,DBLP:journals/jacm/KoutrisUBHS15,koutris2013toward,DBLP:journals/jsac/ZhengPWTC17,DBLP:journals/tmc/ZhengPWTC20,DBLP:conf/infocom/SunCLH22,DBLP:journals/isci/JiangXWYWL23}. Data sellers can create data products by answering arbitrary queries or selecting data subsets. For versioning a data set, one can provide data slices or statistical results by custom queries (views). Queries that create views are a natural way to create multiple versions of datasets. For versioning a model, one can select data subsets to train different models.
        
    \item Noise-based versioning \cite{DBLP:conf/icdt/LiLMS13, DBLP:journals/tods/LiLMS14, DBLP:journals/cacm/LiLMS17,DBLP:conf/sigecom/GhoshR11,DBLP:conf/kdd/NiuZWTGC18,DBLP:conf/mdm/ZhengCY20,DBLP:conf/ec/AgarwalDS19,DBLP:conf/sigmod/ChenK019}. Data sellers can create data products with varying degrees of accuracy or variance by adding noise. Typically, a high-quality version is produced first, then downgraded by adding noise to get a low-quality version without incurring too many additional costs. For versioning a query, one can perturb it with noise to produce query answers with varying levels of approximation. For versioning a model, one can add noise to the optimal model instance, or add noise to the original data set to train a low-quality model.

    \item Hybrid versioning \cite{DBLP:journals/pvldb/LiuLL0PS21}. Data sellers can produce data products using view/subset-based versioning and noise-based versioning together.
\end{itemize}

\partitle{View/subset-based versioning}
In dataset-based markets, \citet{DBLP:conf/birthday/BalazinskaHKSU13} discuss a framework for trading relational data, suggesting that a workable scheme is to provide views of data instances. Views can be considered as analogs to versions in data markets. A view may be a slice of data or may contain more coarse-grained information. More deeply, they describe a vision for choosing and expressing views from three dimensions for sellers who sell relational data: relational view (query), increasing/decreasing accuracy, and user-defined functions. 
\nop{
\begin{enumerate}
    \item Relational view: Sellers are allowed to use arbitrary relational views to define versions of the data.
    
    \item Increasing/Decreasing accuracy: Sellers can produce views with varying degrees of accuracy. For example, sellers can increase the accuracy of their data products by performing data cleaning and can decrease the accuracy of their data products by adding noise to the data.
    
    \item User-defined functions: Sellers should be able to define views with user-defined functions. For example, sellers can produce more valuable new data products from the raw data by using data mining algorithms.
\end{enumerate}
}

Subsequently, \citet{DBLP:journals/isr/MehtaDJM21} explore price-quantity schedules, a versioning strategy for datasets where different quantities of data are available at different prices according to buyer demands. 
This versioning strategy can be likened to a selection query, allowing for the selection of data subsets. Moreover, it can be extended to more complex queries involving aggregation. \citet{DBLP:journals/jsac/ZhengPWTC17} propose a mobile crowdsensing marketplace that sells posterior distribution about multiple data acquisition points. Each version is created by selecting some observation random variables from these points, the quality of which can be measured by its conditional entropy. To minimize expenditure following the principle of cost minimization, exploiting the submodularity of the entropy function, they propose to greedily select the data acquisition point with the smallest ratio of expected payment to marginal entropy until the normalized joint entropy of the selected points meets the required quality. The follow-up work by \citet{DBLP:journals/tmc/ZhengPWTC20} defines the quality of data products as the average posterior variance. They model the versioning problem as a basic reward minimization problem, aiming to find the subset of data sellers that satisfies the variance requirement and minimizes the basic data acquisition cost. Since the variance function is a monotonic submodular function, they greedily select the data seller with maximum variance reduction until the variance requirement is satisfied.

Rather than choosing from predetermined views, \citet{koutris2012query,DBLP:journals/jacm/KoutrisUBHS15,koutris2013toward} allow data buyers to customize queries. They propose a pricing system for queries called QueryMarket. Given the current database instance and past purchases of the data buyer, QueryMarket automatically computes the cheapest set of queries to purchase to answer the query. Thus, the data buyer is not forced to purchase a superset of the data she really needs.

In model-based markets, \citet{DBLP:journals/isci/JiangXWYWL23} version classifiers by employing progressive sampling to select data subsets of different sizes to train multiple classifiers, each with different accuracy. In a similar context, \citet{DBLP:conf/infocom/SunCLH22, DBLP:journals/pacmmod/SunWWLLJ24} factor in the privacy costs of data sellers when selecting data subsets. To train each model version, they propose a greedy algorithm that selects the data seller with the smallest privacy cost per unit privacy budget.

\partitle{Noise-based versioning}
In dataset-based markets, \citet{DBLP:conf/ec/AgarwalDS19} design an allocation function for selling training data 
that allocates dataset features 
to the data buyer based on the received bid $b_{n}$ and the current price $p_{n}$. More specifically, it collectively adjusts the quality of dataset features 
by adding noise that is proportional to the difference between $p_n$ and $b_n$. 

In query-based markets, several studies \cite{DBLP:conf/icdt/LiLMS13, DBLP:journals/tods/LiLMS14, DBLP:journals/cacm/LiLMS17,DBLP:conf/sigecom/GhoshR11,DBLP:conf/kdd/NiuZWTGC18} propose that in response to answering aggregation queries with varying precision requirements from data buyers, a data broker collects corresponding data from data sellers, computes true query answers, adds random noise (e.g., Laplace noise) to true query answers, and returns the perturbed answer to data buyers. This approach can result in uniform privacy losses across all data sellers. To cope with data sellers with varying privacy sensitivities, \citet{DBLP:conf/mdm/ZhengCY20} propose to inject different levels of noise to different data sellers and return perturbed query results derived from noisy data to data buyers.

In model-based markets, \citet{DBLP:conf/sigmod/ChenK019} propose a scenario where data brokers release a model instance through a randomized mechanism that adds noise to the optimal model instance. 
Given a dataset and a specific machine learning model, the data broker first trains the optimal model instance. Then, given a noise control parameter $\delta$, 
the data broker generates multiple noisy models and computes a curve that plots the prices together with the expected error for every $\delta$. 

\partitle{Hybrid versioning}
\citet{DBLP:journals/pvldb/LiuLL0PS21} use a hybrid versioning strategy for models. They propose an end-to-end model marketplace called Dealer. The data broker in Dealer releases multiple versions of models following the principle of privacy preservation and revenue maximization. To protect the privacy of data sellers, the data broker adopts enhanced objective perturbation called approximate minima perturbation to train models satisfying varying levels of privacy budget, which perturbs the objective function of the model by quantified noise according to differential privacy. To maximize revenue, the data broker selects dataset subsets to optimally train each model version under the manufacturing budget of compensating participating data sellers. 

Versioning is a critical aspect of data markets. By offering multiple versions of data products at different prices, data sellers can capture a larger share of the market and generate more revenue. It is worth noting that versioning strategies and pricing strategies are closely linked. Their interplay can have a significant impact on the success of data products in the market. At the same time, the data pricing strategy will depend on various factors such as the value of the data product, the cost of production, the demand for the data product, and the market competition, and consider various desiderata for profitability.

\subsection{Discussion on Actual Data Marketplaces}
A variety of data products are available on actual data marketplaces, including tabular datasets, images, videos, and generative AI. Data sellers offer data in multiple formats, such as CSV, JSON, and Parquet, at various frequencies (monthly, quarterly, annually), and with tiered access.  
For example, similar to subset-based versioning, SchoolDigger offers U.S. K-12 school data through API at three subscription levels: the basic level (\$13) provides core data including school address, phone number, student body makeup, free-lunch program recipients, and SchoolDigger ranking; the pro level (\$110) adds standardized test scores to the basic level; and the enterprise level (\$170) offers the most comprehensive access \cite{schooldigger}. Similar to noise-based versioning, Coresignal offers an AI-enriched version of its company data set on Datarade \cite{Coresignal}.

\section{Data Transaction}\label{sec:transaction}
Data transaction is a commitment between data sellers and data buyers to exchange data products and transfer payments. We survey solutions for signaling schemes of data sellers in Section \ref{subsec:signal} and data acquisition of data buyers in Section \ref{subsec:dataacq}.

\subsection{Data Selling and Information Signaling}\label{subsec:signal}
Information signaling (also known as statistical experiments \cite{kamenica2019bayesian}) refers to the practice of data sellers selectively revealing partial information/data to influence the beliefs and decisions of the data buyers, thereby increasing revenue \cite{liu2021optimal, hoxha2024selling, DBLP:conf/sigecom/BabaioffKL12,bergemann2018design,bergemann2022selling,bonatti2023coordination,yang2022selling}.
Moreover, signaling can also be employed to reveal a small amount of free information about the data quality in order to update the beliefs of both parties about the quality/appropriateness of the underlying data, leading to more transparent transactions and often also better revenue. 

One widely studied format of data selling is through the sale of partially revealed information, also known as signaling schemes or statistical experiments \cite{kamenica2019bayesian}. Intuitively, information could be viewed as a special form of data that is processed, e.g., certain statistics distilled from data or parameter values estimated from data. To the best of our knowledge, the work by \citet{DBLP:conf/sigecom/BabaioffKL12} is the first to design mechanisms for selling information. In their setting, there is a single seller and a single buyer. The seller knows the state of the world but the buyer does not, and the buyer has a private payoff type that the seller does not observe. The utility of the buyer depends on her actions and the random state of the world. Upon receiving the signal sent by the seller, the buyer updates its posterior beliefs and chooses an action to further his utility, during which the buyer  extracts some amount of the surplus as revenue. Under different conditions regarding the correlation between the world state and the buyer's payoff type, they develop algorithms to compute the revenue-optimal mechanism. A key observation they find is that mechanisms with multiple rounds of interaction with buyers can possibly gain more revenue than any single-round mechanism. Subsequently, \citet{DBLP:conf/soda/ChenXZ20} study the same setup but extend it to the setting where the buyer is budget-constrained. Perhaps interestingly, despite studying a more general problem, they discover a simpler algorithm for finding the optimal mechanism by solving linear programs, thanks to the discovery of a more succinct form of revelation principle. \citet{liu2021optimal} develop the first closed-form optimal information selling mechanism for a fundamental special case of the model pioneered by \cite{DBLP:conf/sigecom/BabaioffKL12}. They prove that when the buyer's decision is between taking an active move with an uncertain state of nature and keeping status quote with a default $0$ utility, and would like to purchase more information to evaluate the active move's return, the optimal information selling mechanism can be characterized in a simple closed form --- it simply sells a threshold test (i.e., a binary classifier) to each buyer, with both the threshold and payment amount tailored toward the buyer type.  
\citet{DBLP:conf/innovations/0001V21a} employ multi-item auction design approaches to develop a polynomial time algorithm for computing the optimal mechanism. These mechanisms   \cite{DBLP:conf/sigecom/BabaioffKL12,DBLP:conf/soda/ChenXZ20} allow the seller to interact with the buyer for multiple rounds which may be difficult to implement. Hence \citet{DBLP:conf/sigecom/Zheng021} design simpler mechanisms with a two-round interaction for selling information. They propose to reveal some free partial information before the sale to change the belief of the buyer, and then sell the full version of the information at some price.  \citet{bergemann2022selling} then study when selling all the data could be approximately optimal. 

\citet{DBLP:conf/icml/ChenLX22} devise a revenue-maximizing mechanism where a data seller sells data to a machine learner who purchases data to train its machine learning model. The proposed pricing scheme consists of two steps: a costly signaling step and a pricing step. In the costly signaling step, a subset of data is shared to train a model and learn a preliminary model accuracy for both the seller and buyer. In the pricing step, the seller prices the remaining data based on the preliminary model accuracy and the shared subset of data to maximize its revenue. 
For homogeneous data, each data point contributes equally to the model performance. Thus, designing an optimal mechanism refers to finding an optimal number of shared data such that the revenue is maximized coming with a little loss caused by sharing data, which can be solved by enumeration. For heterogeneous data, they employ quality vectors to measure the usefulness of features and prove that selling the entire dataset based on common prior can achieve $\frac{1}{k}$-approximate, where $k$ is the number of the machine learner's private types. 

\citet{DBLP:journals/mansci/DrakopoulosM23} outline the conditions under which offering free sample strategies can lead to significantly higher revenues compared to static strategies. They also demonstrate the advantages of providing free data samples through a series of illustrative examples.

\partitle{Economics literature}
There has been significant interest in the recent economics literature on information selling. \citet{bergemann2018design} identify various properties of the optimal mechanism in the information selling environment where the goal of the data buyer is to take action to match the state of the world. Motivated by online advertising,  \citet{yang2022selling} studies how a data broker can sell consumer demand information to certain producers with private production costs and characterizes the revenue-maximizing mechanisms for the data broker. 
\citet{bonatti2023coordination} study how a data seller can sell information to two competitive data buyers, whose utility will both be affected by the data of the seller as well as the actions of other players. Given the intricacy of the general problem, they focus on a special class of symmetric games with quadratic payoffs including Cournot competition, Bertrand competition, and Keynesian beauty contest. They study a class of mechanisms called the Gaussian mechanisms (where the joint distribution of player actions and private signals is a multivariate normal distribution) and identify the welfare-optimal and revenue-optimal mechanisms within this class. For games of strategic complements (e.g., Bertrand competition), the optimal mechanisms maximally correlate the player actions, and conversely maximally anti-correlate them for games of strategic substitutes. Recently, \citet{hoxha2024selling} studies information selling with motivations from consulting services where consultant sells information products to clients to assist decision making of clients. This work also identifies conditions under which the consultant can extract full (socially efficient) surplus from the buyer. Similar to \cite{liu2021optimal}, the author also find that the middle buyer type enjoys the most surplus from purchasing information, which is in sharp contrast to typical results in mechanism design where it is the buyer with the lowest and highest type who enjoys the most surplus. 

\subsection{Data Acquisition}\label{subsec:dataacq}

During transactions between data sellers and data buyers, data sellers provide descriptions of their data products, including architectural details, pricing mechanisms, and access patterns. Data buyers with limited budgets then look for the optimum purchase strategy to achieve their specific targets by leveraging this available information. When the data seller provides highly integrated data products, such as machine learning models, buyers only need to specify some parameters to get the desired product. In this case, the data seller is considered to offer buyer-tailored services, thus buyers carry a smaller exploration burden. However, when the data seller provides buyers with datasets, data buyers always need to discover and integrate relevant data to get cost-effective products. In this section, we review several works that focus on addressing the latter data acquisition problem for data buyers in data markets. These works can be generally categorized based on data usage: dataset construction in Section \ref{subsec:dataset}, statistical estimation in Section \ref{subsec:stas}, and model training in Section \ref{subsec:modeltrain}.

\subsubsection{Dataset Construction}\label{subsec:dataset}
Dataset construction refers to building a dataset by selecting, acquiring, and integrating data from multiple sources, with the goal of maximizing the dataset utility within a budget or minimizing the cost while meeting the dataset utility requirements. The dataset utility can be defined in terms of data distribution \cite{DBLP:journals/pvldb/NargesianAJ21}, accuracy \cite{DBLP:journals/pvldb/DongSS12,DBLP:conf/nips/ChenSZ20}, freshness \cite{DBLP:conf/sigmod/RekatsinasDS14}, and relevance to the applications of data buyers \cite{DBLP:conf/infocom/SunXXGZ23}.

\citet{DBLP:conf/cidr/RekatsinasDGS15} outline a vision of a data source management system and report a demonstration system SourceSight in \cite{DBLP:conf/sigmod/RekatsinasDDGS16}. Considering multiple data quality metrics (e.g., coverage, accuracy, timeliness, and position bias), they cast the data source selection as a multi-criteria optimization problem under budget constraints such as money and design an interactive query engine system to find Pareto optimal solutions.

\citet{DBLP:journals/pvldb/NargesianAJ21} study how to acquire data samples from multiple data sources to achieve a desired target distribution based on predefined groups for fairness requirements. They consider the cost of sampling in each data source and the desired number of data points for each group. For the case where the group distributions in each data source are known, they propose two approaches: an exact dynamic programming algorithm and an approximate algorithm inspired by the solution to the coupon collector’s problem. For the case where the group distributions in each data source are unknown, they propose an exploration-exploitation-based algorithm using the Upper Confidence Bound, where each data source is modeled as an arm and the reward function captures the cost and approximations of group distributions. Follow-up work by \citet{asudeh2022towards} presents a vision of a distribution-aware cost-effective query answering engine for user-provided schema and distribution requirements in data markets. The system consists of a pipeline of data view discovery, get-next operation, and get-next cost estimation. Solutions to the multi-arm bandit or coupon collector’s problem apply to the data view selection problem. 

Inspired by the Marginalism principle in economic theory, which posits that the incremental benefit derived from each additional unit of resource will eventually diminish, \citet{DBLP:journals/pvldb/DongSS12} introduce the problem of source selection named the marginalism problem. They terminate integrating a new data source when the marginal gain is less than the marginal cost. Formally, given a set of data sources, an integration model, and a budget on cost, the marginalism problem is to find a subset of sources that maximizes the difference between the gain and the cost of integrating the sources by the model, while keeping the cost within the given budget. They focus on data fusion tasks by defining the gain as the accuracy of the integration result and apply greedy randomized adaptive search procedure meta-heuristic to solve the marginalism problem. 

Building on the above work \cite{DBLP:journals/pvldb/DongSS12} for static sources, \citet{DBLP:conf/sigmod/RekatsinasDS14} study the source selection problem in dynamic scenarios where the contents of data sources are updated over time. They propose time-dependent metrics, including coverage, freshness, and accuracy, to quantify the quality of integrated data. Since the coverage and global freshness are proved to be non-decreasing submodular functions, they develop a greedy algorithm to output the near-optimal solution with theoretical guarantees for time-aware source selection, varying frequency source selection, and slice time-aware source selection.

\citet{DBLP:conf/nips/ChenSZ20} study a scenario where a data buyer has a budget but no test dataset to verify the truthfulness of the collected data. They adopt peer prediction to guarantee truthfulness, which encourages data sellers to report their datasets truthfully. Assuming that data from all data sellers is generated by the same distribution model following a standard Bayesian process, a data seller will receive a higher payment if her reported dataset is more consistent with the statistical model estimated using the reported datasets of her peers. They apply log-PMI (point-wise mutual information) payment rule \cite{DBLP:conf/sigecom/KongS18} to ensure truthfulness. The above efforts seek to allocate the budget efficiently for purchasing data, which also involves data pricing.

\citet{DBLP:conf/infocom/SunXXGZ23} focus on the problem of trading crowdsourced data in a data market where data buyers and data sellers have unknown preferences and privacy concerns. Their goal is to find a match between data collection tasks (data buyers) and mobile users (data sellers) for data trading in a manner that maximizes data quality while ensuring stability and privacy preservation. Since data sellers are reluctant to disclose their data qualities that may contain some private sensitive information, both seller data qualities for tasks and buyer preference sequences for sellers are unknown. They tackle this problem as a multi-player multi-armed bandit problem. Data sellers are treated as arms, the market platform as players, and data qualities as rewards. In each trading round, noise is added to the data qualities using a differential privacy mechanism to protect seller privacy. Preference sequences based on estimated data qualities are formed to compute a stable matching between buyers and sellers using the Gale-Shapley algorithm. This process is iteratively refined as more data quality information is collected and preference sequences are updated.

\subsubsection{Statistical Estimation}\label{subsec:stas}
Statistical estimation involves acquiring data to derive statistical insights under budget constraints, such as calculating the population mean \citet{DBLP:conf/sigecom/ChenILSZ18,DBLP:conf/ec/ChenZ19}, estimating an underlying parameter \cite{DBLP:conf/sigecom/0001MMO22}, performing correlation analysis \citet{DBLP:journals/pvldb/LiSDW18}, and answering queries \cite{DBLP:journals/ton/RenLZW18,DBLP:journals/pvldb/LiG16,DBLP:conf/ecai/LiZZL23}.

The problem of an analyst purchasing data from individuals with private costs subject to a budget constraint is first formulated by \citet{roth2012conducting} and further explored by \citet{DBLP:conf/sigecom/ChenILSZ18,DBLP:conf/ec/ChenZ19}. Their target is to obtain an optimal unbiased estimation of the population mean with minimum worst-case variance by incentivizing individuals to reveal their data through monetary payments. In addition, \citet{cai2015optimum} aim at incentivizing individuals to obtain high-quality data for regression while minimizing the weighted sum of payments and estimation error. \citet{cummings2015accuracy} wish to achieve a certain desired accuracy level of the aggregated estimator while minimizing the total cost. \citet{abernethy2015low} consider an active data procurement for general supervised learning tasks where data arrives online.

The purchasing decision problem of spatiotemporal information is first studied by \citet{DBLP:journals/tsas/AlyKRH19}. They investigate the value of location data without access to its specific coordinates to guide the purchases of the data buyer based on purchase history. Three cases are studied, including home-targeted advertisements, traffic congestion inference, and location prediction. 

\citet{DBLP:journals/pvldb/LiSDW18} study the data acquisition problem for data buyers who would like to perform correlation analysis. They design a middleware called DANCE. DANCE aims to identify a join result of data instances from the data marketplace, which maximizes the correlation between source attributes and target attributes considering data quality, join informativeness, and data purchase budget. It builds a two-layer join graph including the instance level and the attribute set level. Then the data acquisition problem is modeled as a graph search problem subject to several constraints. They prove that the search problem is NP-hard and design a heuristic algorithm based on the Markov chain Monte Carlo.

\citet{DBLP:journals/ton/RenLZW18} consider a geo-distributed data cloud market where a data cloud purchases data of varying quality levels from data sellers, stores the data in chosen data centers, and answers queries from data buyers that have requirements for data sellers and quality levels. The total cost of the data cloud is the sum of the operation cost incurred in transferring data of required quality levels from data sellers to data centers, the execution cost incurred in transferring data of required quality levels from data centers to data buyers, and the purchasing cost incurred in purchasing data from data sellers. They study a cost minimization problem for the data cloud, which is formulated as a joint optimization of data purchasing and data placement. By a reduction to and from an uncapacitated facility location problem, the problem is proved to be NP-hard. For the case of a single data center, they relax integer constraints of the original integer linear programming problem to obtain a linear programming problem and find its optimal solution in polynomial time. For general cases, they decompose the joint problem into two subproblems. Specifically, they first use the algorithm for the case of a single data center, and then easily get data placement decisions.

For querying data that changes over time, \citet{DBLP:journals/pvldb/LiG16} study a sequential data acquisition problem. That is, upon receiving a query, an online decision is made whether to use the current version of data blocks in the database or to acquire the latest data blocks from the data source and update the database at a cost, with a goal of minimizing the long term costs. They adapt reinforcement learning and tailored locality-sensitive hashing to solve the decision-making problem.

Taking privacy into consideration, \citet{DBLP:conf/sigecom/0001MMO22} examine the design of mechanisms for acquiring data from users who have privacy concerns but also benefit from reduced estimation errors. They consider two architectures: a central privacy setting, where users share their raw data with the platform, and a local privacy setting, where users share a differentially private version of their data with the platform. Furthermore, they develop a unified framework to analyze the design of data acquisition mechanisms for users who have both local and central privacy concerns and vary in how they prioritize these privacy considerations \cite{DBLP:conf/nips/0001MMO22}.

\citet{DBLP:conf/ecai/LiZZL23} introduce a framework for private data trading that integrates the procurement and query processes, allowing data seller selection and data perturbation to occur simultaneously. They propose two data seller selection methods: Greedy Private Query Mechanism (GPQM) selects data sellers with lower valuations in a greedy manner until the budget is exhausted, while Neural-based Private Query Mechanism (NPQM) parameterizes the probability of each data seller being selected with a neural network, with parameters learned using the dual-ascent algorithm.

\subsubsection{Model Training}\label{subsec:modeltrain}
Model training involves acquiring data to train a high-performance model or improve the performance of a model within a budget constraint, using performance metrics like model utility \cite{DBLP:conf/sigmod/GershteinMNR22}, accuracy \cite{DBLP:journals/pvldb/LiYK21,DBLP:journals/corr/abs-2012-08874,DBLP:conf/icml/ZhaoLFK23,DBLP:journals/corr/abs-2403-13893}, fairness \cite{DBLP:conf/sigmod/TaeW21}, and confidence \cite{DBLP:journals/pacmmod/LiYN24}.

\citet{DBLP:conf/sigmod/GershteinMNR22} study an optimization problem called the Budgeted Classifier Construction problem (BCC). BCC involves selecting a set of classifiers to cover a set of queries, maximizing classifier utility while satisfying a budget constraint. They show that BCC is NP-hard and observe that BCC can be partitioned into subproblems: the Knapsack problem and the Quadratic Knapsack problem. 

\citet{DBLP:journals/pvldb/LiYK21} consider a scenario where a data buyer queries data from a data seller to enhance a machine learning model trained on an initial dataset. They make an assumption that the data seller maintains a collection of data records that follows the same distribution as the target distribution (the distribution of the initial dataset and the test dataset), while the distribution is invisible to the data buyer. Limited by the maximum number of records that can be obtained from the data seller (also as the fixed budget of the data buyer), the goal of the data buyer is to issue an optimal series of queries by selecting predicates, such that the obtained data improves the accuracy of the machine learning model. They use a notion of novelty to quantify the predicate utility as the anticipated accuracy improvement. Novelty measures the difference between acquired data and existing data. The higher the difference, the more information the predicate produces. Then, they propose two strategies based on the exploration-exploitation trade-off. The first solution, EA, focuses on exploration in the estimation stage (querying to estimate the predicate utility with a part of the budget first) and on exploitation in the allocation stage (then allocating the remaining budget based on the utility estimation), respectively. The second solution, SPS, fuses exploration and exploitation in each round of interactions and sequentially allocates the budget for the next round. Both EA and SPS do not require re-training the machine learning model.
 
\citet{DBLP:journals/corr/abs-2012-08874} consider a data purchasing problem in real-world data marketplaces and design ``Try Before You Buy'' greedy data purchasing approaches.
They consider two versions of the data purchasing sequence: stand-alone version and assisted version. In stand-alone version, the accuracy achievable on the buyer-specified AI/ML task trained by each dataset held by data sellers is provided. They propose to buy datasets in descending order of expected profit. In the assisted version, the accuracy achievable on the AI/ML task trained by each dataset held by data sellers together with the dataset already purchased by the data buyer is provided. They propose to buy datasets in descending order of marginal profit.

\citet{DBLP:conf/icml/ZhaoLFK23} focus on the scenario where data from data sellers is not provided to data buyers directly. The data buyer purchases gradient updates from data sellers to train a model. To allocate the budget effectively, they use an adaptive sampling method based on an online stochastic mirror descent sampler to ensure that data sellers with a higher-quality history will be accessed more often. At each training iteration, the algorithm updates the probability distribution of data sellers being selected for gradient updates. The number of iterations is determined by the budget. They show that the average regret of the proposed method compared to spending all budget on the most valuable providers is asymptotically zero as the total budget goes to infinity. 

\citet{DBLP:journals/corr/abs-2403-13893} present a model where the platform is responsible for acquiring data in decentralized data marketplaces. The setting includes a data buyer with an unlabeled test dataset and data sellers with individual data points priced by marginal costs. They formulate the data acquisition problem as assigning a selection weight to each data point to minimize the prediction error on the unlabeled test dataset instead of a validation dataset within a budget, which is inspired by linear experimental design. The Frank-Wolfe algorithm is used to iteratively update the selection weights of data points. 

In addition to the model accuracy, the model fairness and model confidence are also considered. Given a budget and multiple data slices with varying acquisition costs, Slice Tuner proposed by \citet{DBLP:conf/sigmod/TaeW21} controls the amounts of data acquired in each slice to optimize the accuracy and fairness of the model across all slices. The key idea of the Slice Tuner is to maintain power-law learning curves, which estimate model loss based on the amount of training data. When slices are assumed to be independent, Slice Tuner solves an optimization problem once with the objective of minimizing loss and unfairness. When slices are assumed to be dependent, and the influence between slices mainly comes from the relative sizes of the slices (i.e., the bias), Slice Tuner adopts an iterative method. It first computes the optimal amount of data to be sampled from each slice. If the result causes the bias to exceed a threshold, the number is adjusted so that the bias does not exceed the threshold, and the learning curve is updated. Otherwise, the result is adopted directly. This process is repeated until the budget is exhausted.

Model confidence is defined based on the geometric distance between evaluation samples and training samples \cite{DBLP:conf/uai/Chouraqui0EL22}. The problem involves selecting data samples from a data pool to maximize the model confidence improvement within a budget of sample size. Given its NP-hardness, \citet{DBLP:journals/pacmmod/LiYN24} identify two polynomial-time cases where optimal solutions can be found. The first is that data samples with the highest confidence improvement are independent and thus produce the optimal solution. In the other way, data samples that dominate a certain number of other samples in terms of model confidence improvement are guaranteed to be part of the optimal solution. Based on these two methods, they further restrict data acquisition to the nearest neighbors of each evaluation sample in the data pool, improving efficiency while providing bounded approximate guarantees.

Some studies assume that data products are invisible before purchase or only reveal partial information, such as data distribution \cite{DBLP:journals/pvldb/NargesianAJ21} or unlabeled data \cite{DBLP:journals/pacmmod/LiYN24}. Data buyers cannot exactly assess the quality of the data product before purchase. Methods based on exploration and exploitation are used to address the issue. Additionally, there are ``try before you buy'' options that allow data buyers to understand the data quality before purchase \cite{DBLP:journals/corr/abs-2012-08874}, or assigning the data acquisition task to a trusted data marketplace \cite{DBLP:journals/corr/abs-2403-13893}. 

\subsection{Discussion on Actual Data Marketplaces}
Actual data marketplaces support keyword search, filtering, and ranking mechanisms. Users enter keywords and the data marketplace will return relevant data products based on their metadata and descriptions. In addition, users can refine search results by applying filters such as dataset types, sellers, pricing models, and pricing units. Different sorting options are available, allowing users to sort the search results by criteria such as relevance, update time, and rating \cite{dawexSort}.

Data sellers provide comprehensive details of their data products, including descriptions, data sources, update frequency, data dictionaries, geographical coverage, categories, and pricing. In some cases, as with signaling schemes, data samples and overall quality scores are also available, giving potential buyers a preview of the data quality and relevance \cite{getrightdata}. To market data products and increase visibility, data sellers can publish blogs, tweets, etc., which will be reposted by the data marketplace \cite{Marketingyourproduct}. Data buyers and sellers can communicate during the data transaction, passing more detailed information about the data product to buyers \cite{dawexDiscussion}.

In AWS Data Exchange, data sellers can receive data feeds and daily, weekly, or monthly reports containing product usage, buyers, billing, and payment information to optimize their business operations \cite{AWSsellerReport}.

\section{Data Pricing}\label{sec:pricing}
\nop{
Pricing in data markets can be divided into two categories: revenue allocation and data pricing. \emph{Revenue allocation} refers to distributing revenue generated from data transactions among data owners, while data product pricing refers to determining prices for data products. An appropriate price requires revisiting economic concepts in terms of data characteristics, bringing profits to data sellers. Additionally, game-theoretic pricing combines revenue allocation and data product pricing to find a price equilibrium in one or multiple rounds of negotiations between data buyers and data sellers. We overview approaches to revenue allocation in Section \ref{subsec:revenueallocation}, approaches to data product pricing in Section \ref{subsec:dataproductpricing}, and game-theoretic pricing approaches in Section \ref{subsec:gametheorypricing}.
}

Data pricing refers to determining prices for data products. In Section \ref{subsec:dataproductpricing}, we overview data product pricing approaches proposed from a traditional economic perspective. These approaches revisit economic concepts in terms of data characteristics to set an appropriate price for the data product. In Section \ref{subsec:gametheorypricing}, we overview game-theoretic pricing approaches, where the data pricing process is designed as a game to find a price equilibrium in one or multiple rounds of negotiations between data buyers and data sellers.

\subsection{Data Product Pricing}\label{subsec:dataproductpricing}
In data markets, data sellers are supposed to make informed decisions on the trading prices of a series of data products. Central to data product pricing is the well-acknowledged economic notions including arbitrage freeness, envy freeness, revenue maximization, privacy preservation, and dynamicity, which have been considered by many researchers \cite{DBLP:journals/dss/LiR14, DBLP:journals/tifs/XuJQZLR17, DBLP:conf/gis/KanzaS15, DBLP:conf/gis/NguyenKS20, DBLP:journals/isci/Parra-Arnau18, DBLP:journals/compsec/ShenGSDDZZJ22, DBLP:journals/isci/JiangNYWL22, heckman2015pricing, DBLP:conf/apweb/DingWZLG15, DBLP:journals/candie/YuZ17, DBLP:journals/toit/CaoPVTLD16, DBLP:conf/nips/ChenSZ20, koutris2012query,DBLP:journals/jacm/KoutrisUBHS15, koutris2013toward, DBLP:conf/webdb/LiM12, DBLP:conf/icdt/LiLMS13, DBLP:journals/tods/LiLMS14, DBLP:journals/cacm/LiLMS17, DBLP:journals/pvldb/LinK14, DBLP:journals/pvldb/UpadhyayaBS16, DBLP:conf/sigmod/DeepK17, DBLP:conf/kdd/NiuZWTGC18, DBLP:conf/soda/GuruswamiHKKKM05, DBLP:journals/pvldb/ChawlaDKT19, DBLP:conf/icde/NiuZWTC20, DBLP:conf/icml/ChenLX22, DBLP:journals/tkde/MiaoGCPYL22, DBLP:journals/tpds/CaiYYZLX22, DBLP:conf/icde/ChenYWWL22, DBLP:conf/mdm/ZhengCY20, DBLP:conf/sigmod/ChenK019, DBLP:journals/pvldb/LiuLL0PS21, DBLP:conf/infocom/SunCLH22}. In this section, we overview studies on data pricing for datasets, queries, and machine learning models.

\subsubsection{Dataset Pricing}

Various types of studies for pricing datasets are proposed, including those based on privacy cost and data quality.

\partitle{Pricing based on privacy cost}
Studies \cite{DBLP:journals/tifs/XuJQZLR17,DBLP:journals/isci/Parra-Arnau18,DBLP:conf/gis/KanzaS15,DBLP:conf/gis/NguyenKS20,DBLP:journals/cn/FengYZ23} focusing on personal dataset pricing consider the degree of privacy disclosure and the privacy sensitivity of data sellers. \citet{DBLP:journals/isci/Parra-Arnau18} mathematically examine the trade-off between privacy disclosure and economic rewards in personal data markets. \citet{DBLP:journals/cn/FengYZ23} allow data sellers with different privacy preferences to control the privacy of their data by combining local differential privacy with contract theory. They propose a set of optimal contracts for data sellers, which specify different levels of local differential privacy and corresponding data prices. \citet{DBLP:journals/tifs/XuJQZLR17} introduce a dynamic pricing model where a data collector adjusts the price paid to sequentially arriving data sellers based on their privacy valuations, which are assumed to follow an unknown distribution. To maximize total revenue, they model the pricing problem as a multi-armed bandit problem. Each arm represents a candidate price, and the reward incorporates the impact of data anonymization on the data value over time. 

A vision of marketplaces for geosocial data is proposed by \citet{DBLP:conf/gis/KanzaS15}. Geosocial data refers to social spatio-temporal data pertaining people on the social network, that is, triples $(l,t,u)$ of location $l$, time $t$, and user $u$. 
\citet{DBLP:journals/gis/Sakr18} develops uniform grids to discretize the whole spatial extent into a finite set of cells. The price of data in an area is the sum of the costs of the cells within that area. Naively iterating over all cells to sum costs is inefficient, so they precompute and cache price aggregation for interior nodes in the quadtree to estimate the price of data in an area quickly. 
\citet{DBLP:conf/gis/NguyenKS20} consider the interplay between privacy, utility, and price in geo-marketplaces and propose a spatial privacy pricing for maximizing the profit of buyers. In their setting, the privacy concern of each data seller includes its overall privacy concern, its concern regarding a particular data point, and its concern about a particular buyer. The purchase strategy of the data buyer is a sequence of buying actions. Each buying action of the buyer is to propose a price for a data point and get a noisy data point with corresponding noise magnitude compared to the privacy valuation and the price. Then, the pricing problem is to decide the next buying action to maximize the expected incremental profit given the noisy datasets currently owned by the buyer. 

\partitle{Pricing based on data quality}
Data quality, a comprehensive and multifaceted notion, is also a key factor affecting dataset prices. \citet{heckman2015pricing} propose a linear model, where the price of a dataset is measured as a linear function with several quality dimensions, including fixed cost, age, periodicity, volume, and accuracy. In a different approach, \citet{DBLP:conf/apweb/DingWZLG15} evaluate data quality using a logarithm function that focuses on consistency and currency, while also offering recommendations to repair inaccuracies and inconsistencies in the dataset. 
To capture the interaction among multiple quality dimensions, \citet{DBLP:journals/candie/YuZ17} consider two kinds of cost functions dependent on data quality: a linearly increasing cost function and an integrated cost function where one quality dimension may impact the other quality dimension. To maximize the total profit, they formulate the problem of designing the quality level and price for each dataset version as a bi-level programming model. In this model, the monopolistic platform owner acts as the leader, with data consumers as followers. 
\citet{DBLP:journals/toit/CaoPVTLD16} suggest designing a data quality analysis service in a marketplace for near-realtime human-sensing data that uses the quality of data models \cite{DBLP:conf/euromicro/DobsonLS05,DBLP:conf/icws/DAmbrogio06} to continuously monitor and analyze the quality of data streams. \citet{DBLP:conf/infocom/XuZWC22} quantify the value of mobile health data for model training by its uncertainty reduction in model predictions. The pricing method, formulated as a contextual multi-armed bandit problem, dynamically adjusts prices to maximize profits based on the reserve prices of data sellers and the measured value of its data.

Instead, \citet{DBLP:journals/tdsc/XiaoLZ23} view the data costs as private information of the data seller and propose an online learning algorithm to figure out the accurate data costs. They assume that a data seller will sell its data if the price covers the cost. The learning process is divided into exploration and exploitation phases. In the exploration phase, the data collector learns about the cost model of data sellers using a modified stochastic gradient descent (MSGD) algorithm. The collector provides a range of prices to gather the responses of owners. These interactions are used as inputs for the MSGD algorithm, enabling it to refine the cost model against the noise added for privacy concerns. In the exploitation phase, the collector uses the learned model to set appropriate prices for maximizing profit.

\citet{DBLP:journals/pvldb/ZhuZZLR24} demonstrate an interactive dataset pricing system that uses a pricing model trained on real datasets collected from actual data marketplaces to predict the price of a dataset based on its metadata.



\subsubsection{Query Pricing}
Given a database instance $D$ and a query $Q$, query pricing assigns a price $p(Q, D)$ to the query answer $Q(D)$. This price reflects the information of the answer in a fine-grained manner, while the computational overhead of processing the query is generally considered negligible. Studies on fine-grained query pricing have mainly focused on defining a well-functioning pricing function. For cost assessment of query execution, one can refer to \cite{DBLP:conf/sigmod/KantereDGA11}. We summarize works on query pricing in Table \ref{tab:querysummary}.

\begin{table}[htbp]
  \centering
  \caption{Research works about query pricing.}\label{tab:querysummary}
  \adjustbox{width=\linewidth}{
  \scriptsize
    \begin{tabular}{|l|l|l|l|l|l|}
    \hline
    \multicolumn{1}{|c|}{\textbf{Reference}} & \multicolumn{1}{c|}{\textbf{Application}} & \multicolumn{1}{c|}{\textbf{Arbitrage Freeness}} & \multicolumn{1}{c|}{\textbf{Privacy Preservation}} & \multicolumn{1}{c|}{\textbf{Dynamicity}} & \multicolumn{1}{c|}{\textbf{Revenue Maximization}} \\
    \hline
    \citet{koutris2012query,DBLP:journals/jacm/KoutrisUBHS15} & chain query, cyclic query & \cmark   & \xmark    & \xmark    & \xmark \\
    \hline
    \citet{DBLP:conf/webdb/LiM12} & linear query & \cmark   & \xmark    & \xmark    & \xmark \\
    \hline
    \citet{koutris2013toward} & query & \cmark   & \xmark    & \cmark   & \xmark \\
    \hline
    \citet{DBLP:conf/icdt/LiLMS13, DBLP:journals/tods/LiLMS14, DBLP:journals/cacm/LiLMS17} & linear query & \cmark   & \cmark   & \xmark    & \xmark \\
    \hline
    \citet{DBLP:journals/pvldb/LinK14} & query & \cmark   & \xmark    & \xmark    & \xmark \\
    \hline
    \citet{DBLP:conf/sigmod/LiuH14} & query & \xmark    & \xmark    & \cmark   & \xmark \\
    \hline
    \citet{DBLP:journals/pvldb/UpadhyayaBS16} & query & \xmark    & \xmark    & \cmark   & \xmark \\
    \hline
    \citet{DBLP:conf/sigmod/DeepK17} & SQL query & \cmark   & \xmark    & \cmark   & \xmark \\
    \hline
    \citet{DBLP:conf/kdd/NiuZWTGC18} & aggregate query & \cmark   & \cmark   & \xmark    & \xmark \\
    \hline
    \citet{DBLP:journals/pvldb/ChawlaDKT19} & query & \cmark   & \xmark    & \xmark    & \cmark \\
    \hline
    \citet{DBLP:conf/icde/NiuZWTC20} & sequential query & \xmark    & \xmark    & \xmark    & \cmark \\
    \hline
    \citet{DBLP:conf/mdm/ZhengCY20} & query on trajectory data stream & \cmark   & \cmark   & \xmark    & \xmark \\
    \hline
    \citet{DBLP:journals/tkde/MiaoGCPYL22} & incomplete data & \cmark   & \xmark    & \xmark    & \xmark \\
    \hline
    \citet{DBLP:conf/icde/ChenYWWL22} & graph query & \cmark   & \xmark    & \xmark    & \xmark \\
    \hline
    \citet{DBLP:conf/dasfaa/HouQYCW23} & incomplete graph data & \cmark   & \xmark    & \xmark    & \xmark \\
    \hline
    \citet{DBLP:journals/tmc/CaiZWH23} & range counting query & \cmark   & \cmark   & \xmark    & \xmark \\
    \hline
    \end{tabular}%
    }
\end{table}%

\partitle{Pricing with arbitrage freeness}
A seminal vision paper proposed by \citet{DBLP:journals/pvldb/BalazinskaHS11} outlines key challenges associated with cloud-based data markets, including fine-grained pricing, arbitrage freeness, fairness, efficiency, usability, and predictability. To enable fine-grained data pricing, the task is to automatically derive the price of any query result given an individual price for each tuple. They propose lineage (i.e., provenance) of data as a possible approach for computing the price of query results. The provenance information for queries with aggregation is studied by \citet{DBLP:conf/pods/AmsterdamerDT11}. In addition, they would like to support submodularity in the pricing function. They are the first to introduce arbitrage freeness into query pricing.

\citet{koutris2012query,DBLP:journals/jacm/KoutrisUBHS15} propose the first formal query-based pricing framework that can automatically derive the appropriate price of any other query given a few assigned prices to views. They identify two important properties for the pricing function, arbitrage freeness and discount freeness. Arbitrage freeness requires the price of a query does not exceed the combined price of any other set of queries that can produce the same output on any database instance. Thus, the key issue in investigating arbitrage freeness is to determine from which given set of views an answer to a query can be derived. The discount freeness requires the pricing function to derive the maximal feasible price, which is similar to revenue maximization. For full conjunctive queries, they show that the price can be computed in PTIME if the explicit price points are restricted to selection queries with equality predicates. Pricing chain queries can be reduced to the maximum flow problem, and pricing cyclic queries can be reduced to the weighted bipartite vertex cover problem. Later, \citet{koutris2013toward} cast the arbitrage-free pricing problem as an integer linear program, the solution of which is the minimum price of views needed to answer the buyer query under the arbitrage-free requirement. Specifically, each binary variable indicates whether to purchase the view. Constraints are set to ensure the combination of purchased views sufficiently answers the buyer query. The objective is to minimize the total price of purchasing views that satisfy constraints. 

\citet{DBLP:conf/webdb/LiM12} focus on linear queries and introduce a pricing function that is non-disclosive, arbitrage-free, and regret-free. A non-disclosive pricing function prevents buyers from deducing query answers by inquiring about query prices; a regret-free pricing function requires buyers not to be penalized for requesting queries interactively rather than all at once. Specifically, they transform a database instance $\mathbf{I}$ into a data vector $\mathbf{x}$, where each element counts the number of tuples in $\mathbf{I}$ that satisfy each condition. The linear query is represented as a vector $\mathbf{q}$, where each element represents the coefficient corresponding to the condition. The query answer for $\mathbf{q}$ on $\mathbf{I}$ is calculated as the vector product $\mathbf{q} \cdot \mathbf{x}$. Thus, the arbitrage opportunity for buyers lies in using linear algebra to infer query answers. Based on the query matrix and the linear span, they provide inductive/deductive pricing functions that are arbitrage-free and regret-free.

In studies \cite{koutris2012query,DBLP:journals/jacm/KoutrisUBHS15,koutris2013toward,DBLP:conf/webdb/LiM12}, pricing functions are proposed for restricted subclasses of queries. \citet{DBLP:journals/pvldb/LinK14} address the problem of arbitrage-free pricing for arbitrary queries from a theoretical perspective. They provide new pricing functions and consider three types of pricing models for query bundles: instance-independent pricing, up-front dependent pricing, and delayed pricing. They also summarize five conditions for avoiding arbitrage in instance-independent and delayed pricing: avoiding price-based arbitrage, separate-account arbitrage, post-processing arbitrage, almost-certain arbitrage, and serendipitous arbitrage. They highlight the tension between flexible pricing and arbitrage freeness, and show how it can result in unreasonable prices.

From a practical perspective, \citet{DBLP:conf/sigmod/DeepK17} present QIRANA, a system that supports real-time pricing of SQL queries with arbitrage freeness. To enable arbitrage freeness, QIRANA regards a query as a database instance reduction problem. A query is considered more informative than the other if it classifies more inconsistent database instances, thus the price of the query can be formulated as a function of how much database instances shrink. To make the computation feasible, they show that it suffices to examine a subset of all possible database instances which is called the support set. QIRANA uses uniform random samples and random neighbors as the support set. QIRANA offers several pricing functions, including the weighted coverage function, the uniform gain function, Shannon entropy function, and q-entropy function, which are information arbitrage-free and bundle arbitrage-free. A prototype demonstration system of QIRANA is reported in \cite{DBLP:journals/pvldb/DeepKB17}.

Since incomplete data is pervasive in real applications, \citet{DBLP:journals/tkde/MiaoGCPYL22} consider the scenario of buying and selling incomplete data. They propose a pricing mechanism, iDBPricer, and present two pricing functions, UCA price and QUCA, to price queries over incomplete data based on data contribution, data completeness, and query quality. The UCA price discounts the base price by the completeness degree of the lineage tuples. The QUCA price considers the missing state and the quality of lineage tuples, and prevents leaking some information about the result set and lineage set size, which is more practical than UCA.

Orthogonal to relational data, some efforts are made to price non-relational data \cite{DBLP:conf/icde/ChenYWWL22,DBLP:conf/dasfaa/HouQYCW23}. In order to price graph data, \citet{DBLP:conf/icde/ChenYWWL22} propose GQP, a framework that derives arbitrage-free prices for graph queries from a set of explicit graph views with prices specified by data sellers. The support of a query is defined as the set of views that can answer the query by graph matching. To avoid arbitrage, the price of a query is the minimum price among all sets of views that cover the query. GQP uses a greedy strategy to approximate the price of a query by iteratively selecting views with the maximal ratio of the number of remaining edges to the price until the query is covered. If the query is not covered by any view, GQP approximates its price by the price of another query subject to coverage by a view. Considering changes in graph views and prices, DyDQP leverages previous results to save computational cost. \citet{DBLP:conf/dasfaa/HouQYCW23} extend to price queries on incomplete graph data. Given that incomplete graph data may degrade query quality, they compute discounted prices for queries based on a query quality metric called the certainty coefficient.

To address the arbitrage problem in pricing queries, these studies calcuate the price of a new query based on the prices of base queries, which can be regarded as a kind of unit prices. This approach captures the combined cost of information in queries and prevents buyers from exploiting price inconsistencies through strategic combinations. Nevertheless, the practice of providing combined offers for more profits, i.e., setting query prices that are favourable but still arbitrage-free, is not the focus of these studies but is worth considering.

\partitle{Pricing with privacy preservation}
The mapping of privacy loss to arbitrage-free pricing is well investigated. \citet{DBLP:conf/icdt/LiLMS13, DBLP:journals/tods/LiLMS14, DBLP:journals/cacm/LiLMS17} present the first query-based marketplace with privacy consideration, where the market maker sells noisy linear queries. To protect the privacy of data sellers, the market maker computes the true answer to a linear query and then adds Laplace noise to ensure differential privacy. The pricing function for noisy linear queries depends on the query itself and the amount of noise added (the variance of the query answer). Variance introduces an additional arbitrage opportunity. That is, a buyer can issue multiple queries with high variance to derive a query answer with low variance. To address this, they propose a class of arbitrage-free pricing functions for query $q$ with variance $v$: $\pi(q,v) = \frac{f^2(q)}{v}$, where $f(q)$ is a semi-norm function depending on the query $q$. This pricing function is adopted for trading aggregate statistics over private correlated data by \citet{DBLP:conf/kdd/NiuZWTGC18}.

\nop{
Focusing on trading aggregate statistics over private correlated data, \citet{DBLP:conf/kdd/NiuZWTGC18} propose a pricing framework ERATO. In ERATO, data buyers are allowed to request its customized service $S = (\mathbf{w}, v)$, where $\mathbf{w}$ is a weight vector associated with the data and $v$ is a tolerable variance of noise added to the returned answer. Arbitrage happens if a data buyer can average multiple answers and get an unbiased answer but with lower variance and a lower price. To this end, \citet{DBLP:conf/kdd/NiuZWTGC18} specify the service determinacy relation between requested services instead of the concept of determinacy relation in queries/views. They prove that an arbitrage-free pricing function $\pi(\mathbf{w}, v)$ cannot decrease faster than $1/v$. To further ensure arbitrage free on weight vector $\mathbf{w}$, they prove that a pricing function $\pi(\mathbf{w}, v) = g(\mathbf{w})^2/v$ for some positive function $g(\mathbf{w})$ that only depends on $\mathbf{w}$ is arbitrage-free iff $g(\mathbf{w})$ is a semi-norm.
}

\citet{DBLP:journals/tmc/CaiZWH23} propose a framework for trading range counting queries in Internet of Things systems, aiming for bandwidth efficiency, privacy preservation, and arbitrage freeness. The framework uses two storage structures: histogram sketches for exact counts in disjoint ranges and sample sets for storing data sampled according to an optimal probability that balances variance and bandwidth consumption. Given a query range, it aggregates exact counts of fully covered histograms and approximate counts from partially covered sample sets, adding minimum noise to ensure differential privacy. They list the requirements of arbitrage-free pricing functions in two scenarios: periodical and query-based data collection. In periodical data collection, pricing is based on variance to prevent buyers from averaging higher variance results to obtain a low variance result. In query-based data collection, pricing is based on the sampling probability and the differential privacy budget to prevent buyers from combining low-accuracy queries into a higher-accuracy query.

While \citet{DBLP:conf/icdt/LiLMS13, DBLP:journals/tods/LiLMS14, DBLP:journals/cacm/LiLMS17,DBLP:conf/kdd/NiuZWTGC18,DBLP:journals/tmc/CaiZWH23} set uniform privacy losses for different data sellers, \citet{DBLP:conf/mdm/ZhengCY20} enable data sellers to control their personalized privacy loss when trading their personal streaming trajectory data. In their setting, data sellers can specify their own privacy loss bound, a compensation rate, and a sliding-window size under personalized $\mathcal{W}-$event privacy. A data buyer can request a histogram query over the trajectory dataset with a specific histogram variance at a time point. 
The pricing function is constructed by using the inverse of a utility function that maps privacy losses to histogram variances. They propose $\phi-$pattern to ensure the existence of the inverse utility function and prove arbitrage freeness of the pricing function by satisfying certain properties of the utility function. 

To ensure privacy preservation, true query answers are perturbed with noise, implementing differential privacy. The noise variance poses an additional challenge in defining a pricing function, requiring it to be arbitrage-free not only at the query level but also at the variance level. These studies give the essential properties that the pricing function should possess.

\partitle{Pricing with dynamicity}
Pricing for dynamicity prevents double charging for overlapping query answers when buyers submit multiple queries over time, regardless of whether the database is static or dynamic. \citet{koutris2013toward} considers the pricing of views. It computes and stores a set of views with a minimum cost that determines the buyer-purchased queries. These views are then recorded as purchased by the buyer and assigned update prices for database updates. If views remain unmodified after a database update, the buyer can access them for free in future queries, otherwise, the update price applies. \citet{DBLP:conf/sigmod/DeepK17} store buyer query history and charge buyers a weighted sum of database instances in the support set where the new query behaves inconsistently with previous queries. These history-based approaches face the disadvantage of the storage cost for buyer history. To address the issue, \citet{DBLP:journals/pvldb/UpadhyayaBS16} introduce a refund-based approach that allows buyers to request refunds for repeated purchased data in query answers. Refunds are implemented through refund coupons generated by the system in every transaction. A coupon contains unique identifiers for data and queries, as well as a cryptographic hash generated using identifiers and a secret key held by the seller. When a buyer submits a refund request with coupons, the seller verifies the coupons by checking the hashes.

To optimize query sharing and efficiently maintain views, \citet{DBLP:conf/sigmod/LiuH14} propose an algorithm of MANAGEDRISK. MANAGEDRISK introduces the notion of regret, which measures the additional cost incurred by not using a subexpression in the previous sharing. MANAGEDRISK measures sharing plans with a function of regret and cost and chooses the sharing plan with the maximum score. Since the price of data in their data market is determined as a function of the monetary value of the data specified by the owner and the operational cost, they further calculate the cost of each sharing plan by building the partial order of sharings and calculating the cost savings of using intermediate results, which satisfies five fairness criteria.

\partitle{Pricing with revenue maximization}
\nop{
\citet{DBLP:conf/soda/GuruswamiHKKKM05} study the optimization problem of revenue maximization with an envy-free guarantee. Envy-free ensures that no individual would prefer to switch their payment and privacy cost with each other. They investigate two cases of inputs: unit demand consumers and single-minded consumers and show the optimization problem is APX-hard for both cases, which can be efficiently solved by a logarithmic approximation algorithm.
}
\citet{DBLP:journals/pvldb/ChawlaDKT19} investigate three types of pricing functions and study the corresponding revenue maximization problems. They consider the setting of unlimited supply and single-minded buyers, where the seller can sell an unlimited number of units of each query, and each buyer will buy only a single query priced below its valuation. Focusing on maximizing revenue under three subclass of monotone and subadditive pricing functions: the uniform pricing function, the additive pricing function, and the XOS pricing function, they empirically evaluate the performance of several approximation algorithms proposed in previous work \cite{DBLP:conf/soda/GuruswamiHKKKM05,DBLP:conf/focs/CheungS08,DBLP:conf/sigecom/BalcanB06} and develop several new heuristics. In addition, \citet{DBLP:conf/icde/NiuZWTC20} explore how the data broker maximizes its cumulative revenue by posting reasonable prices for sequential queries. They propose a contextual dynamic pricing mechanism with the reserve price constraint, which supports both linear and non-linear market value models and is robust to uncertainty. 

\subsubsection{Model Pricing}
Due to the increasing pervasiveness of machine learning-based analytics, there is an emerging interest in studying the price of machine learning models. \citet{DBLP:conf/sigmod/ChenK019} introduce the first framework for model pricing, considering the varying amounts of noise added to the models. They formulate an optimization problem to find the arbitrage-free price that maximizes the data broker revenue and prove this problem is coNP-hard. To address this challenge, they replace the subadditivity constraint with the condition that the unit price of models, defined as the model price divided by the model noise, decreases as the noise decreases, and propose a dynamic programming algorithm. A prototype demonstration system called Nimbus is reported in \cite{DBLP:conf/sigmod/ChenWCK019}. However, 
they assume there is only one (averaged) survey price point for each model, which is oversimplified; and they do not have an explicit privacy analysis and guarantee for their proposed Gaussian mechanism that adds noise to the models. 

Further, \citet{DBLP:journals/pvldb/LiuLL0PS21} propose the first end-to-end model marketplace with differential privacy Dealer, which formalizes the abilities and restrictions of all market participants (i.e., data sellers, data broker, and model buyers) and studies their abbreviated interactions. Dealer depicts the full marketplace dynamics through two important functions, including model pricing and model training. Before releasing model versions and their prices, the data broker conducts a market survey to collect price points (budget for a target model) from model buyers. Based on survey price points, the data broker prices models with the aim of maximizing revenue and at the same time following the market design principle of arbitrage freeness, which is formalized as a revenue maximization problem of model pricing. A special case of the revenue maximization problem, where each model version has one survey price point, has been studied by \citet{DBLP:conf/sigmod/ChenK019}. To cope with the general case where each model has multiple survey price points, they relax the revenue maximization problem and propose an efficient dynamic programming algorithm having a lower bound with respect to the maximum revenue. Specifically, they first construct a complete price space with unit prices determined by survey price points, and propose a recursive equation to find the optimal solution of the relaxed revenue maximization problem. For model training with the aim of maximizing Shapley coverage given a manufacturing budget, they propose efficient dynamic programming, greedy, and guess-based algorithms with approximation guarantees. A prototype demonstration system of Dealer is reported in \cite{DBLP:journals/pvldb/LiuLZR0L0PS21}. 

In addition, \citet{DBLP:journals/isci/JiangXWYWL23} propose an online pricing mechanism for data classification services, named DIVINE. DIVINE allows buyers to choose from different versions of classifiers with different prices based on their accuracy requirements. To maximize revenue, DIVINE infers the valuations of buyers from their responses, dynamically adjusts the probability of candidate prices being selected, and sets prices online for each buyer.

\nop{
Federated learning provides a promising approach to mitigating privacy concerns by enabling data owners to train machine learning models collaboratively without directly sharing their raw data with data brokers. Based on this, \citet{DBLP:conf/infocom/SunCLH22} propose a machine learning model marketplace with differentially private federated learning from a broker-centric perspective. They follow the revenue maximization principle to decide the price of each model version.
}

\citet{DBLP:journals/isci/JiangNYWL22} propose RARIEA, a framework for trading private data generators that augments data while protecting privacy. RARIEA adopts the GAN as the data generator and adds noise during GAN training to ensure privacy protection. RARIEA uses Rényi differential privacy to quantify the privacy loss of each data seller and compensates them accordingly. The price of the GAN is calculated by multiplying the total privacy compensation by a profit ratio $1+\lambda$, where $\lambda$ represents the desired profit margin set by the data broker.

\subsection{Game-theoretic Pricing}\label{subsec:gametheorypricing}
Game theory \cite{T1982Dynamic} offers a widely adopted mathematical framework for reasoning about strategic behaviors of agents, which is natural in markets. In data markets, game theory as well as its celebrated branches such as auction theory \cite{roughgarden2010algorithmic}  and bargaining  \cite{muthoo1999bargaining} have all be used to formulate data pricing problems. 
\citet{vision} present a blueprint for applying game-theoretic modeling to data pricing via a four-dimensional framework. By identifying four key dimensions (\emph{Participant}, \emph{Object}, \emph{Action}, and \emph{Information}) for  data pricing, they examine granular challenges and propose corresponding methods within each dimension, which enables a practical data pricing problem to be composed and solved by the outlined methodologies and also provides a lens for analyzing related works on game-theoretic pricing. 
The studies on game-theoretic pricing in data markets \cite{DBLP:conf/www/GrubenmannBMS18, DBLP:conf/sigecom/GhoshR11, wang2016strategy, DBLP:conf/ec/AgarwalDS19, DBLP:journals/corr/abs-2003-08345, DBLP:journals/corr/abs-2202-08780, DBLP:journals/jsac/ZhengPWTC17,
DBLP:journals/jsac/ZhangAWB21, DBLP:journals/pacmmod/Fernandez23, DBLP:conf/uai/ZhangBL20, DBLP:journals/ton/ZhangYHLZ21, DBLP:conf/dasfaa/AnX0GZ19, DBLP:journals/isci/XiongX21, zheng2022fl, DBLP:journals/tmc/DuGJHR20, DBLP:conf/sigmod/Fernandez22, DBLP:conf/icassp/CaoCL17, DBLP:journals/cj/CaiZLY19, DBLP:journals/corr/abs-2107-08630, 
an2021crowdsensing, DBLP:journals/tmc/XuXWZG23, DBLP:conf/infocom/XiaoXZWZZ23, DBLP:conf/icc/XuJWRYG15, 9322221, DBLP:journals/jsac/ZhangAHP21, DBLP:journals/tmc/YuCH22, DBLP:conf/bigdataconf/JungP19, DBLP:journals/isr/RayMM20, share, 10037216, DBLP:conf/nips/RavindranathJP23, DBLP:conf/infocom/JinXLG19} can be generally divided into three categories as follows: auction-based pricing, Stackelberg game-based pricing, and bargaining-based pricing.

\subsubsection{Auction-based Pricing}
Auction design is a subfield of game theory that formulates market mechanisms by designing payment and allocation rules during market transactions. An ideal auction \cite{roughgarden2010algorithmic} should satisfy several natural properties, 1) incentive compatibility (truthfulness) which means bidding true value is an optimal strategy for bidders, 2) individual rationality which indicates that the bidder's utility is no less than the opt-out choice of not participating in the trading, 3) optimizing certain objectives such as the social welfare or the seller's revenue as often considered, 4) budget balance which refers to no incurred deficit to the data broker, and 5) computationally efficiency. However, these desired requirements can be hard to meet simultaneously. In terms of data markets, various data auction frameworks \cite{DBLP:conf/www/GrubenmannBMS18, DBLP:conf/sigecom/GhoshR11, wang2016strategy, DBLP:conf/ec/AgarwalDS19, DBLP:journals/corr/abs-2003-08345, DBLP:journals/corr/abs-2202-08780, DBLP:journals/jsac/ZhengPWTC17,
DBLP:journals/jsac/ZhangAWB21, DBLP:journals/pacmmod/Fernandez23, DBLP:conf/uai/ZhangBL20, DBLP:journals/ton/ZhangYHLZ21, DBLP:conf/dasfaa/AnX0GZ19, DBLP:journals/isci/XiongX21, zheng2022fl, DBLP:journals/tmc/DuGJHR20, DBLP:conf/sigmod/Fernandez22, DBLP:conf/icassp/CaoCL17, DBLP:journals/cj/CaiZLY19, DBLP:journals/corr/abs-2107-08630, DBLP:conf/nips/RavindranathJP23, DBLP:conf/infocom/JinXLG19} have been proposed to facilitate data trading under certain rules, which differ in application scenarios, optimization goals, desired properties, and underlying mechanisms. The related works are summarized in Table \ref{tab:auctionsummary} based on the application scenario, applied method, and satisfied desiderata. We describe these related works in more detail as follows which are reorganized by the specific approaches used.

\begin{table}[htbp]
  \centering\footnotesize
  \caption{Research works about auction-based pricing.}\label{tab:auctionsummary}
\adjustbox{width=\linewidth}{
    \begin{tabular}{|c|c|c|c|}
    \hline
    \textbf{Method} & \textbf{Reference} & \textbf{Scenario}  & \textbf{Desiderata}  \\
    \hline
    \multirow{3}{*}{VCG auction} & \citet{DBLP:conf/www/GrubenmannBMS18} & the Web of Data  & \makecell[c]{truthfulness,\\revenue maximization/welfare maximization}\\
     \cline{2-4}
    & \citet{DBLP:conf/sigecom/GhoshR11} & private data trading  & \makecell[c]{truthfulness, individual rationality}\\
    \cline{2-4}
    &\citet{wang2016strategy} & data auction with negative externalities  & \makecell[c]{truthfulness, individual rationality,\\welfare maximization}\\
    \hline
    \multirow{6}{*}{Myersonian auction} &
    \citet{DBLP:conf/ec/AgarwalDS19} & \makecell[c]{data markets for\\ machine learning tasks} &  \makecell[c]{truthfulness,\\revenue maximization/welfare maximization}\\
    \cline{2-4}
    &\citet{DBLP:journals/corr/abs-2003-08345} & data trading with externalities  & \makecell[c]{truthfulness, individual rationality,\\revenue maximization/welfare maximization}\\
    \cline{2-4}
    &\citet{DBLP:journals/corr/abs-2202-08780} & \makecell[c]{selling information in\\ competitive environments}  & \makecell[c]{truthfulness, individual rationality,\\revenue maximization/welfare maximization}\\
    \cline{2-4}
    &\citet{DBLP:journals/jsac/ZhengPWTC17}& data acquisition for mobile crowdsensing & truthfulness, individual rationality\\
    \cline{2-4}
    &\citet{DBLP:journals/jsac/ZhangAWB21} & Fresh Data Acquisition & truthfulness, individual rationality\\
    \cline{2-4}
    &\citet{DBLP:journals/pacmmod/Fernandez23} & data sharing markets & \makecell[c]{truthfulness, individual rationality, \\revenue maximization}\\
    \hline
    \multirow{2}{*}{knapsack auction} &
    \citet{DBLP:conf/uai/ZhangBL20} & private data trading  & \makecell[c]{truthfulness, individual rationality, \\budget balance}\\
    \cline{2-4}
    &\citet{DBLP:journals/ton/ZhangYHLZ21} & crowdsensing data trading & \makecell[c]{truthfulness, individual rationality, \\computational efficiency}\\
    \hline
    \multirow{2}{*}{decentralized auction} 
    &\citet{DBLP:conf/dasfaa/AnX0GZ19} & crowdsensing data trading & truthfulness, individual rationality\\
    \cline{2-4}
    &\citet{DBLP:journals/isci/XiongX21} & anti-collusion data trading & $/$\\
    \hline
    \multirow{2}{*}{DL-powered auction} 
    &\citet{zheng2022fl} & private data model trading  & \makecell[c]{(approximate) truthfulness, individual rationality}\\
    \cline{2-4}
    &\citet{DBLP:conf/nips/RavindranathJP23} & information trading with externalities &(approximate) truthfulness, individual rationality\\  
    \hline
    \multirow{2}{*}{sequential auction} 
    &\citet{DBLP:journals/tmc/DuGJHR20} & \makecell[c]{data allowance transaction \\in mobile networks} &  $/$\\
    \cline{2-4}
    &\citet{DBLP:conf/sigmod/Fernandez22} & data markets with strategic buyers & $/$\\
    \hline
    iterative auction &\citet{DBLP:conf/icassp/CaoCL17} & \makecell[c]{three-sided data markets \\with selfish agents} & \makecell[c]{individual rationality, budget balance, \\welfare maximization}\\
    \hline
    \multirow{2}{*}{double auction} &
    \citet{DBLP:journals/cj/CaiZLY19} & \makecell[c]{data markets with preferences \\and conflicts of interest}  &\makecell[c]{truthfulness, individual rationality,\\ budget balance, computational efficiency,\\(approximate) welfare maximization}\\
    \cline{2-4}
    &\citet{DBLP:journals/corr/abs-2107-08630} & data sharing markets & \makecell[c]{truthfulness, individual rationality, \\budget balance, welfare maximization}\\
    \hline
    \end{tabular}%
    }
\end{table}%

\partitle{VCG Auction}
A Vickrey-Clarke-Groves (VCG)  auction \cite{roughgarden2010algorithmic} (named after its inventors Vickrey, Clarke, and Groves) is a generalization of the single-good second price auction of \citet{10.2307/2977633}, which sells the single item to the buyer with the highest bid and charges her a price equaling the second highest bid. VCG generalizes this mechanism to the sale of multiple goods with possibly correlated item values. In this case, the buyer may be asked to bid one value for each subset of the items, hence sometimes also referred to as combinatorial auctions. The VCG auction admits a clean and elegant design with strong incentive guarantees. At an intuitive level, the auction will allocate available items in a socially optimal manner and simply charge each bidder her ``marginal harm'' to other bidders  
The VCG auction is incentive compatible and individual rational for each bidder \emph{regardless} what bids any other bidder submits, and maximizes social welfare at equilibrium. Despite these nice incentive properties, several challenges remain when applying VCG  to practice, e.g., the difficulty of obtaining (possibly exponentially) many bids from bidders for each possible combination of items, the computational complexity even for approximate welfare optimization, and the weak performance in terms of the seller's revenue.
Some works \cite{DBLP:conf/sigecom/GhoshR11, DBLP:conf/www/GrubenmannBMS18, wang2016strategy} have introduced VCG auction into data pricing, especially for the guaranteed truthfulness which is desired in data markets just as in traditional auction markets. 

To create incentives to subsidize data providers in the context of the Web of Data (a massive network of interlinked data where data is freely available), \citet{DBLP:conf/www/GrubenmannBMS18} propose that sponsors should pay the data providers to promote sponsored data which would be privileged over non-sponsored data. They introduce a delayed-answer auction where sponsors act as bidders to prioritize their data which appears in batches with different delays. A click model for SPARQL query answers, Batch Link Model, is designed to formulate the probability of a user selecting a link (answer), and the VCG mechanism \cite{roughgarden2010algorithmic} is adapted into a weighted form to decide the payment that sponsors should pay when their sponsored data is chosen, which can ensure the truthfulness. They also extend the auction model to take the user's experience into account by assigning part of the data randomly to batches instead of deciding their sequence in strict accordance with the bids. 

Instead of the Web of Data, \citet{DBLP:conf/sigecom/GhoshR11} design auction-based mechanisms for a more regular two-sided data market where private data is traded between a data analyst who aims to query private data to estimate statistics and data sellers who want compensation for their privacy loss. Based on differential privacy, they formulate the problem into variants of multi-unit procurement auctions where each data seller submits a bid reflecting their privacy valuation and the data analyst decides on a level of privacy to be purchased and produces a noisy query output.  Two auction models are designed which are optimal up to small constant factors for two settings respectively. In one model,  they apply the VCG auction \cite{roughgarden2010algorithmic} which can guarantee truthfulness and individual rationality to minimize the total payment with a fixed accuracy goal.  In another, a FairQuery mechanism is proposed to maximize accuracy with a fixed budget.

Rather than the privacy concerns of data sellers, \citet{wang2016strategy} pay attention to the negative externalities among data buyers, as mentioned in Section \ref{subsec:difference}. The authors argue a simple but effective binary valuation for buyers in the context of competition scenarios, which include the externality of data into account, i.e., buyer $i$ has value $v_i$ if the number of its competitors who can get the data from the given set $S_i$ is no more than $t_i$. By parameterizing the buyer type as a triple $\theta_i=\left(S_i, t_i, v_i\right)$ notifying the private information including the set of competitors, the tolerance bound, and the valuation, \citet{wang2016strategy} adapt the VCG auction \cite{roughgarden2010algorithmic} to determine the payment after using an efficient algorithm to compute the optimal allocation for social welfare. The use of VCG auction guarantees the desired strategy-proof property, i.e., incentive compatible and individual rational. They also consider partial competition markets where the optimal allocation is proved to be NP-hard and present a greedy approximation allocation algorithm combined with a payment scheme using critical bid, i.e., the payment $p_i$ is the bid of the critical buyer who is the first buyer failing to get the data but would have been allocated the data without buyer $i$. The strategy-proofness is also proved for the data auction mechanism with partial competition. 

Some assumptions are set in the above VCG auction-based works  which, however, can be somewhat strong. For example, Batch Link Model in the work \cite{DBLP:conf/www/GrubenmannBMS18} requires that all the available query answers have the same relevance to the user. \citet{DBLP:conf/sigecom/GhoshR11} initiates the study of markets for private data but considers a simple market, e.g., only a single analyst to buy data from a population. 
The important characteristic of data, i.e., externality, is considered by \citet{wang2016strategy} but is simply modeled as binary. 

\partitle{Myersonian Auction}
The revenue-optimal auction designed by \citet{DBLP:journals/mor/Myerson81} (also widely referred to as Myerson's or \emph{Myersonian auction}) aims at maximizing total expected revenue in single-parameter environments where each bidder has a private value (hence a ``single parameter'') for the to-be-sold item. Fundamental to Myerson's design is the so-called \emph{Myerson's lemma} which characterizes the incentive-compatible payment as a uniquely determined  function of the item allocation rule. Armed with this lemma, Myerson transforms the expected revenue into the expected virtual welfare, i.e., the welfare under carefully defined ``virtual value'' of each bidder. Perhaps surprisingly, this intricate analysis leads to a surprisingly simple (and widely used) auction format which turns out to be revenue-optimal. Specifically, when bidders' valuations are drawn i.i.d., Myerson's optimal auction (under mild regularity assumption) is simply the second price auction, but with a carefully designed reserve payment. This optimal auction can be naturally generalize to independent yet non-identical values, where each bidder will have a personalized reserve price.  Perhaps even more surprisingly, despite this elegant solution from Myerson for the single-parameter auction design, the design problem turns out to become dramatically more difficult once we enter multiple parameter regime, in fact even two-parameter regime \cite{daskalakis2015multi}.
Some work \cite{DBLP:conf/ec/AgarwalDS19, DBLP:journals/corr/abs-2003-08345, DBLP:journals/corr/abs-2202-08780, DBLP:journals/pacmmod/Fernandez23, DBLP:journals/jsac/ZhengPWTC17, DBLP:journals/jsac/ZhangAWB21} have introduced Myersonian auctions including its payment formula and the concept of virtual value into data markets.

As an asset, data is freely replicable, combinatorial, having no prior on values, and with varying values for varied tasks, which is pointed by \citet{DBLP:conf/ec/AgarwalDS19}.  Due to the proposed four characteristics of data, they claim that the traditional auction markets (e.g., online ad auctions) do not suffice. For example, the second price auction of  \citet{10.2307/2977633} cannot directly apply to data pricing because data can be freely replicated, and bids cannot be directly made on raw data because buyers have no prior on the accurate value of data. Therefore, they propose a truthful auction mechanism to trade machine learning models (a specific type of data goods) and the training data. Specifically, the buyer's valuation of data is defined based on the increase in prediction accuracy of data models. 
The revenue (payment from buyers) function is formulated based on Myerson's payment function \cite{DBLP:journals/mor/Myerson81} which ensures the truthfulness of buyers, and the allocation (the data quality allocated to the buyer) function is set to collectively adjust the data quality by degradation according to the bids and market prices. Moreover, Shapley value \cite{shapley1953value} is adapted to take the freely replicable property of data into account for revenue division among data sellers.  

In their follow-up works, \citet{DBLP:journals/corr/abs-2003-08345} further extend the Myersonian auction format \cite{DBLP:journals/mor/Myerson81} to the setting of data auctions with externalities. Instead of the binary model for the data externality in the work \cite{wang2016strategy}, the authors assume that the externalities are additive and formulate the valuation of data as a linear function of the allocation vector.
Based on the valuation model in the work  \cite{DBLP:conf/ec/AgarwalDS19}, they reduce the combinatorial problem of allocating and pricing multiple datasets to the auction of a single digital good by using the increase in prediction accuracy to measure utility for data. Two scenarios differing in the form of buyers' private information are considered. Both welfare and revenue maximizing mechanisms are designed which satisfy incentive compatibility and individual rationality. For the former, the VCG mechanism \cite{roughgarden2010algorithmic} is used, and for the latter, virtual values proposed in Myersonian auction \cite{DBLP:journals/mor/Myerson81} reduce the problem of maximizing revenue to maximizing virtual welfare. 

The externality of data is also taken into account in \cite{DBLP:journals/corr/abs-2202-08780}. The authors study a different setting, i.e., a data seller needs to determine how to sell information to buyers in competitive environments. They formulate the competitive relationship among buyers as a binary game (with binary actions) and derive the general recommendation rule (how the seller allocates the information) for a general optimization object subject to specific constraints which can satisfy individual compatibility, individual rationality, and a well-defined property, obedience, to ensure that buyers prefer to follow the seller's recommendation. Similar to the work \cite{DBLP:journals/corr/abs-2003-08345},  the externalities among downstream buyers are considered and the mechanism objectives of both welfare maximization and revenue maximization are considered. Specifically, the formulation of revenue maximization is extended from the welfare maximization problem through virtual values introduced from Myersonian auction \cite{DBLP:journals/mor/Myerson81}. 

\citet{DBLP:journals/jsac/ZhengPWTC17} focus on a special scenario of data trading, crowd-sensed data markets. By building a statistical model upon raw data as the commodity, they capture the uncertainty and correlation of data, and cater to the diverse demands via versioning. Based on this designed data trading format, a framework, VENUS, for profit maximization of the data broker is proposed, which is composed of two mechanisms, VENUS-PRO and VENUS-PAY. Taking the payment determined by VENUS-PAY as input, VENUS-PRO applies a greedy algorithm to select data acquisition points for profit maximization. Specifically, the posting price for each version candidate is determined to maximize the revenue and the best version is then derived for which the necessary data points with minimum expenditure are selected. 
VENUS-PAY describes a data procurement auction between the data broker and data providers, for which the Myersonian auction \cite{DBLP:journals/mor/Myerson81} with virtual values is adopted to provide truthfulness and individual rationality as well as to realize the payment minimization (the reserve revenue maximization). 

Similarly, \citet{DBLP:journals/jsac/ZhangAWB21} model a data procurement auction to buy data from various sources yet take the fressness of data, age-of-information (AoI) into account. A destination (data buyer) wants to acquire data updates from multiple strategic sources (data sellers) with private sampling costs. Data sources as bidders aim to maximize their payoff composed of the attained payment and sampling cost while the data destination as the auctioneer aims to minimize its payments and age-related costs. A Myersonian auction \cite{DBLP:journals/mor/Myerson81} is established which can incentivize data sources to report truthful costs and ensure their individual rationality while meantime optimizing the goal of the destination. A quantized mechanism with asymptotic optimality is further proposed to resolve the computational overheads of the optimal mechanism based on the Myersonian auction. Compared with benchmarks of second-price auction and complete-information auction, both optimal and quantized mechanisms are verified great benefits.

The aforementioned works typically distinguish data sellers from data buyers clearly and focus on the one-sided data flow (from sellers to buyers). Instead, \citet{DBLP:journals/pacmmod/Fernandez23} approaches data markets from a different way, i.e., Data-Sharing Consortia. In the proposed data-sharing markets, participants constitute a seller and a buyer at the same time and share data to accomplish a common task. A four-stage protocol  is designed to implement sharing dominance (individual rationality) and entitlement stake (fair revenue allocation) incentives for participants which hence incentivizes the creation of data-sharing markets and unleashes the value of data. The first stage is Contract Agreement which stipulates rules chosen by participants. The second stage is Signalling, where the trusted platform performs the task on the data and signals the value of the data to each participant. A robustness algorithm is included to protect the consortium from bad inputs. The third stage is Value Extraction where the platform is meant to elicit truthful value from participants and to maximize total payments (revenue). Myerson's optimal auction \cite{DBLP:journals/mor/Myerson81} is adopted since in such scheme participants do not pay their bid and are thus incentivized to bid truthfully as that maximizes their chances of winning. Besides this homogeneous price auction mechanism which charges each winning buyer the same price, the author also presents a discriminatory price auction mechanism which borrows from the sequential second auction \cite{DBLP:conf/stoc/SyrgkanisT13}. 
The last stage is Value allocation, aiming at allocating the revenue fairly among participants. By analyzing the inefficiency of Shapley value based revenue allocation, an alternative efficient algorithm is proposed, of which the intuition follows closely that of leave-one-out. Integrating four stages into the data-sharing protocol, the consortium is evaluated viable and sustainable.

The studies by \citet{DBLP:conf/ec/AgarwalDS19,DBLP:journals/corr/abs-2003-08345} have opened up new research directions on some critical issues in data markets, i.e., how to deal with the special properties of data when designing trading mechanisms and how to take the externality into account. However, there are limitations to be further solved, for example, the replicability of data is only embodied in revenue division for sellers and the model for an externality is simply assumed linear.
In terms of the work \cite{DBLP:journals/corr/abs-2202-08780}, the proposed framework requires binary states, binary actions, and linear payoff structure, which needs to be further extended for practical applicability. 
While Myersonian auction is combined in \cite{DBLP:journals/jsac/ZhengPWTC17, DBLP:journals/jsac/ZhangAWB21} to minimize the purchaser's expenditures with desirable economic properties, only one provider is considered in \cite{DBLP:journals/jsac/ZhengPWTC17} which impairs the advantage of mobile crowdsensing and the quality of the acquired data except the fressness is not considered in \cite{DBLP:journals/jsac/ZhangAWB21}.  
The sharing model established by \citet{DBLP:journals/pacmmod/Fernandez23} comprehensively codes the framework and procedure of data sharing, yet requires a trusted platform to implement the task on every participant's data as well as the pooling data, which may be costly. Alternative ways of signaling data value to buyers or enabling buyers to discover data value proactively are worth further investigation.

\partitle{Knapsack Auction}
Knapsack auction \cite{10.5555/1109557.1109677} models the allocation problem where the number of items is limited. Each bidder $i$ has a private value $v_i$ measuring the gained utility if acquiring the desired $\omega_i$ items, and the total number of items, $W$, is publicly known. The problem is to determine the allocation rule $x_i\in \{0,1\}$ in order to maximize social welfare $\sum_{i}v_i x_i$ with the constraint of $\sum_i \omega_i x_i \le W$, which is yet NP-hard. The approximation algorithms for knapsack auction \cite{10.5555/1109557.1109677} have been studied. Some settings in data markets can be reduced to a knapsack auction, which has been studied by researchers \cite{DBLP:journals/ton/ZhangYHLZ21,DBLP:conf/uai/ZhangBL20}.

Private data query has been a hotspot in data markets. \citet{DBLP:conf/uai/ZhangBL20} study how to query private data from multiple data sellers to produce aggregated statistics. Single-minded data sellers defined in the work would sell data only when their distinctive privacy requirements are met which are formulated based on personalized differential privacy mechanisms. SingleMindedQuery (SMQ) mechanism is proposed to select data sellers and determine their compensations to derive accurate outputs for commonly-used queries while preserving every data owner’s declared privacy requirement.  Specifically, the accuracy goal is transformed into the purchased privacy expectation maximization problem, which is then turned into the knapsack problem \cite{10.5555/1109557.1109677} where the budget can be treated as the capacity of the knapsack, the privacy requirement as the value, and the data valuation as the weight of items. They solve the problem by finding an optimal threshold based on the Lagrange function. They justify that the proposed mechanism satisfies incentive compatibility, individual rationality, and budget balance.

A similar private data trading problem is studied by \citet{DBLP:journals/ton/ZhangYHLZ21}, which targets the mobile crowdsensing situation. In their work, the platform acts as an auctioneer running an auction to recruit workers who are allowed to report noisy versions of their data for privacy protection.  Privacy-passive scenarios (where workers participate if adequately compensated) and privacy-proactive scenarios (where workers have privacy requirements) are considered respectively. For the former, they formulate the problem as allocating the sensing task to a set of workers that can minimize the total payment to the workers subject to the accuracy constraint of the aggregated result, which can be reducible to a reverse binary knapsack problem \cite{10.5555/1109557.1109677}. Despite the NP-hard nature of the problem, by exploring the problem structure, the authors formulate the payment rule based on the critical value approach in auction theory \cite{milgrom2004putting}, and the allocation (winner determination) problem is solved as a linear program.
For the latter, additional constraints on workers’ privacy levels are added, and similarly the problem is reduced to a knapsack problem which is further relaxed to a solvable linear program using the integer variable condition. 
They justify that their proposed mechanisms are truthful, individually rational, and achieve cost minimization approximately in a computationally efficient manner.
The work by \citet{DBLP:conf/infocom/JinXLG19} also focuses on mobile crowdsensing but formulates the trading problem more than a standard knapsack problem. The trading object is set as the location privacy of workers which is measured using the Bayesian attack model. A combinational reverse auction is designed to allow each worker to select a bundle of tasks with a bid, i.e., the minimum willing payment, considering the sensing cost and the privacy cost. The objective of the auction mechanism for the platform is to achieve desirable service accuracy which is modeled as to minimize the geo-information loss via a geo-information quality model. Defined as the desired properties of the  mobile crowdsensing market, budget feasibility, truthfulness, individual rationality, and constraints for privacy and accuracy are satisfied along with polynomial-time complexity and bounded optimality gap derived by a heuristic algorithm.

The works \cite{DBLP:journals/ton/ZhangYHLZ21,DBLP:conf/uai/ZhangBL20} both formulate data trading as knapsack auction forms but focus on different scenarios. While distinctive solving approaches are adopted which are suitable to their specific models, some desiderata are both satisfied, e.g., incentive compatibility, individual rationality, and the balance between privacy and accuracy. Despite the complexity over the normal knapsack problem, the work \cite{DBLP:conf/infocom/JinXLG19} formalizes the auction problem first with desiderata and underlying models which in principle aligns with the knapsack auction modeling.

\partitle{Decentralized Auction}
Traditional auction platforms \cite{chen2018blockchain} are centralized and depend on third-party intermediaries for transaction coordination between buyers and sellers, which raises critical concerns on trust problems. To eliminate intermediaries and enhance security, a decentralized approach using blockchain technology has been adopted, i.e., decentralized auction \cite{OMAR2021120786}. Smart contract, a technique widely used in blockchain network, enables trading rules to be programmed in the contract and automatically executed in the auction process, which can verify credible transactions without third parties. Some researchers \cite{DBLP:journals/isci/XiongX21,DBLP:conf/dasfaa/AnX0GZ19} have introduced decentralized auction into data trading scenarios to promote efficient and trusted data flow without third parties.

Decentralized auction \cite{OMAR2021120786} has been introduced to crowdsensing data trading for the trust problem between buyers and sellers without a third party. \citet{DBLP:conf/dasfaa/AnX0GZ19} combine smart contract and reverse auction (where sellers act as bidders) together to build a  crowdsensing data trading system where mutually untrusted parties are enforced to participate in the data trading honestly. Specifically, a greedy strategy is used to determine winners by taking the reliability of each seller into consideration, and a pricing algorithm based on critical value \cite{milgrom2004putting} is designed to determine payments for winners. 
The truthfulness and individual rationality of the auction mechanism are justified by the monotonic winner selection and critical payments. 
The truthfulness of the whole data trading process is further enhanced by deposit and modifier settings which can resist certificate forgery and tamper attacks. Bid privacy is protected in the auction procedure by a two-stage bid strategy which can blame untruthful sellers and ensure the immutability of bids while the confidentiality of data is preserved using symmetric and asymmetric encryptions for the data delivery process.

Besides the lack of trust among multiple parties, \citet{DBLP:journals/isci/XiongX21} concern the potential collusion behaviors of buyers in the data transaction, i.e., buyers as bidders form coalitions to manipulate the prices for benefit. They propose a decentralized auction \cite{OMAR2021120786} framework based on smart contracts to satisfy the defined anti-collusion feature of data trading which is an extension of fairness and truthfulness. An anti-collusion data auction protocol is designed to determine the highest bid and the corresponding winners, which guarantees the anti-collusion nature by storing the true valuations of data for both the seller and buyers on the blockchain and taking advantage of the tamper-resistant property of information on the blockchain. They implement the anti-collusion data auction mechanism by adopting the webpack in the Truffle Boxes and verify that collusion can be effectively prevented.

These decentralized data auction frameworks \cite{DBLP:conf/dasfaa/AnX0GZ19,DBLP:journals/isci/XiongX21} promote truthful data trading without third parties and provide a new paradigm for data pricing based on smart contract. It may be promising to further exploit the vast potential of blockchain technology in data markets.

\partitle{Deep-Learning-enpowered Auction}
While Myerson solved the single-parameter expected revenue maximization problem,  multi-item revenue-optimal auction design  remains unsolved. Following a line of bringing in tools from machine learning to mechanism design, \citet{DBLP:conf/icml/Duetting0NPR19} provide a general deep-learning-enpowered framework for solving the multi-item auction design problem, i.e., to maximize expected revenue subject to constraints for incentive compatibility. Multi-layer neural networks are used to encode auction mechanisms with bids being the inputs and allocation and payment results being the outputs. Deep-learning-enpowered auction has been adopted to empower data acquisition by \citet{zheng2022fl}. Another deep learning driven data market framework \cite{DBLP:conf/nips/RavindranathJP23} focuses on signaling scheme design rather than data allocation rules to handle obedience constraints with externality taken into consideration.

FL-Market, proposed by \citet{zheng2022fl}, refers to a private model marketplace with untrusted data brokers where data sellers give locally perturbed gradients to collaboratively train a machine learning model. Trustworthy data acquisition problem where both privacy and utility can be satisfied is studied based on federated learning and local differential privacy.
The local gradients' perturbation levels are decided through a deep learning-empowered auction mechanism, i.e., Deterministic Multi-Unit RegretNet (DM-RegretNet), which adapts the multi-item auction framework RegretNet \cite{DBLP:conf/icml/Duetting0NPR19} to support trading multi-unit items, i.e., a portion of the privacy budget. Specifically, two networks, a payment network and an allocation network, take data sellers' privacy budget bids and the buyer's financial budget as inputs, and output the payments and  privacy losses (deterministic allocation results) for data sellers. Besides the expected error bound of the perturbed global gradient, the training objective function also consists of a regret penalty indicating the violation degree of truthfulness and individual rationality in order to approximately guarantee these two desiderata. Combined with a proposed aggregation mechanism for aggregating the perturbed gradients, the auction mechanism can maximize the global gradient's accuracy (and thus buyers' utility) while incentivizing data sellers to data sharing with their preferred privacy levels and compensations. 

\citet{DBLP:conf/nips/RavindranathJP23} focus more on the information design problem in data markets rather than the traditional resource allocation problem, where information is traded and its value hinges on the specific utilization envisioned by the buyer as well as other buyers embodied in the externality. Scenarios with one buyer and multiple buyers are abstracted and the incentive compatibility is studied by extending the RochetNet architecture and RegretNet framework respectively.  

These works \cite{zheng2022fl, DBLP:conf/nips/RavindranathJP23}  prove the effectiveness of deep learning as a tool in solving data auction problems, which is worth to be further explored.

\partitle{Sequential Auction}
Different from the aforementioned auctions which model the one-shot transaction, a sequential auction \cite{bernhardt1994note} is an auction process in which items are sold one after the other. Considering the future transactions, the strategies of participants would differ from those in one auction. An important practical question for sellers selling several items is how to design a sequential auction that maximizes their revenue, e.g., selling in which order and in what amount each time. In terms of data transaction, it is practical to consider the scenarios of trading data over time slots which have been studied in a few works \cite{DBLP:journals/tmc/DuGJHR20, DBLP:conf/sigmod/Fernandez22}.

In the work \cite{DBLP:journals/tmc/DuGJHR20}, a data transaction system is proposed where mobile users can sell their redundant data allowance to other users who demand extra data resources. Instead of designing auction models which mainly aim at truthful bidding, they focus on describing a traditional and practical auction process to model a series of successive data transactions.
Using the knowledge of stochastic process and queueing theory, they establish a basis auction model for a single data seller and a networked auction model for multiple data sellers each operating their own basic data auction based on two automated auction models introduced in \cite{DBLP:journals/toit/Gelenbe09}. Three data allocation mechanisms are further designed on the basis of the proposed networked data transaction system for three different optimizing objectives, respectively, i.e., to maximize the income in each time slot, each auctioneer's income, or the entire income.

Different from researching the data allowance transaction in mobile networks, \citet{DBLP:conf/sigmod/Fernandez22} gets closer to data market paradigms with data trading and focuses on the strategic behaviors of data buyers in light of sequential data transactions. A buyer-seller market is focused on where a data set is sold multiple times and interested buyers offer bids in a streaming fashion. Three strategic behaviors are considered, for each of which a protection technique is designed and integrated into the pricing algorithm to shield the market from buyers' manipulated bids. The first strategic behavior is that buyers send artificially low bids to affect the price in their profitable direction since the price is generated based on submitted bids. To prevent such manipulation, \citet{DBLP:conf/sigmod/Fernandez22} proposes Epoch-Shield, which separates the steaming bids into epochs and generates the price based on the bids in an epoch instead of individual bids. The key insight is to reduce a single buyer's capability of influencing the price. While Epoch-Shield can limit the impact of a single bid, buyers may bid low continuously to drive the price down, which is considered the second strategic behavior. The corresponding protection technique is Time-Shield, which sets a waiting period as the penalty for low-bid buyers. Since deadline-patience (i.e., only by getting the data before a certain deadline can buyers gain utility) is introduced into the utility model of buyers, the wait-period penalty poses the risk of losing utility to buyers and can thus prevent buyers from strategizing. The waiting period is carefully designed by predicting the time when future bids become competitive to disincentivize strategic behavior while guaranteeing no harm to truthful buyers. The last strategic behavior is from boundedly-rational actions of buyers. Uncertainty-Shield is designed, aiming to further decouple the price and past bids to protect against buyers guessing the price by adding noise to the posting price. The major challenge of Uncertainty-Shield is to balance the protection and the revenue which supposes to be optimized using past bids. Combining three protection techniques, a pricing algorithm is established based on Multiplicative Weights, where the price is chosen from candidate ones based on weights that are updated according to the relative revenue difference. 

While most works on auction-based data pricing concern one-shot trading settings and properties of bidding strategies, \citet{DBLP:journals/tmc/DuGJHR20} and \citet{DBLP:conf/sigmod/Fernandez22} put emphasis on sequential data auction which makes sense in practical data trading scenarios where multiple requests for access to data appear as time goes by. While \citet{DBLP:journals/tmc/DuGJHR20}, deviating from typical data trading agendas, focus on data allowance, \citet{DBLP:conf/sigmod/Fernandez22} emphasizes the strategic behaviors of data buyers in a two-sided data market. Despite a user study and several simulations presented justifying the motivations and effectiveness of the protection techniques, some performance lacks theoretic guarantees, e.g., to what extent the Epoch-Shield disincentivizes low bids and which epoch size optimizes the effect. Moreover, although the ex-post valuation is mentioned, more study is needed to extend the protection framework to fit the ex-post scenarios.   

\partitle{Iterative Auction} 
The above sequential auction extends the one-shot auction and captures the procedure of trading data multiple times. Iterative auction \cite{parkes2006iterative} also supports multiple bids but concerns one transaction with participants repeatedly bidding for one sale. The iterative auction allows iterative interaction with bidders which can elicit information about their preferences and guide them to a desired direction, e.g., welfare maximization. While mostly applied in combinatorial auctions to resolve the problem of costly preference elicitation, the iterative auction has also been introduced into data markets by \citet{DBLP:conf/icassp/CaoCL17} to guide the selfish data agents to trade data efficiently in terms of social welfare without direct access to the agents' private information.

\citet{DBLP:conf/icassp/CaoCL17} consider a data market with multiple data sellers, collectors, and users (uniformly treated as data agents who submit bids) and aim at maximizing social welfare from the perspective of a system designer (auctioneer). An iterative auction framework is proposed, where the auctioneer announces the data allocation and payment rules (as functions of bids) and the selfish bidders submit appropriate bids to maximize their own utilities in each iteration. The proposed mechanism is theoretically proved to converge to the socially optimal point, and individual rationality and (weakly) balanced budget are satisfied.  While data agents can be guided to achieve social optimum without reporting private information,  they are assumed as price-takers, e.g., only taking the announced prices as fixed and not expecting their impact on pricing. More research is worth extending the iterative auction to scenarios with strategic agents.

The surveyed works above design one-sided auction frameworks which consider either multiple buyers or multiple sellers bidding for transactions, or approaching buyers and sellers as agents in the same way. The following works adopt double auction frameworks which allow both buyers and sellers to bid to participate in data trading and focus on their mutual interactions.

\partitle{Double Auction}
A double auction \cite{friedman2018double} is a process of buying and selling goods between multiple sellers and multiple buyers. In the basic double auction model, buyers submit their bids and sellers submit their ask prices to the data broker, and then the data broker chooses some price that clears the market, i.e., all the sellers who ask less than such price sell and all buyers who bid more than such price buy at this price. A double auction is also possible without any exchange of currency, which is referred to barter double auction \cite{tagiew2009towards} where every participant has a demand and an offer consisting of multiple attributes and no money is involved. Various mechanisms with distinctive (dis)advantages \cite{babaioff2001concurrent} have been applied to formulate auction rules including k-double auction \cite{wilson1985incentive}, McAfee's mechanism \cite{mcafee1992dominant}, and several one-sided auction adaptations, e.g., VCG auction \cite{roughgarden2010algorithmic}. 
Some work \cite{DBLP:journals/cj/CaiZLY19,DBLP:journals/corr/abs-2107-08630} use double auction to formulate data trading which takes the bidding power of both data buyers and data sellers into consideration.

Considering the strategic behaviors of both data sellers and data buyers, \citet{DBLP:journals/cj/CaiZLY19} propose a truthful double auction mechanism for data trading with preferences and complex conflicts of interest (CoI) relations among data buyers. Through the truthful double auction mechanism, both winning data sellers and winning data buyers along with the corresponding payments could be determined.  The mechanism is composed of a group rule which generates conflict-free virtual groups of buyers based on the CoI graph, a data trading rule which adopts the group buying to share data and expense, and a payment rule which determines each winning buyer's charged price and each winning seller's payment by finding the critical value \cite{milgrom2004putting}.  Through the proposed mechanism, social welfare is approximately maximized while the affordability rule and conflict-free rule are guaranteed. The authors also justify four desired properties of the proposed auction, truthfulness, budget balance, individual rationality, and computational efficiency.

Similar to the data sharing market studied by \citet{DBLP:journals/pacmmod/Fernandez23}, \citet{DBLP:journals/corr/abs-2107-08630} focus on cases where each participant in data markets can be both buyer and seller. Two types of data exchange are considered, bilateral (trading data with data, similar to barter double auction \cite{tagiew2009towards}) and unilateral (trading data with money). For the former, they model the bilateral sharing as a network formulation game, show the conditions where the strongly stable outcome exists, and design an algorithm to find such outcomes in quadratic time. For the latter, they improve the standard VCG mechanism \cite{roughgarden2010algorithmic} to satisfy budget requirements. Specifically, Mixed-VCG mechanism is proposed which uses data distortions as data money to achieve budget balance while truthfully implementing social welfare maximizing outcomes to the exact level of budget imbalance of the standard VCG mechanism. Distorted-mixed-VCG is further proposed in the case of relaxing zero cost data distortion assumption by adding fixed distortion to the data at the initiation.

Although both the supply side and demand side of data trading are taken into account, the roles of buyers and sellers are designated uniquely in the work \cite{DBLP:journals/cj/CaiZLY19} while participants can both be buyers and sellers in the work \cite{DBLP:journals/corr/abs-2107-08630}, which both have a basis in practice and incur further implications on the use of double auction in data pricing. 

\partitle{Other Incentive Mechanisms} Instead of explicitly adopting the auction mechanisms where the bids are commonly the private valuation of data, another strand of research \cite{DBLP:journals/isr/MehtaDJM21, DBLP:journals/dss/LiR14} focuses on incentive-compatible data pricing mechanisms to extract the true information of data buyers which is not necessarily the willingness to pay for the data.

\citet{DBLP:journals/isr/MehtaDJM21} model the private information of data buyers as a multidimensional vector comprised of their ideal records and the decay rate of their valuation for the records differing from the ideal ones, which makes Myersonian auction is not applicable to this multidimensional mechanism design problem for revenue maximization. By investigating the price-quantity schedule (e.g., the two-part tariffs), they formulate the conditions for the optimal pricing for the data seller and analyze the worst-case revenue guarantee without the conditions.
Standing from the perspective of data sellers as well, \citet{DBLP:journals/dss/LiR14} develop an incentive-compatible mechanism to price private data with the tradeoff between data privacy and data utility. By adapting the two-part tariff pricing scheme for the amount of data with sensitivity level, the data seller can maximize its profit with the data buyer incentivized to reveal the true purpose of data usage which is assumed within two types. 

Anchored in raw data pricing, these works \cite{DBLP:journals/isr/MehtaDJM21, DBLP:journals/dss/LiR14} propose incentive-compatible pricing mechanisms via different valuation models for data buyers. More complex mechanisms are required when faced with multifaceted data products where the essential incentive guarantee is worth investigating. 

\subsubsection{Stackelberg Game-based Pricing}
Stackelberg game \cite{von2011market}, a leader-follower two-stage game with complete information, can be used to formulate the determining process for agents with hierarchical organizations or certain orders of play. One agent acts as a leader who makes strategy first by predicting the other's response and maximizing their own profit, and the other agent as a follower chooses the best strategy for profit given the leader's behavior. Stackelberg equilibrium (also known as subgame perfect Nash equilibrium) is a stable state where both the leader and the follower wouldn't change their strategies since no other strategies can bring them higher profits.
In terms of data markets, Stackelberg game which originally consists of one leader and one follower has been further extended to formulate the interactions among various parties in data trading \cite{share, 10037216, DBLP:journals/tmc/XuXWZG23, an2021crowdsensing, DBLP:conf/infocom/XiaoXZWZZ23, DBLP:conf/icc/XuJWRYG15, 9322221, DBLP:journals/jsac/ZhangAHP21, DBLP:journals/tmc/YuCH22}.

\citet{share} combine Stackelberg games with simultaneous-move games into data pricing to first realize the incentive data market construction with absolute pricing for a demand-driven market. Stackelberg game \cite{von2011market} is applied to model the trading dynamics among three parties which not only optimizes the profits of all selfish participants but also adapts to the common buyer-broker-sellers market flow. Nash equilibrium \cite{nash1950equilibrium} is adopted to solve the seller selection problem (the simultaneous-move part) based on sellers' inner competition where sellers can strategize their data quality by local differential privacy \cite{DBLP:journals/siamcomp/KasiviswanathanLNRS11} to balance their compensations and privacy loss. Stackelberg-Nash Equilibrium (SNE) is defined where all the participants would not change strategies since the equilibrium strategies bring them the highest profits.
To derive the SNE, backward induction is used, and a novel mean-field approximation with provable guarantees is proposed for complicated cases where direct derivation fails. 

While the work \cite{share} presents a pricing framework applicable to general data products from query answers to machine learning models, \citet{10037216} study the big data market where data is directly measured and traded. The Stackelberg game is similarly adopted for price determination with the interactions between data providers and users formulated, where the data provider leads the game by setting the price and the data user follows by deciding the optimal demand. Following a similar backward induction approach to solving the Stackelberg equilibrium, a gradient iteration algorithm is applied, and variational inequality (VI) theory is used to prove the equilibrium. Besides the pricing module, they also make contributions to the evaluation of data value and privacy protection. A data valuation model based on both qualitative and quantitative methods is proposed which comprehensively and fairly represents the value of the data. A local differential privacy based protocol is used to limit the leakage
of private information in the pricing process.
\citet{DBLP:conf/icc/XuJWRYG15} adopt a different way of dealing with privacy issues in data trading by applying anonymization on data. They focus on a typical data collecting-publishing-mining scenario. A data user buys data for data mining tasks, the data collector collects data from providers and performs k-anonymity on the data to balance privacy protection for data providers and data utility for data users, and data providers sell data for compensation.  Similar to the work \cite{share}, the interactions among three parties are modeled as a Stackelberg game \cite{von2011market} with data user, data collector and data providers acting in order. Subgame perfect Nash equilibrium is also derived by backward deduction.

Specific data trading markets, e.g., crowdsensing data markets, have also been investigated. 
Stackelberg games \cite{von2011market} have been introduced to transactions for crowdsensing data collected by mobile users with sensing devices to price data and data aggregations \cite{an2021crowdsensing}. 
The authors formulate the interactions among the consumer, the platform, and the mobile users (sellers) as a three-stage Stackelberg game. Specifically, the data aggregation price, the data price, and the sensing time are the strategies of the consumer, the data broker, and each seller, respectively, which they decide to maximize their own profit by considering the mutual effects of other agents’ decisions.
The authors derive an optimal strategy profile by a backward deduction approach, which constitutes a unique Stackelberg equilibrium so that no one can improve the profit by singly deviating from this strategy profile. They also address the problem of optimizing the data aggregation quality without the knowledge of sellers' sensing quality by adopting the combinatorial multi-armed bandit mechanism.
In their follow-up work \cite{DBLP:journals/tmc/XuXWZG23}, they further take social networks into consideration, namely, a social-aware worker can benefit from shared information through social relations. An extra term representing social benefit is added to the utility of workers, which leads to the interrelationship between each worker's optimal strategy in the third stage of Stackelberg game. Under some assumptions, a unique Nash equilibrium is built in this stage and further solved in a matrix form. Another work \cite{DBLP:conf/infocom/XiaoXZWZZ23} adopts a similar way of formulating the social network effects among workers, which takes incomplete information into account and models the inter-worker interactions as a Bayesian sub-game. Besides the social effects, the Age of Information (AoI) is introduced to the framework and the strategy of workers is set as the data update frequency while additional constraints are attached to ensure the AoI values of all data uploaded to the platform are not larger than a given threshold. Different from the three-sided trading framework in \cite{an2021crowdsensing, DBLP:journals/tmc/XuXWZG23}, \citet{DBLP:conf/infocom/XiaoXZWZZ23} focus on the interplay between the platform and workers, and propose a two-stage Stackelberg game where the platform is the leader and the workers are the followers. Karush-Kuhn-Tucker (KKT) conditions and the backward deduction approach are applied to derived the unique Stackelberg equilibrium. 


Instead of crowdsensing data trading, Stackelberg games \cite{von2011market} have been adapted by \citet{9322221} to data pricing in the car-sharing data markets with vast vehicle data generated from car-sharing activities. Similarly, a three-layer Stackelberg game among a data seller, a service provider, and a data buyer is proposed. Different in this work, the data seller is the leader in determining the price of raw data while the service provider in the middle layer processes the raw data and prices data with different data accuracy levels, and the data buyer selects the data to buy and determines the size of purchased data. Moreover,  blockchain is combined into the designed framework as a trusted trading environment without the centralized intermediary to support P2P trading and enhance security.
For dynamic data scenarios where data freshness, captured by the age of information (AoI) metric, is taken into account, \citet{DBLP:journals/jsac/ZhangAHP21}  study pricing fresh data based on Stackelberg game \cite{von2011market}. They propose a two-sided fresh data trading framework where the destination user requests and pays for fresh data updated from a source provider. Based on a two-stage Stackelberg game \cite{von2011market}, the source as the leader designs a pricing mechanism to maximize its profit and the destination as the follower chooses a data update schedule considering its payment and age-related cost. Three pricing models are studied which focus on different dimensions including time, quantity, and subscription. For each model, the pricing solutions are derived under both finite-horizon and infinite-horizon 
assumptions using various algorithms including bilevel optimization, fractional dynamic programming, and bisection search. The profitability of different pricing models is analyzed and compared with the social optimum. 
Besides the data in the normal sense, data quato can also be traded in the mobile data trading market. \citet{DBLP:journals/tmc/YuCH22} propose a Stackelberg game \cite{von2011market} based data quato trading framework where the mobile network operators deploy the market and users trade mobile data quato with each other. They formulate the two-sided interactions as a three-stage Stackelberg game where the operator optimizes its operation fee imposed on sellers for each unit of sold data, each user chooses his operator given the operation fee, and users then decide their trading decisions, i.e., their roles as sellers or buyers and the trading prices and quantities. The closed-form expressions of the unique equilibrium are derived and the optimal operation fee is computed. Several conclusions on the dynamics and benefits of mobile data trading markets are drawn based on numerical simulations.

These works \cite{share, 10037216, an2021crowdsensing,DBLP:conf/icc/XuJWRYG15,9322221,DBLP:journals/jsac/ZhangAHP21,DBLP:journals/tmc/YuCH22} leverage Stackelberg games for its widely-adopted formulations for sequential interactions and well-defined mathematic models, and adapt it into data pricing despite the difference in detailed modeling. However, the assumption of complete information in Stackelberg game may not be met in practical data markets, which requires further investigation. 

\subsubsection{Bargaining-based Pricing}
Bargaining game \cite{muthoo1999bargaining} refers to the process where two or more agents reach an agreement regarding how to distribute an object or monetary amount, which can be used to achieve fair allocation among agents. Some works \cite{DBLP:conf/bigdataconf/JungP19,DBLP:journals/isr/RayMM20} have introduced several bargaining models into data pricing to formulate the interaction between data supply and demand.

\partitle{Rubinstein Bargaining}
Rubinstein bargaining \cite{rubinstein1982perfect} features alternating offers through an infinite time horizon to reach an agreement on allocation. The participants take turns making offers and the bargaining continues until an acception while every rejection leads to a costly delay. A discount factor is used to reflect the diminishing future affect on the present, which is determined based on the patience of participants. A no-delay solution is derived, which is the unique allocation result satisfying the subgame perfect equilibrium. Rubinstein bargaining has been introduced into data pricing \cite{DBLP:conf/bigdataconf/JungP19} since its solution can prevent repeated tedious proposals in the negotiation. 

Aiming to capture both the usefulness and the cost of data, \citet{DBLP:conf/bigdataconf/JungP19} propose a data pricing framework by taking both the data provider and data consumer into account. After being matched by the market manager, a provider and a consumer participate in a negotiation where the amount of noise measured by differential privacy and the corresponding data price are determined by applying the Rubinstein bargaining approach \cite{rubinstein1982perfect}. Specifically, the requirements of both sides are turned into the discount factors which formulate a weight for each side, and the data price is then decided based on the weights and the reported prices. Moreover, to incentive participants to report their required prices truthfully, the manager records their loss or profit as credits to be reflected in future trading. The authors also propose a social welfare based approach to further solve the profit imbalance and enhance fairness between data providers and consumers.

\partitle{Nash Bargaining}
Nash bargaining \cite{nash1950bargaining} studies how the surplus should be shared by formulating appealing axioms that the solution should satisfy. Nash considers the desired axioms as Pareto Optimality (PAR), Individual Rationality (IR), Independent of Expected Utility Representations (INV), Independence of Irrelevant Alternatives (IIA), and Symmetry (SYM). He proves that a solution satisfying these four axioms is exactly the one that maximizes the product of the payoff of participants. Nash bargaining solution, considered as a proportionally fair approach,  has been adopted into data pricing 
\cite{DBLP:journals/isr/RayMM20}. 

The value of data can be uncertain to both buyers and sellers. On one hand, buyers have limited knowledge about the data sellers own. On the other hand, sellers are often unclear on the buyers' needs. Considering the two-sided uncertainty of data value, 
\citet{DBLP:journals/isr/RayMM20} propose that a negotiation process is needed to achieve a mutually acceptable price for data. They represent the negotiation as a generalized Nash bargaining process \cite{harsanyi1972generalized} which incorporates the information asymmetry. An incentive-compatible decision mechanism from  \citet{myerson1979incentive} is adopted to solve the bargaining problem, which maps the participants’ information about the value of the data to bargaining outcomes. Moreover, they assume that the seller can provide a demonstration to reduce uncertainty and analyze the effect of the demonstration in the case of outside options. However, the value of data is simply assumed to be binary based on a binary distribution model, which makes it challenging to apply to practice.

While \citet{DBLP:conf/bigdataconf/JungP19} and \citet{DBLP:journals/isr/RayMM20} have taken a step towards introducing bargaining theory into data markets, more research are needed to compare and analyze different bargaining models when applied into data pricing and take full use of the high practicality and other potentials of bargaining models.

\subsection{Discussion on Actual Data Marketplaces}
In actual data marketplaces, common pricing models include subscription-based pricing, usage-based pricing, negotiation-based pricing, and dynamic pricing. Under the subscription-based pricing model, data buyers pay regular fees to access data products. The fees vary according to the subscription level. Usaged-based pricing allows data buyers to pay based on actual data usage, such as the number of requests issued. Negotiation-based pricing determines the price of the data product through private negotiations between data sellers and buyers. In addition, data sellers may dynamically adjust prices and offer private discounts, e.g., for large enterprises \cite{awsPrivateoffer}. Despite distinctive pricing methods in practice, Dawex lists $10$ criteria for estimating the price of data products: reputation, completeness, rareness, reliability, organization, frequency and time period, sustainability, volume, clarity, and dependence \cite{dawexDataValue}. Recently, \citet{DBLP:conf/icde/AzcoitiaIL23} conduct a comprehensive study of the prices of data products listed in commercial data marketplaces. They identify that data product prices range from a few to hundreds of thousands of US dollars, highlighting that data products about telecommunications, manufacturing, automotive, and gaming have the highest prices.

\section{Revenue Allocation}\label{sec:revenueallocation}

Data markets involve multiple data sellers who contribute their data. Revenue generated from data transactions is then distributed among data sellers. Fairness is a desirable property in revenue allocation. In this section, we discuss how to ensure fairness in revenue allocation among data sellers based on their privacy cost, Shapley values, and others.

\subsection{Privacy Cost-Based}
Studies \cite{DBLP:conf/icdt/LiLMS13, DBLP:journals/tods/LiLMS14, DBLP:journals/cacm/LiLMS17,DBLP:conf/kdd/NiuZWTGC18,DBLP:conf/iwqos/Cai0YZL19,DBLP:journals/pvldb/LiuLL0PS21} allocate revenue based on the privacy costs incurred by data sellers during trading their personal data. These can be categorized into query-based markets and model-based markets.

\partitle{Query-based markets}
\citet{DBLP:conf/icdt/LiLMS13, DBLP:journals/tods/LiLMS14, DBLP:journals/cacm/LiLMS17} present the first theoretical framework for pricing noisy query answers based on accuracy, and for dividing the revenue among data sellers who deserve compensation for their privacy loss. Following differential privacy, they define the privacy loss of a data seller in the database instance for a query $\mathbf{Q}$ and its perturbing mechanism $\mathcal{K}$ as the maximum ratio between the probability of returning fixed output with and without the data seller's data, denoted by $\varepsilon(\mathcal{K}_\mathbf{Q})$. Then, the data broker compensates data seller $i$ who answers query $\mathbf{Q}$ with micro-payments $\mu_{i}(\mathbf{Q}) \ge W_i(\varepsilon(\mathcal{K}_\mathbf{Q}))$, where $W_i$ is a contract between the data seller and the data broker. 

\citet{DBLP:conf/iwqos/Cai0YZL19} propose a scenario where data sellers have diverse privacy preferences in different categories, such as health, social, and financial information. They develop a perturbation scheme to ensure that the privacy loss defined the same as differential privacy in each category of data sellers does not exceed its specified upper bounds. The compensation for data sellers is then calculated based on privacy loss and the cost of privacy loss reported by the data seller in each category while ensuring individual rationality, truthfulness, and budget balance.

\citet{DBLP:conf/kdd/NiuZWTGC18} further consider that data correlations exist among data sellers and extend to reach general dependent fairness. In dependent fairness, data sellers who are correlated to data sellers involved in the service can also receive privacy compensation. They use $\epsilon-$dependent differential privacy to measure individual privacy loss and propose two privacy compensation mechanisms. 
Similar privacy compensation mechanisms are used by \citet{DBLP:journals/compsec/ShenGSDDZZJ22}, termed positive pricing and reverse pricing in their approach. Follow-up work by \citet{niu2019making} trades personal time-series data which has temporal correlations and adopts Pufferfish privacy \cite{DBLP:conf/pods/KiferM12} to measure privacy loss of data sellers.

\partitle{Model-based markets}
\citet{DBLP:journals/pvldb/LiuLL0PS21} define a monotonic compensation function $c_i$ for the data seller $i$ to model its compensation for using its data with $\epsilon$-differential privacy: $c_i(\epsilon)=b_i\cdot s_i(\epsilon)$, where function $s_i(\epsilon)$ reflects the data seller’s belief of privacy in price, $b_i$ is the base price and should be proportional to Shapley value of its data with regard to the whole dataset contributed to building models. Similarly, \citet{DBLP:journals/isci/JiangNYWL22} define a compensation function for the data seller $t$ to model its compensation for using its data with $(\alpha,\frac{2\alpha L^2}{\sigma^2})$-Rényi differential privacy: $\rho(\sigma^2, t) = c \times \frac{2\alpha L^2}{\sigma^2(n+1-t)}$, where $c$ is a coefficient and $n$ is the total number of data sellers.

While the compensation is based on the amount of noise added according to different differential privacy settings and the resulting privacy loss, it is worth noting that this compensation still reflects the contributions of data sellers to the query answer or the model. The reason is that the accuracy of the query answer or the model is ultimately influenced by the amount of noise added.

\subsection{Shapley Value-Based}\label{subsubsec:Shapley-Value-Based-Revenue-Allocation}
Studies \cite{DBLP:conf/ec/AgarwalDS19, DBLP:journals/pvldb/JiaDWHGLZSS19, DBLP:journals/pvldb/LiuLL0PS21, DBLP:journals/pvldb/LiuLZR0L0PS21,DBLP:conf/icml/GhorbaniZ19,DBLP:journals/corr/abs-2110-14049,DBLP:journals/tmc/ZhengPWTC20,DBLP:journals/pvldb/XiaL0Q00023,DBLP:journals/pvldb/Xia24,DBLP:conf/nips/SchochXJ22,DBLP:journals/tc/HeC24,DBLP:conf/aaai/TianSFL24} model interactions among data sellers as a cooperative game and allocate revenue among data sellers using the Shapley value. 

The Shapley value \cite{shapley1953value} is a weighted average of the marginal utilities of the player to coalitions. It is general and flexible to support various utility functions. In query-based markets, by treating data as players cooperating to generate query answers, \citet{DBLP:journals/tmc/ZhengPWTC20} define the utility function of a coalition as the sum of the rewards of all versions that the coalition can produce. Then, the reward of a data seller is directly its Shapley value. \citet{DBLP:journals/tc/HeC24} set the utility function as accuracy of answers to aggregate queries and define a truthful, individually rational compensation function based on Shapley values. In model-based markets, by treating data as players cooperating to train a machine learning model, the utility function of a coalition is defined as the machine learning model performance trained by the coalition, such as prediction accuracy for classification and $1$-$RMSE$ (root-mean-squared-error) for regression. Then, the Shapley value applies to formulate the payment allocation function. \citet{DBLP:conf/ec/AgarwalDS19} build a data market to sell models and use the Shapley value in building payment-division functions to distribute payment among data sellers fairly. 
\citet{DBLP:journals/pvldb/LiuLL0PS21} propose a monotonic compensation function, where the base price is proportional to the Shapley value of its data with regard to the whole dataset contributed to building models.
Additionally, \citet{DBLP:journals/dcg/CabelloC22} use the Shapley value as a payment rule for auctions and exchanges by treating winning bidders as a grand coalition and setting the surplus as the utility function. The Shapley value has also been proposed as a principled solution for royalty sharing in recent AI copyright discussion \citep{deng2024economic}.

Computing the exact Shapley value involves the computation of $O(2^n)$ utility functions, which is prohibitively expensive. A series of techniques are proposed to address the challenge. We summarize these works in Table \ref{tab:svsummary}, and further details are provided below. Generic algorithms apply to all data markets, algorithms for databases can support query-based markets, and algorithms for machine learning can support model-based markets.

\begin{table}[htbp]
  \centering
  \caption{Research works about Shapley value computation.}\label{tab:svsummary}
    \resizebox{\linewidth}{!}{
    \begin{tabular}{|c|l|l|l|l|}
    \hline
    \multicolumn{2}{|c|}{\textbf{Application}} & \multicolumn{1}{c|}{\textbf{Reference}} & \multicolumn{1}{c|}{\textbf{Type}} & \multicolumn{1}{c|}{\textbf{Method}} \\
    \hline
    \multicolumn{1}{|l|}{\multirow{10}[20]{*}{generic games}} & static & \citet{DBLP:journals/cor/CastroGT09} & appro &  Monte Carlo permutation sampling \\
\cline{2-5}          & static & \citet{DBLP:journals/jmlr/MitchellCFH22} & appro &  quasi Monte Carlo permutation sampling \\
\cline{2-5}          & static & \citet{DBLP:journals/corr/MalekiTHRR13} & appro &  stratified sampling with Hoeffding inequality  \\
\cline{2-5}          & static & \citet{DBLP:journals/cor/CastroGMT17} & appro &  stratified sampling with Neyman allocation \\
\cline{2-5}          & static & \citet{DBLP:conf/ijcai/BurgessC21} & appro &  stratified sampling with a stratified empirical Bernstein bound \\
\cline{2-5}          & static & \citet{DBLP:conf/icpr/OkhratiL20} & appro &  multilinear extension \\
\cline{2-5}          & static & \citet{DBLP:journals/pacmmod/0006SL0P023} & appro &  complementary contribution \\
\cline{2-5}          & static & \citet{DBLP:conf/aaai/KolpaczkiBMH24} & appro & coalition sampling \\
\cline{2-5}          & static & \citet{li2023faster} & appro & least squares optimization \\
\cline{2-5}          & dynamic & \citet{DBLP:conf/icde/0006XSL0P023} & appro & permutation sampling, heuristic \\
    \hline
    \multicolumn{1}{|l|}{\multirow{5}[10]{*}{database}} & query & \citet{DBLP:conf/sigmod/DeutchFKM22} & exact, appro & exact \\
\cline{2-5}          & query & \citet{DBLP:journals/pvldb/LuoPCX22} & exact & exact \\
\cline{2-5}          & query & \citet{DBLP:journals/pacmmod/LuoPXZX24} & exact & decomposition \\
\cline{2-5}          & convex hull & \citet{DBLP:journals/dcg/CabelloC22} & exact & exact \\
\cline{2-5}          & convex hull & \citet{DBLP:conf/isaac/Tan21} & exact & exact \\
    \hline
    \multicolumn{1}{|l|}{\multirow{7}[14]{*}{centralized ML}} & data valuation & \citet{DBLP:conf/icml/GhorbaniZ19} & appro & permutation sampling with truncation or one-point gradient descent \\
\cline{2-5}          & data valuation & \citet{jia2019towards} & appro & permutation sampling with group testing or compressive sensing \\
\cline{2-5}          & KNN   & \citet{DBLP:journals/pvldb/JiaDWHGLZSS19} & exact, appro & Shapley value differences \\
\cline{2-5}          & weighted KNN & \citet{wangj2024aistats} & exact, appro & counting problem \\
\cline{2-5}          & ML    & \citet{DBLP:journals/corr/abs-2107-06336} & appro & model training \\
\cline{2-5}          & feature explanation & \citet{DBLP:conf/nips/LundbergL17} & appro & least squares optimization \\
\cline{2-5}          & feature explanation & \citet{DBLP:conf/iclr/JethaniSCLR22} & appro & model training \\
    \hline
    \multirow{3}[6]{*}{federated learning} & client valuation & \citet{DBLP:conf/bigdataconf/SongTW19}  & appro & gradient reconstruction \\
\cline{2-5}          & federated Shapley & \citet{DBLP:series/lncs/0013RZJS20} & appro & permutation sampling, group testing \\
\cline{2-5}          & extended fairness & \citet{DBLP:conf/icde/FanFZPFLZ22} & appro & factorization-based low-rank matrix completion \\
    \hline
    \end{tabular}%
    }
\end{table}%

\partitle{Computing Shapley values in generic games}
Studies in \cite{DBLP:journals/cor/CastroGT09,DBLP:journals/jmlr/MitchellCFH22,DBLP:journals/corr/MalekiTHRR13,DBLP:journals/cor/CastroGMT17,DBLP:conf/ijcai/BurgessC21,DBLP:conf/icpr/OkhratiL20,DBLP:journals/pacmmod/0006SL0P023,DBLP:conf/icde/0006XSL0P023,DBLP:conf/aaai/KolpaczkiBMH24,sun2024shapley,DBLP:journals/pacmmod/PangPXLL24} approximate the Shapley value in generic games. These proposed methods can be generally categorized based on their sampling mechanisms: simple random sampling and stratified random sampling. Simple random sampling is designed as sampling random permutations and computing average marginal contributions as Shapley values \cite{DBLP:journals/cor/CastroGT09,DBLP:conf/icml/GhorbaniZ19,DBLP:journals/jmlr/MitchellCFH22}. Alternatively, stratified random sampling is designed as stratifying marginal contributions based on coalition cardinality and computing the expectation of the strata average marginal contributions as Shapley values~\cite{DBLP:journals/corr/MalekiTHRR13,DBLP:conf/ijcai/BurgessC21,DBLP:journals/cor/CastroGMT17}. 

\emph{(1) Simple random sampling}

The most representative work proposed by \citet{DBLP:journals/cor/CastroGT09} develops a Monte Carlo permutation sampling method that estimates the Shapley value as the expectation of marginal contributions. The algorithm first samples a random permutation of players, scans the permutation from the first player to the last, and calculates the marginal contribution of each player. Repeating the procedure over multiple permutations, the Shapley value is approximated as the average of all calculated marginal contributions. The Monte Carlo sampling gives an unbiased estimate of the Shapley value. The more permutations, the more accurate the estimated Shapley value tends to be.

\citet{DBLP:journals/jmlr/MitchellCFH22} improve the permutation sampling via Quasi Monte Carlo techniques. They investigate new approaches based on reproducing kernel Hilbert spaces and hyperspheres. Specifically, they embed permutations into reproducing kernel Hilbert spaces and propose two greedy algorithms to select quadrature samples with low discrepancy relative to these kernels. The first algorithm follows kernel herding \cite{DBLP:conf/uai/ChenWS10} and selects the one yielding the maximum separation from a fixed number of samples at each iteration. The second algorithm follows the Bayesian quadrature \cite{DBLP:conf/uai/HuszarD12} and selects the sample yielding the minimum variance from a fixed number of samples at each iteration. Besides reducing the discrepancy of the sampled permutations, they map permutations of length $d$ to the Euclidean sphere $\mathbb{S}^{d-2}=\left\{x \in \mathbf{R}^{d-1}:\|x\|=1\right\}$ and propose two sampling algorithms. The first algorithm uses Gram-Schmidt process to generate permutations and the second algorithm uses transformed high-dimensional Sobol sequences to sample well-distributed permutations.

\emph{(2) Stratified random sampling}

As stated in \cite{mcbook}, the Monte Carlo estimator typically has an error variance of $\sigma^2/m$, where $\sigma^2$ is the variance and $m$ is the number of samples. A better approximation can be obtained by sampling with a larger value of $m$, but the computing time grows with $m$. Thus, reducing $\sigma$ instead is a widely adopted way. Stratified sampling is one of the classic variance reduction techniques. In approximating Shapley values, the stratification design is to stratify all marginal contributions into $n$ strata such that $k^{th}$ ($0\leq k\leq n-1$) stratum contains all marginal contributions $\mathcal{U}(\mathcal{S}\cup\{\bm{z}_i\})-\mathcal{U}(\mathcal{S})$ with $|\mathcal{S}|=k$. By sampling marginal contributions among strata, the Shapley value is approximated as the average of marginal contribution averages across all strata.  If strata and the total sample size are determined, there exists a further question of how to allocate the number of samples among the strata. Several approaches \cite{DBLP:journals/corr/MalekiTHRR13,DBLP:journals/cor/CastroGMT17,DBLP:conf/ijcai/BurgessC21} are developed to address the problem as follows.

\citet{DBLP:journals/corr/MalekiTHRR13} propose a stratified sampling algorithm that relies on an assumption about the range of utilities and give the sample size of each stratum based on the Hoeffding bound~\cite{hoeffding1994probability}. Denote by $m$ ($m\ge 0$) the total number of samples and $m_k$ ($0\leq k\leq n-1$) the number of samples allocated to $k^{th}$ stratum. By bounding the range of utilities as $a|\mathcal{S}| \leq \mathcal{U}(\mathcal{S}) \leq b|\mathcal{S}|$, they formulate an optimization problem about error bound by Hoeffding inequality \cite{hoeffding1994probability}: $\min \sum_{k=0}^{n-1} \frac{(k+1)}{\sqrt{m_{k}}}$, subject to $\sum_{k=0}^{n-1} m_{k} \leq m$. Then, $m_{k}=\frac{m(k+1)^{\frac{2}{3}}}{\sum_{j=0}^{n-1}(j+1)^{\frac{2}{3}}}$.

\citet{DBLP:journals/cor/CastroGMT17} propose a two-stage sampling algorithm inspired by Neyman allocation \cite{neyman1992two}. Denote by $m$ ($m\ge 0$) the total number of samples and $m_{i k}$ ($0\leq k\leq n-1$,$0\leq i\leq n$) the number of samples allocated to $k^{th}$ stratum of player $\bm{z}_i$. The optimal allocation is derived as $m_{i k}=\frac{\sigma_{i k}^{2}}{\sum_{j=1}^{n} \sum_{\ell=0}^{n-1} \sigma_{j \ell}^{2}} m{\tiny }$, where $\sigma_{i k}^{2}$ is the variance of $k^{th}$ stratum of player $\bm{z}_i$. Since $\sigma_{i k}^{2}$ is unknown, the algorithm takes $\frac{m}{2n^2}$ samples from each stratum and calculates unbiased sample variance of each stratum instead in the first stage. Then in the second stage, the algorithm assigns the remaining sample size to each stratum in proportion to the estimated variance based on the optimal allocation.

\citet{DBLP:conf/ijcai/BurgessC21} further extend to employ empirical Bernstein inequality and derive the Stratified Empirical Bernstein Bound (SEBB). In addition to the range of marginal contributions, SEBB also considers the sample variance, which helps monitor the estimator of each stratum and dynamically allocate samples among strata. Given a fixed number of samples, the algorithm sequentially selects a stratum to sample from in order to minimize SEBB. The stratified sample distribution needs to be determined on the fly and thus requires higher computational cost.

Furthermore, a multilinear extension of the Shapley value is represented in a probabilistic form as $\mathcal{SV}_i = \int_{0}^{1} e_{i}(q) d q$, where $e_{i}(q)=\mathbb{E}\left[\mathcal{U}\left(\mathcal{S} \cup\left\{z_{i}\right\}\right)-\mathcal{U}(\mathcal{S})\right]$. $q$ can be viewed as the parameter of a Bernoulli distribution. Based on the multilinear extension of the Shapley value, \citet{DBLP:conf/icpr/OkhratiL20} propose new sampling methods. They first propose Owen Sampling algorithm, which segments $q$ to discretize the integral and generate Bernoulli random sequences with parameter $q$ to estimate $e_i(q)$. The accuracy of Shapley value estimation is controlled by the discretization level of $q$ and the number of Bernoulli random sequences. They then improve Owen Sampling algorithm through symmetry and propose Halved Owen Sampling algorithm, which discretizes $q$ between $0$ and $0.5$ and generates Bernoulli random sequences with parameter $q$ and parameter $1-q$ simultaneously. Since antithetic variables are used, Halved Owen Sampling algorithm reduces the variance of the estimated Shapley value compared to Owen Sampling algorithm.

Since utility evaluation in many applications is costly, maximizing the use efficiency of sampled utility is necessary. The major bottleneck of sampling based on marginal contributions is that one sample of marginal contributions $\mathcal{U}(\mathcal{S}\cup \{\bm{z}_i\})-\mathcal{U}(\mathcal{S})$ is only used to update the Shapley value estimate for one player $z_i$, although coalition $\mathcal{S}$ may contain many other players. Therefore, \citet{DBLP:journals/pacmmod/0006SL0P023} propose a novel stratification design based on complementary contributions defined as $\mathcal{U}(\mathcal{S}\})-\mathcal{U}(\mathcal{N}\setminus\mathcal{S})$. Benefiting from the unique advantage that a complementary contribution can be used to update the estimate of the Shapley value for every player, the sample size and utility evaluation can be significantly reduced to achieve a good approximation. In addition to a simple stratified sampling algorithm based on complementary contributions, they also develop a sample allocation method based on Neyman approach to improve the performance. The algorithm based on the Neyman approach aims to minimize the sample variance using a sample allocation scheme. They divide the sampling process into two stages, get the approximate sample variance in the first stage, and allocate the last sample size based on the approximate sample variance in the second stage. But the two-stage sampling is heavily controlled by the results of the first-stage sampling, which limits the potential of dynamic updates. They further adopt the empirical Bernstein-Serfling inequality \cite{bardenet2015concentration} to design a dynamic sample allocation method that adaptively selects samples based on the sample variance.

\citet{DBLP:conf/aaai/KolpaczkiBMH24} approximate Shapley values by sampling coalitions and use these samples to update Shapley value estimates of multiple players simultaneously. Compared with marginal contributions, it reduces the number of samples required to achieve a certain level of approximation. They enhance this method by stratifying the coalitions based on their sizes. This stratification accelerates the convergence of Shapley value estimates and ensures better efficiency in the approximations. In addition, \citet{li2023faster} adjust the coalition coefficients to convert the Shapley value from the weighted difference of two utilities to a weighted sum of utilities. They then formulate the Shapley value approximation problem as a least squares optimization problem and sample coalitions to solve it, which enhances efficiency.

\emph{(3) Dynamic datasets}

Previous works assume that all data points are stable and immutable. They mainly attempt to estimate the Shapley value efficiently on a fixed dataset. In practice, a dataset can be continuously changed with new or removed data points. \citet{DBLP:conf/icde/0006XSL0P023} present several algorithms to efficiently get new Shapley values on a dynamic dataset without computing from scratch. In the case of adding data points, they propose the pivot-based algorithm and the delta-based algorithm. The pivot-based algorithm focuses on reusing computation, while the delta-based algorithm focuses on reducing the number of sampled permutations. In the case of deleting data points, they define and propose the YN-NN algorithm of polynomial complexity achieving full accuracy. They then present the delta-based algorithm, based on Shapley value differences similar to adding data points. Inspired by empirical observations, they propose similarity-based heuristic algorithms, which can quickly update Shapley values but with a little sacrifice on accuracy.

The above techniques are agnostic to game settings in the sense that users can provide their own definitions of the utility function. Instead, these techniques provide for the selection of samples and the allocation of sample sizes.

\partitle{Computing Shapley values in database}
In the context of relational databases, database $D$ is partitioned into a set $D_x$ of exogenous facts and a set $D_n$ of endogenous facts. Given a Boolean query $q(\bar{x})$ and a fact $f\in D_n$, the Shapley value is used to measure the importance of $f$ in $D$ for query $q$, which is defined as $Shapley\left(q, D_{\mathrm{n}}, D_{\mathrm{x}}, f\right) \stackrel{\text { def }}{=}\sum_{E \subseteq D_{\mathrm{n}} \backslash\{f\}} \frac{|E| !\left(\left|D_{\mathrm{n}}\right|-|E|-1\right) !}{\left|D_{\mathrm{n}}\right| !}\left(q\left(D_{\mathrm{x}} \cup E \cup\{f\}\right)-q\left(D_{\mathrm{x}} \cup E\right)\right)$, where $E$ is a subset of $D_n$. 
\citet{DBLP:conf/icdt/LivshitsBKS20,DBLP:conf/pods/ReshefKL20}, and \citet{DBLP:conf/sigmod/DeutchFKM22} provide an exact computation algorithm and heuristic for computing Shapley values of facts in query answering. For exact computation, they represent the lineage of query $q$ as a formula in disjunctive normal form, which is then compiled into a deterministic and decomposable Boolean circuit $C$. Rewrite Shapley value into $\text { Shapley }\left(q, D_{\mathrm{n}}, D_{\mathrm{x}}, f\right)=\sum_{k=0}^{\left|D_{\mathrm{n}}\right|-1} \frac{k !\left(\left|D_{\mathrm{n}}\right|-k-1\right)}{\left|D_{\mathrm{n}}\right|}(\#SAT_k(C_1)-\#SAT_k(C_2))$, where $C_1$ (resp., $C_2$) be the Boolean circuit obtained from $C$ by replacing all variable gates corresponding to the fact $f$ by a constant 1-gate (resp., by a constant 0-gate) and $\#SAT_k(C)$ ($|C|=k$) is the size of the set of all assignments satisfying $C$. Since they prove that the quantity $\#SAT_k(C)$ ($k \in \{0,\ldots,|Vars(C)|\}$) can be computed in polynomial time \cite{DBLP:conf/aaai/ArenasBBM21}, the exact Shapley value can be computed by computing all $\#SAT_k(C)$. For heuristic, they represent the query lineage as a formula in conjunctive normal form and approximate the Shapley value as the sum (instead of the product) of the Shapley value in each literal disjunction. Recently, the complexity of computing Shapley values in boolean query answering has been rigorously analyzed through a reduction to either weighted or unweighted model counting \cite{DBLP:journals/corr/abs-2306-14211,DBLP:journals/corr/abs-2312-14529,DBLP:journals/corr/abs-2401-06493}.

Under the assumption of independent utility, where the utility of a data set is the sum of the utilities of its independent tuples, \citet{DBLP:journals/pvldb/LuoPCX22} study efficiently computing the exact Shapley value for sharing revenue among multiple data sellers. In a data assemblage task, data sets of data sellers are assembled to produce a coalition (data) set. The independent utility makes the Shapley value of the data seller's data set decomposed to the sum of its Shapley values measured on each individual tuple in the coalition data. They introduce the concept of minimal syntheses, which are the smallest coalitions of data sellers that can collectively produce a particular tuple. They find that the time complexity of computing the exact Shapley value for each data seller's data set can be reduced to enumerating combinations of these minimal syntheses. They further improve their approach with a monotone Boolean function decomposition strategy and the corresponding computations, enabling efficient handling of scenarios with dense minimal syntheses \cite{DBLP:journals/pacmmod/LuoPXZX24}.

Additionally, \citet{DBLP:journals/dcg/CabelloC22} study the problem of computing the Shapley value in the plane. They provide polynomial-time algorithms to compute exact Shapley values for various geometric shapes, such as convex hulls ($O(n^{2})$), minimum enclosing disks ($O(n^{3})$), and anchored rectangles ($O(n^{3/2})$), where the utility function for a point coalition is set to the corresponding area or perimeters. For axis-parallel problems, they use algebraic methods. For non-axis-parallel problems, they decompose Shapley values and group permutations. \citet{DBLP:conf/isaac/Tan21} explore in higher dimensions and extend to computing Shapley values in 3D when adopting the mean width of the convex hull as the utility function.

\partitle{Computing Shapley values in centralized machine learning}
In the context of machine learning, the performance of a model trained using a subset of the training data and tested on another test set is often used as the utility function. Studies \cite{DBLP:conf/ec/AgarwalDS19,DBLP:conf/icml/GhorbaniZ19,DBLP:journals/pvldb/JiaDWHGLZSS19,DBLP:conf/icml/GhorbaniK020,DBLP:conf/aistats/KwonRZ21,DBLP:journals/corr/abs-2104-08312,DBLP:conf/nips/GhorbaniZ20,DBLP:conf/sigmod/Fernandez22,DBLP:journals/pvldb/XiaL0Q00023,DBLP:journals/pvldb/LiuLZR0L0PS21,DBLP:journals/pvldb/Xia24} draw on the characteristics of machine learning models and propose specific Shapley value approximation algorithms.

\citet{DBLP:conf/icml/GhorbaniZ19} propose two algorithms to accelerate the estimation of Shapley values by reducing the computational cost of training models. They consider that the change in model performance caused by adding one more training data point gets smaller and smaller as the dataset size increases. Thus, the computational overhead of the near-zero marginal contributions can be truncated. The first algorithm they proposed, named Truncated Monte Carlo Shapley, samples a random permutation, scans the permutation, and computes the marginal contribution until the marginal contribution is less than a performance tolerance. Also, they consider that the fully trained model which is trained with stochastic gradient descent can be approximated by a model trained by the new training data with one pass/epoch. The second algorithm they proposed, named Gradient Shapley, updates the model by performing gradient descent on one data point at a time to approximate the marginal contributions of the sampled permutation. 

\citet{jia2019towards} propose a group testing-based approach, which estimates Shapley value difference $\mathcal{SV}_i - \mathcal{SV}_j
$ and then derives Shapley values from the Shapley value differences by solving a feasibility problem. 
The technique is further being refined by \citet{wang2023notegroup}. 
Exploiting the sparsity of Shapley values for large datasets, they propose a compressive permutation sampling algorithm, which utilizes compressive sensing to compute the deviation between Shapley values and the average of Shapley values and then derives Shapley values by adding the average of Shapley values and the deviation. In addition, they propose a heuristic algorithm, inspired by the result in \cite{DBLP:conf/icml/KohL17}, to use influence to infer the effect of adding a data point on the model.

Follow-up works by \citet{DBLP:journals/pvldb/JiaDWHGLZSS19} and \citet{wang2023noteknn} focus on one family of models relying on $k$-nearest neighbors ($K$NN). They define the utility function of a $K$NN classifier as the proportion of training examples whose class label matches the test label among the nearest training examples, i.e., $\mathcal{U}(\mathcal{S})=\frac{1}{K} \sum_{k=1}^{\min \{K,|\mathcal{S}|\}} \bm{1} \left[y_{\alpha_{k}(\mathcal{S})}=y_{\text {test }}\right]$, where $\alpha_{k}(\mathcal{S})$ is the index of training feature that is $k^{th}$ closest to $x_{test}$ among the training examples in $\mathcal{S}$. Based on the utility function, they have the Shapley value difference between two data points as $\mathcal{SV}_{i}-\mathcal{SV}_{i+1}=\frac{\bm{1}\left[y_{i}=y_{\text {test }}\right]-\bm{1}\left[y_{i+1}=y_{\text {test }}\right]}{K} \frac{\min \{K, i\}}{i}$ and Shapley value of the data point that is farthest from the test data as $\mathcal{SV}_n = \frac{\bm{1}\left[y_{N}=y_{\text {test }}\right]}{N}$. Thus, the exact Shapley value can be computed in a recursion. For scalability to larger and higher-dimensional datasets, they employ Locality Sensitive Hashing to find approximate nearest neighbors, which improves the efficiency. Besides, \cite{DBLP:conf/cvpr/Jia0SXDK00S21} explore the superiority of $K$NN-Shapley in scalability. Although unweighted $K$NN-Shapley can be computed efficiently in \cite{DBLP:journals/pvldb/JiaDWHGLZSS19}, the complexity of computing weighted $K$NN-Shapley (W$K$NN Shapley) is still $O(n^K)$, resulting in that its computation becomes impractical even for modest $K$, such as $5$. To fasten the computation of W$K$NN Shapley, \citet{wangj2024aistats} employ the accuracy of hard-label $K$NN with discrete weights as the utility function and consequently propose a quadratic-time algorithm. Furthermore, a deterministic approximation algorithm is developed to improve the efficiency in computing W$K$NN-Shapley while maintaining the key fairness properties.

\citet{DBLP:journals/corr/abs-2107-06336} use a parametric model learned by the utilities of sampled coalitions to predict the utilities of previously unsampled coalitions, thereby accelerating the estimation of Shapley values for data points.

\citet{DBLP:conf/nips/LundbergL17} derive the Shapley kernel 
which yields optimal regression coefficients equal to the Shapley value. Based on the Shapley kernel, they develop KernelSHAP, which approximates Shapley values via linear regression. \citet{DBLP:conf/aistats/CovertL21} further improve KernelSHAP by introducing an unbiased version. They claim unbiased KernelShap has both consistency and unbiased properties while KernelShap is only proven to be consistent. Moreover, they propose a paired sampling strategy for variance reduction where each sample $\bm{z}_i$ is paired with its complement $1-\bm{z}_i$. 
The follow-up work by \citet{DBLP:conf/iclr/JethaniSCLR22} proposes FastSHAP, which estimates Shapley values fast based on learning. FastSHAP studies a task of model explanation. In this task, the Shapley value assigns a value to each feature to represent the effect of that feature on the model prediction. The features of each data tuple are treated as players, and the utility function of a subset of features (players) is the model prediction of these features. Inspired by the weighted least squares characterization of the Shapley value, FastSHAP trains a neural network based on sampling coalition utilities on the training dataset to predict the Shapley value on the test dataset. Experimental results show that FastSHAP performs well on image explanations.

\citet{DBLP:conf/icml/GhorbaniK020} propose distributional Shapley values to measure the value of data points where the dataset is drawn in an independent and identically distributed (i.i.d.) manner from the underlying distribution. On the basis of this work, \citet{DBLP:conf/aistats/KwonRZ21} derive the analytic expressions for distributional Shapley for linear regression, binary classification, and non-parametric density estimation. As an alternative extension of Shapley value, \citet{kwon2021beta} propose Beta-Shapley, which relaxes the efficiency axiom, i.e., one of the cooperative game-theory axioms used in the Shapley value, demonstrating promising performance in mislabeled data detection. Along similar lines, \citet{DBLP:conf/aistats/WangJ23} propose DataBanzhaf, which assigns equal weight for every marginal contribution and relaxes the efficiency axiom like Beta-Shapley. They theoretically show that DataBanzhaf achieves the largest safety margin among semivalues. The safety margin is a measure of robustness, indicating how much noise or perturbation in the model performance scores can be tolerated before the data value rankings change. This ensures that DataBanzhaf can reliably rank the importance of data points even under noisy conditions.

\partitle{Computing Shapley values in federated learning}
\citet{DBLP:conf/bigdataconf/SongTW19} apply the Shapley value to quantify the contributions of clients in federated learning and propose two approximation algorithms. The first algorithm estimates Shapley values in one round by reconstructing models trained with subsets of clients using their gradients from all rounds, while the second algorithm aggregates Shapley values in each round. In their setting, all clients should participate in each training round.

However, federated learning usually involves the sequential training process and client selection to improve efficiency. Thus, the assumption in applying the Shapley value that the order of training datasets has no impact on the model performance becomes invalid. To address the issue, \citet{DBLP:series/lncs/0013RZJS20} introduce the federated Shapley value, which is the sum of the Shapley values of the selected clients in all training rounds. They propose a permutation sampling-based approximation algorithm and a group testing-based approximation algorithm for estimating the federated Shapley value. Furthermore, \citet{DBLP:conf/icde/FanFZPFLZ22} point out that assigning zero Shapley values to unselected clients in the federated Shapley value may lead to unfairness, as two clients with the same local data could receive different valuations. Thus, they propose the completed federated Shapley value, which is the sum of the Shapley values of all clients in all training rounds. They design a utility matrix to store all possible utilities of client coalitions overall training rounds. It is necessary to complete the utility matrix to compute the completed federated Shapley value. Under the mild assumption about the loss function, the utility matrix is proved to be approximately low-rank, so they propose the factorization-based low-rank matrix completion problem to complete the utility function. They further adapt the Monte Carlo sampling method to cope with the exponential size of the utility matrix.

\subsection{Potpourri}
In addition to the privacy cost-based and Shapley value-based approaches, there are other ways to allocate revenue among data sellers.

In query-based markets, \citet{koutris2013toward} design a policy called FAIRSHARE to fairly distribute revenue among data sellers who contribute to answering the query. FAIRSHARE calculates the maximum revenue that each seller can contribute among all minimum-cost solutions for a query. This revenue is determined by the maximum price that a seller's views can achieve in any of these minimum-cost solutions. The trading price of the query is then allocated to each seller in proportion to their maximum revenue.

In model-based markets, Leave-One-Out (LOO) \cite{DBLP:journals/pr/CawleyT03} is a well-known method for assessing the importance of individual data points for a model. LOO compares the difference between the performance of the model trained by the entire dataset and the performance of the model trained by the entire dataset minus one data point. If a model's performance decreases significantly when a particular data point is left out, that data point can be considered important for the model. \citet{DBLP:conf/icml/KohL17} adopt influence functions \cite{hampel1974influence,jaeckel1972infinitesimal} to evaluate the change in test loss after removing a single training data point. However, the computation of the influence function requires the inverse Hessian-vector product (iHVP), which is infeasible for large models from both computational and memory perspectives. Several follow-up works propose different ways to scale up the influence function \citep{guo2020fastif, schioppa2022scaling, grosse2023studying}. 
There are also works that extend the influence function to capture higher-order interactions \citep{basu2020second}. 


While Shapley-based valuation methods provide a principled revenue allocation, their computational costs often cause practical challenges in real-world applications. This is because estimating the Shapley value requires training multiple models, making it infeasible to apply to large-scale datasets. 
To address this problem, \citet{wang2024data} modifies the utility function and tracks the Data Shapley scores during the training process. \citet{kwon2023data} propose Data-OOB which leverages an ensemble model and does not require training multiple models. \citet{DBLP:conf/iclr/JustK0ZK0J23} propose a learning algorithm-agnostic data valuation method based on class-wise Wasserstein distance which avoids the learning costs. Although vast works of literature consider that data is distributed among different data sellers with shared features, it is possible that some data sellers only hold a data fragment including partial features. To realize the data valuation for data fragments, \citet{DBLP:conf/icml/LiuJCCJ23} propose 2D-Shapley to measure the contribution of each data fragment towards a specific machine learning task.

\subsection{Discussion on Actual Data Marketplaces}
The Shapley value is widely used to assess individual contributions in complex real-world systems where multiple participants contribute to a common outcome. For instance, Bittensor applies the Shapley value to determine and reward the contribution of each node in achieving consensus and making accurate predictions in the network \cite{Bittensor}. Google Ads Data Hub utilizes the Shapley value to model the contribution of advertising channels and touchpoints toward a conversion \cite{google_ads_data_hub_shapley}.

Reklaim \cite{Reklaim} positions itself as a privacy-preserving platform that allows data sellers to regain control of their personal data while potentially monetizing it. Data sellers can decide how much data they want to share with Reklaim, and Reklaim compensates data sellers based on the amount of data they share. In calculating the value of data from data sellers, Reklaim considers several metrics, including (1) the value of data to Reklaim, (2) the revenue generated from data by Reklaim, (3) the expenses related to the services Reklaim provides, and (4) profit generated from data by Reklaim. 


\section{Privacy, Security, and Trust}\label{sec:untrusted}
The establishment of data markets that enable the creation of new commercial benefits is a promising prospect, but it is also susceptible to data inference risks and deception by curious or dishonest entities. These entities may attempt to infer the original data from the data products or engage in fraudulent activities to gain additional profits. For instance, malicious data sellers might manipulate their private data to receive more revenue than they deserve, while curious 
data buyers could exploit purchased data products to obtain private information. Malicious data brokers might also abuse personal data and create fake data products. 
Therefore, extensive studies have been conducted to develop defense mechanisms against potential attacks by curious or dishonest entities in untrusted data markets. We describe techniques for ensuring privacy preservation in Section \ref{subsec:privacy}, fairness in Section \ref{subsec:fairness}, profitability in Section \ref{subsec:profitability}, and traceability in Section \ref{subsec:traceability}.

\subsection{Ensuring Privacy Preservation}\label{subsec:privacy}
Data owners face a serious risk of privacy disclosure when trading their data in an untrusted data market where dishonest entities attempt to learn the original input given the resulting data product. We summarize several privacy leakage attacks in Section \ref{subsec:privacy-leakage-attacks}, introduce prominent tools and techniques for neutralizing the attacks in Section \ref{subsec:tools-and-techniques}, and detail how to apply these privacy-preserving methods in Section \ref{subsec:preserve-privacy}. 

\subsubsection{Privacy Leakage Attacks}\label{subsec:privacy-leakage-attacks}
We review several types of privacy leakage attacks against data owners in different data markets.

\partitle{Privacy attacks in dataset- and query-based markets}
In dataset-based markets and query-based markets, preserving the privacy of data owners is about protecting all entries in the database contributed by them. Privacy is violated if an adversary can infer any confidential entry. Prominent attacks in the database include record linkage attacks \cite{DBLP:conf/infocom/GramagliaFTB17}, reconstruction attacks \cite{DBLP:conf/pods/DinurN03}, and membership inference attack \cite{DBLP:conf/ndss/PyrgelisTC18}.

\begin{definition}(Record Linkage Attack \cite{DBLP:conf/infocom/GramagliaFTB17})
A record linkage attack aims at univocally distinguishing a record of an individual in an anonymous database by achieving a linkage with non-anonymous records in public datasets.
\end{definition}

\nop{
\begin{definition}(Reconstruction Attack  \cite{DBLP:conf/infocom/GramagliaFTB17})
A probabilistic attack allows an adversary with partial information about an individual to enlarge its knowledge of that individual by accessing the database.
\end{definition}
}

\begin{definition}(Reconstruction Attack \cite{DBLP:conf/pods/DinurN03})
A reconstruction attack partially reconstructs a private dataset from public aggregate information (e.g., query answers). A private database can be modeled as a sequence of bits $D=(d_1, \ldots, d_n)$, where each bit is the private information of a single entry. Define the answer $a$ to a query $q$ specified by a subset $S$ ($S \subseteq \{1, \ldots, n\}$), $a_{q_S} = \sum_{i\in S}d_i$. Given perturbed answers $a_1, \ldots, a_m$ to queries specified by subsets $S_1, \ldots, S_m$, such that $|a_i - a_{q_{S_i}}| \le \epsilon$ for all $i \in \{1, \ldots, m\}$, an adversary can reconstruct most of $D$ with sufficiently small $\epsilon$ and sufficiently large $m$.
\end{definition}

\begin{definition}(Membership Inference Attack against aggregate data \cite{DBLP:conf/ndss/PyrgelisTC18}) 
A membership inference attack in datasets aims to determine whether a given data instance is included in a dataset.
\end{definition}

\partitle{Privacy attacks in model-based markets}
In model-based markets, preserving the privacy of data owners is about preventing attacks on machine learning models. Prominent attacks in machine learning include membership inference attacks \cite{DBLP:conf/sp/ShokriSSS17}, gradient leakage attacks \cite{DBLP:conf/ccs/HitajAP17, DBLP:conf/nips/ZhuLH19}, model inversion attacks \cite{DBLP:conf/ccs/FredriksonJR15}, and secret memorization attacks \cite{DBLP:conf/uss/Carlini0EKS19}.

\begin{definition}(Membership Inference Attack against models \cite{DBLP:conf/sp/ShokriSSS17}) 
A membership inference attack aims to determine whether a given data instance is in the training dataset of a specific model or not.
\end{definition}

\citet{DBLP:conf/ccs/FredriksonJR15} propose model inversion attacks that exploit confidence information of the machine learning classifiers returned at inference time to reconstruct the victim training samples. In particular, they show the model inversion attack can successfully extract images in the training dataset from facial recognition models and expose the identity of the victim person. 
\begin{definition} (Model Inversion Attack \cite{DBLP:conf/ccs/FredriksonJR15})
    A model inversion attack reconstructs the victim sample in the training dataset of the machine learning model corresponding to a target label based on the confidence information returned by the model during inference time.
\end{definition}

\citet{DBLP:conf/uss/Carlini0EKS19} propose secret memorization attacks that can memorize rare or unique training data sequences utilized for training generative sequence models. In particular, it reveals the phenomenon of unintended memorization, where deep learning models often memorize rare details about the training data.
\begin{definition} (Secret Memorization Attack \cite{DBLP:conf/uss/Carlini0EKS19})
    A secret memorization attack extracts rare and unique secret sequences in the training dataset from a sequence generative model.
\end{definition}

In decentralized machine learning including federated learning, 
\citet{DBLP:conf/ccs/HitajAP17} propose gradient leakage attacks that further exploit the interactive process (i.e., gradients and parameters exchange process) and enable the attacker to recover private prototype samples of the target training set. \citet{DBLP:conf/nips/ZhuLH19} show that we can obtain the training data reversely with the gradients $\nabla w$ w.r.t a specific data instance consisting of a pair of input and label.
\begin{definition}(Gradient Leakage Attack)
A gradient leakage attack (also known as client privacy leakage attack \cite{DBLP:conf/esorics/WeiLLCGTW20}) is a feature reconstruction attack, where the attacker can obtain the gradient update $\nabla w_k(t)$ of client $k$ at round $t$, and then use the reconstruction learning algorithm to reconstruct the private data used in the local training process.
\end{definition}

\subsubsection{Tools and Techniques}\label{subsec:tools-and-techniques}
Tools and technologies designed to neutralize privacy leakage attacks have been widely studied in a variety of scenarios over the years. The most prominent branches with applications in data markets are differential privacy, secure computation, federated learning, and machine unlearning.

\partitle{Differential privacy}
Differential privacy \cite{DBLP:conf/tcc/DworkMNS06,DBLP:conf/tamc/Dwork08,DBLP:journals/fttcs/DworkR14} gives the state-of-the-art model of quantifying and limiting individual information disclosure. The principal idea is to define privacy leakage as the degree of indistinguishability, which requires that a change of one entry in the database causes a small change in the query answers from the view of the adversary. Recalling Definition \ref{def:DP}, differential privacy is quantified by two critical parameters $\epsilon$ and $\delta$. $\epsilon$ controls the amount of noise to be added and captures the degree of indistinguishability, which intuitively can be seen as the upper bound of the privacy leakage for each data sample. The smaller $\epsilon$ is, the more indistinguishable the outputs are, and the less the privacy leakage is. $\delta$ captures the probability that the privacy leakage is out of the upper bound. Given a specified $\epsilon$, the closer $\delta$ is to 0, the less the probability of privacy breach is, and the better the algorithm is. Furthermore, \citet{DBLP:conf/tcc/DworkMNS06} point out that privacy guarantees degrade as additional information is published about the same data.
In practice, for a meaningful differential privacy guarantee, the parameters are chosen as $0<\epsilon \leq 10, \delta \ll \frac{1}{n}$, where $n$ is the number of data owners \cite{DBLP:conf/sp/IyengarNSTTW19}. $\mathcal{S}$ and $\mathcal{S}'$ correspond to adjacent datasets, where $\mathcal{S}'$ can be obtained from $\mathcal{S}$ by adding or subtracting all the records of a single data owner.

Perturbation techniques for differential privacy can be classified into two basic categories based on the presence or absence of a trusted data broker. 
\begin{itemize}
    \item Centralized differential privacy \cite{DBLP:conf/tcc/DworkMNS06}: perturbation occurs at the output stage. The correct answer to a query is first computed exactly from real data, and then a noisy version of it is reported. Centralized differential privacy requires a trusted data broker to collect raw data from data owners and apply perturbations by injecting random noise for privacy concerns before realizing final data products.
    
    \item Local differential privacy \cite{DBLP:journals/siamcomp/KasiviswanathanLNRS11}: perturbation occurs at the input stage. The real data is first randomly modified, and then the answer to a query is computed on the modified data and reported. Local differential privacy does not require a trusted data broker. It assumes that the distribution of the dataset consisting of all data of data owners is always stable even when a data owner can change its response suddenly. However, local differential privacy can be more challenging to have a satisfactory privacy-utility tradeoff for certain tasks when compared to centralized differential privacy \cite{DBLP:conf/colt/Gopi0KNWZ20}, which is mainly due to the stricter privacy requirement and more perturbations introduced.
\end{itemize}

\partitle{Secure computation}
Secure computation enables multiple entities to jointly evaluate functions in a way that only the results of computation are revealed to the intended parties, and no additional information is revealed.

Secure multi-party computing (MPC) is a subfield of cryptography. It can help multiple participants to compute the intended output of a specified function without exposing their own private inputs. This area was kicked off by \citet{DBLP:conf/focs/Yao86} in the 1980’s. Thanks to theoretical and engineering breakthroughs \cite{DBLP:conf/eurocrypt/FurukawaLNW17}, MPC has moved from theory to industrial deployment.
Trusted execution environment (TEE, also known as a secure enclave) enables data operation in an isolated enclave inaccessible from the rest of the host, where code can be proven and verified \cite{DBLP:conf/ccs/Subramanyan0LDS17}. It provides an opportunity to move parts of the data evaluation and transaction process to a trusted environment in the cloud.

Although secure multi-party computation and trusted execution environment provide general solutions for private computing problems in data markets, the resource overhead it brings cannot be ignored. In the case of focusing on a specific data product, targeted optimizations can be made to reduce computational resource overhead. For example, secure aggregation is an effective approach for handling count queries in query-based markets, where each data owner submits their value in an encrypted form and the data buyer computes an aggregate function based on the submitted values.

\partitle{Federated Learning}
Federated learning is introduced by \citet{mcmanhan17}, which refers to ``a machine learning setting where multiple entities collaborate in solving a machine learning problem, under the coordination of a central server or service provider. The raw data of each client is stored locally and not exchanged or transferred; instead, focused updates intended for immediate aggregation are used to achieve the learning objective.'' \cite{DBLP:journals/ftml/KairouzMABBBBCC21}. 

Federated learning enables data owners to jointly learn a machine learning model without revealing the raw data. It aims to prevent attacks on the privacy of data owners by sharing local training model parameters rather than the raw data and minimizing the communication complexity of model training over distributed data \cite{DBLP:conf/iclr/HouTFO22}. However, recent work has shown that there are still some privacy threats in transferring locally trained model parameters. The adversary can partially reconstruct the raw data based on the uploaded parameters (i.e., model gradients), which is well-known as the client privacy leakage attack. To mitigate this privacy threat, different privacy-preserving enhancement techniques, such as differential privacy and secure computation, are applied in federated learning.
For example, \citet{DBLP:conf/ccs/BonawitzIKMMPRS17} design a secure aggregation protocol for federated learning based on secret sharing, which can operate on high-dimensional vectors efficiently. \citet{DBLP:journals/tifs/PhongAHWM18} propose a novel asynchronous federated learning framework in combination with additively homomorphic encryption for the protection of the gradients. However, the computational overhead of secret sharing and homomorphic encryption makes training on large-scale data more difficult. \citet{DBLP:journals/corr/abs-1812-00984} firstly introduce differential privacy into federated learning. They design new optimal locally differentially private mechanisms and enable federated learning to fit large-scale models with little degradation in utility. \citet{DBLP:conf/eurosys/Truex0CGW20} propose a novel federated learning system using local differential privacy and theoretically analyze the privacy guarantee. But when the privacy budget is small, both the approaches in \cite{DBLP:journals/corr/abs-1812-00984} and \cite{DBLP:conf/eurosys/Truex0CGW20} will significantly degrade model performance. \citet{DBLP:conf/kdd/SunL00LLQ023} design a gradient aggregation algorithm that assigns higher weights to high-quality local model parameters, which are evaluated by the server using the Shapley value. \citet{DBLP:conf/ijcai/SunQC21} further reduce the variance of estimated model weights by considering the range differences of weights in different deep learning model layers and achieving superior deep learning performance. In practice, clients often have diverse privacy requirements, driven by unique policies or individual preferences. \citet{DBLP:journals/pvldb/LiuLXLM21} focuses on explicitly modeling and accommodating these heterogeneous privacy demands across clients. Furthermore,  \citet{DBLP:conf/ccs/Liu24} study the cross-silo federated learning with record-level personalized differential privacy, aiming to solve the strong assumption of uniform privacy budget across all records.

\partitle{Machine unlearning} Machine unlearning~\cite{DBLP:conf/sp/CaoY15} refers to the process of removing previously learned data from a well-trained machine learning model.
Recent regulations including GDPR~\cite{GDPR}, CCPA~\cite{CCPA}, and PIPL~\cite{PIPL}, stipulate \emph{the right to be forgotten}~\cite{GDPR}, which mandates users shall have the right to require service providers (e.g., companies and institutions) to erasure their personal data. While it may be straightforward to delete data from back-end databases, this is insufficient in the context of machine learning because machine learning models often memorize the training data. Existing adversarial attacks~\cite{DBLP:journals/csur/ZhouLYZZY23} on trained machine learning models, such as membership inference attacks~\cite{DBLP:conf/sp/ShokriSSS17} and model inversion attacks~\cite{DBLP:conf/ccs/FredriksonJR15}, have proven that we can determine whether a given instance or attribute belonged to the training data and even restore the training data from the trained models. This phenomenon necessitates a new paradigm, machine unlearning, to ensure machine learning models forget particular data effectively and flexibly. 

In model-based markets, machine unlearning is envisioned as an infrastructure since unlearning enables a data owner to cancel a data transaction cleanly or rent out the use rights of its data without giving up ownership. Retraining machine learning models from scratch without deleted data is naive and inefficient.  While effective at unlearning, retraining large models (e.g., large language models) can take months and cost millions. To reduce the cost of unlearning, researchers proposed several approaches which can be categorized as exact or approximate unlearning~\cite{DBLP:journals/corr/abs-2209-00939} by the difference in their goals. 

Exact unlearning requires that the model after unlearning is identical to the model retrained over the remaining data after data removal. Most exact unlearning algorithms limit the influence of each training data instance in the training process to enable efficient retraining. \citet{DBLP:conf/sp/CaoY15} come up with the concept of machine unlearning for the first time and develop the exact unlearning algorithms for statistical query learning models. Then, researchers further consider the unlearning problem for other prevalent machine learning models. \citet{DBLP:conf/nips/GinartGVZ19} develop an exact unlearning algorithm applied to the $k$-means algorithm by recording states and sharding the training data. Some researchers~\cite{DBLP:conf/alt/Neel0S21,DBLP:conf/icml/BrophyL21,DBLP:conf/sigmod/SchelterGD21} develop tree-based unlearning algorithms over randomized decision trees. Motivated by ensemble learning and distributed training, \citet{DBLP:conf/sp/BourtouleCCJTZL21} introduce a sharding methodology SISA to realize exact unlearning for neural networks. \citet{DBLP:conf/www/Chen0ZD22} and \citet{DBLP:conf/ccs/Chen000H022} further study the data sharding mechanism and extend SISA to recommendation systems and graph neural networks, respectively.

In contrast, approximate unlearning relaxes the requirement of generating an identical model to the retrained one to accelerate unlearning~\cite{DBLP:conf/nips/GinartGVZ19}. \citet{DBLP:conf/nips/GinartGVZ19} propose approximate unlearning based on the concept of differential privacy and aims to make the model after unlearning and the model trained over the remaining data after data removal indistinguishable. 
\citet{DBLP:conf/cvpr/GolatkarAS20} develop an approach by modifying the weights and minimizing the KL divergence to forget the training data. \cite{DBLP:conf/icml/GuoGHM20,DBLP:conf/aaai/MarchantRA22,DBLP:conf/aaai/WuHS22} develop approximate unlearning algorithms by estimating the impact of specific data by influence function~\cite{DBLP:conf/icml/KohL17} and removing it from parametric models. Further, \citet{DBLP:conf/ndss/WarneckePWR23} improve the influence-function-based unlearning for the scenarios where a large group of labels and features need to be removed. Moreover, in the model training process of model markets, the training dataset is not always accessible, such as federated learning~\cite{DBLP:journals/corr/abs-2111-12056}. Therefore, federated unlearning~\cite{DBLP:journals/corr/abs-2012-13891} and subsequent unlearning algorithms for specific federated tasks, including federated recommendation~\cite{DBLP:journals/corr/abs-2210-10958} and federated clusters~\cite{DBLP:conf/iclr/0003SPRM23}, emerge. Although approximate unlearning enjoys an impressive speed-up compared to exact unlearning, an intractable task for approximate unlearning algorithms is to verify and prove the specific data has been erased from the model.

\subsubsection{Preserving Privacy of Data Owners}\label{subsec:preserve-privacy}
We detail the application of privacy-preserving methods in data markets from three aspects: producing data products with differential privacy, concealing raw data with federated learning and secure computation, and preventing data product abuse.

\partitle{Producing data products with differential privacy} 
Many research works design data marketplaces with differential privacy. Most of these adopt centralized differential privacy and require a trusted broker who is authorized to access and control the raw data. In dataset-based markets, \citet{DBLP:journals/tpds/CaiYFXZ23} propose a near-optimal attribute clustering scheme to reduce the dimensionality of high-dimensional datasets. This reduction in dimensionality allows for less noise to be added in clusters during data perturbation, which enhances the data utility. In query-based markets, studies \cite{DBLP:conf/icdt/LiLMS13, DBLP:journals/tods/LiLMS14, DBLP:journals/cacm/LiLMS17, DBLP:conf/kdd/NiuZWTGC18,DBLP:conf/sigecom/GhoshR11,DBLP:conf/uai/ZhangBL20} inject random noise to the exact query answer and return this noisy version to data buyers. \citet{DBLP:conf/mdm/ZhengCY20} consider personalized differential privacy, that is, the data broker injects different degrees of noise to the raw data and returns query answers computed based on noisy data. \citet{DBLP:journals/tpds/CaiYYZLX22} process multiple correlated queries at once and propose a matrix mechanism that minimizes the noise added to ensure correlated differential privacy compared to methods that handle each query independently. In model-based markets, \citet{DBLP:conf/sigmod/ChenK019} inject noise to the parameters of machine learning models trained on raw data sets. Considering protecting the privacy of data owners not only against buyers but also against data brokers, several studies adopt local differential privacy to construct data marketplaces, without the assumption of a trusted data broker. \citet{DBLP:journals/pvldb/LiuLL0PS21} inject noise to raw data and use noisy data sets to train machine learning models. \citet{DBLP:journals/tc/HeC24} let data owners perturb their data locally before sending it to the data broker using local differential privacy. \citet{zheng2022fl} develop a locally private model marketplace by combining federated learning with local differential privacy, where the data broker receives local perturbed gradients from data owners to aggregate a model. In addition, \citet{DBLP:conf/nips/Wang00JM23} identify the privacy risk in data valuation and propose Threshold KNN-Shapley to offer a superior privacy-utility tradeoff as well as enable the incorporation of differential privacy guarantee.

\partitle{Concealing raw data with federated learning and secure computation}
Techniques including federated learning and secure computation alleviate a certain degree of privacy breaches by concealing raw data. Federated learning allows data owners to collaboratively train models under a data broker without exposing their raw data. There have been some works proposed to build data marketplaces based on federated learning \cite{zheng2022fl,DBLP:conf/infocom/SunCLH22,DBLP:conf/icde/FanFZPFLZ22}. To defend against potential gradient leakage attacks, a Trusted Execution Environment (TEE) can be used as a computation environment for both data owners and the data broker due to the attestation and integrity of TEE. However, it may be unrealistic to implement TEE in the practical data market because TEE is sensitive to computational resources \cite{DBLP:conf/iclr/TramerB19}. And some data owners may only hold terminal devices such as mobile phones, but not hold TEE. Protection similar to TEE can be achieved by designing a communication protocol using MPC. Specifically, the data broker can use the key generated by herself to encrypt the data sent to data owners, and then data owners perform computational operations based on the homomorphic encryption system, and finally send it back to the data broker. In order to prevent the data broker from inferring private information, data owners need to complete the data aggregation before sending the data to the data broker, that is, build a secret channel for calculation or resort to TEE. Recently, \citet{DBLP:conf/icde/Ma0021} propose a blockchain-based federated learning framework that includes an MPC-compatible method for transparently evaluating each participant's Shapley value. Moreover, \citet{DBLP:journals/corr/abs-2210-08723} study the problem of data valuation without accessing original data sources. Their method efficiently computes Shapley value using MPC circuits along with a performance predictor. \citet{xu2022data} present an SMC-based approach for addressing the private data trading challenge by incorporating a data appraisal phase that employs influential functions. These functions are calculated on private data through multi-party computation, eliminating the need for data sharing between data owners and model owners. \citet{DBLP:journals/pvldb/ZhengCY23} propose SecSV, a privacy-preserving two-server protocol for computing Shapley values in cross-silo federated learning. By employing a hybrid privacy scheme and strategically skipping some test samples, SecSV strikes a good balance between accuracy and runtime. \citet{DBLP:journals/corr/abs-2210-08723} introduce a data marketplace that values data sources based on the Shapley value in a privacy-preserving manner and ensures fair payments, enabled by a suite of innovations on both secure multiparty computation and blockchain.

\partitle{Preventing data product abuse with access control and accountability}
To eliminate or reduce threats, access to data products needs to be carefully designed. A zero trust security model \cite{gilman2017zero} can be used. For example, metadata of the data owner's private data can be published on the blockchain, a data buyer can request an encrypted version of the data owner's data, and the decrypted key is sent after verification \cite{DBLP:journals/pvldb/HynesDYCS18,DBLP:journals/tifs/DaiDCCZ020}. However, this model greatly restricts the possible transaction forms, that is, it can only match the transactions between buyers and sellers and cannot process the data. Furthermore, when the data product is a machine learning model, buyers can only access the model in a black-box manner within a limited number of invocations. Even when the adversary only has black-box access to the machine learning model, extracting the target model with near-perfect fidelity is possible \cite{DBLP:conf/uss/TramerZJRR16}. Therefore, if buyers are allowed to use a model an unlimited number of times, the model is at risk of being stolen. Restricting access to data products can make it harder for malicious entries to steal private information.

In addition, fines can be imposed for misuse of data products by data buyers to combat malicious behavior. \citet{DBLP:conf/ec/NaghizadehS19} assume that adversarial buyers and their privacy attacks can ultimately be detected. Based on the contract theory, they propose a contract-theoretical framework for data sellers to address the pricing and privacy issues in query-based markets. A contract includes the privacy level measured by differential privacy, a price for a bundle of queries, and a fine for misuse of query answers. They formalize the optimal contract design problem that optimizes the data seller revenue while considering the presence of both honest and adversarial buyers. They prove that the required number of contracts is limited to the number of honest buyer types and provide an approximate algorithm for computing contracts.

\partitle{Limitations of Existing Solutions}
Central differential privacy assumes the data broker is trusted, which is by no means guaranteed. Federated learning largely limits the ability of a data broker to obtain private information from data owners, but still faces privacy leakage risks from sensitive local gradients if used without further integration of differential privacy techniques. The techniques without the assumption of a trusted data broker also have limitations in different aspects. TEE or MPC inevitably brings massive overhead of hardware resources. In addition, they only protect the input data during computation and do not guarantee the privacy of the computed results, hence need to be used in combination with output protection mechanisms such as differential privacy.  On the other hand, it is challenging for local differential privacy to preserve privacy while keeping the utility of data \cite{DBLP:conf/focs/DuchiJW13,DBLP:journals/jmlr/KairouzOV16,DBLP:conf/icml/KairouzBR16,DBLP:journals/tit/YeB18,DBLP:conf/sigmod/CormodeKS18,DBLP:journals/jmlr/BassilyNST20}. The zero trust security model only provides the function of matching transactions between data owners and data buyers, degenerating the data marketplace into a data display platform. Furthermore, existing works focus on the privacy of data owners and neglect the privacy of other entities such as data buyers and data brokers.

\subsection{Ensuring Fairness}\label{subsec:fairness}
In trusted data markets, an honest but curious data owner can inspect all information received in data transactions, but not tamper with their data during the data exchange. Conversely, in untrusted data markets, a malicious data owner can inspect all information received in data transactions and tamper with its data during the data exchange. This malicious act typically aims to increase the compensation and will undermine the fairness of the revenue allocation among data owners, which we call a revenue competition attack. We present several revenue competition attacks in Section \ref{subsec:revenue-comp-attack} and a robust valuation method to enhance fairness in Section \ref{subsec:robust-valuation}.

\subsubsection{Revenue Competition Attacks}\label{subsec:revenue-comp-attack}
One of the principal revenue competition attacks that researchers focus on is the data replication attack \cite{DBLP:conf/icml/EngstromISTSM20, DBLP:conf/nips/XuWFL21}. That is, a dishonest data seller attempts to exploit larger revenue from a single buyer by duplicating its data at zero cost. Replication manipulation is first introduced by \citet{DBLP:journals/corr/abs-2006-14583}, where a malicious data seller replicates its data and operates under multiple identities. \citet{DBLP:conf/nips/XuWFL21} extend the attack into a more common scenario where a data seller duplicates its private data before valuation. They formalize the concept of replication robustness and design a replication robust data valuation method based on the robust volume measure. However, the line between data duplication attacks and data augmentation is unclear and requires to be clarified by further research.

\begin{definition}(Data Replication Attack \cite{DBLP:journals/corr/abs-2006-14583})
In a cooperative game $G=(N, v)$, a (malicious) player (i.e., data owner) $i$ executes a replication action $k$ times on its private $D_i$ and acts as $k+1$ players $\mathcal{C}^R=\left\{i_0, i_1, \ldots, i_k\right\}$ each holding one replica of $D_i$. Denote the induced game as $G^R=\left(N^R, v^R\right)$, where the induced set of players are $N^R=N \backslash\{i\} \cup \mathcal{C}^R$, and the induced characteristic function $v^R$ satisfies $\forall \mathcal{C} \subseteq N \backslash\{i\}, \forall i_k \in$ $\mathcal{C}^R: v^R\left(i_k \cup \mathcal{C}\right)=v(i \cup \mathcal{C})$ and $v^R(\mathcal{C})=v(\mathcal{C})$. By replicating, player $i$ receives $a$ total compensation which is the sum of the compensation of all its $k+1$ replicas, i.e., $\varphi_i^{\mathrm{tot}}(k)=\sum_{\kappa=0}^k \varphi_{i_\kappa}\left(N^R, v^R\right)$.
\end{definition}

A poisoning attack happens when an adversary injects fake data to corrupt the agreed-upon state of a system, which decreases the total revenue of data owners by reducing the overall data quality. \citet{DBLP:conf/esorics/TolpeginTGL20} divide poisoning attacks into two types: data poisoning and model poisoning. 

In data poisoning, malicious data sellers manipulate their private data, e.g., by adding poison instances or adversarially changing existing instances. A label flipping attack is a typical data poisoning attack \cite{DBLP:conf/ccs/HitajAP17, DBLP:conf/uss/SuciuMKDD18}.

\begin{definition}(Label Flipping Attack \cite{DBLP:conf/ccs/HitajAP17, DBLP:conf/uss/SuciuMKDD18}) 
Label flipping attack is a type of targeted data poisoning attack where the attackers poison their training data instances by flipping real labels $y_{src}$ to fake labels $y_{dst}$, e.g., flipping ``spams'' into ``nonspams''. This attack has a significant negative impact on model predictions or statistical results on flipped labels.
\end{definition}

Model poisoning occurs in marketplaces that sell models and use federated learning for model training. In model poisoning, malicious data owners modify the local learning process to create adversarial parameter updates for aggregation. Specifically, an adversary (i.e., the compromised clients) sends the modified local models to the server for aggregation, so that the global model deviates the most from the pre-attack global model update direction. \citet{DBLP:conf/icml/BhagojiCMC19} and \citet{DBLP:conf/uss/FangCJG20} demonstrate that the model poisoning attack can substantially increase the error rates of the models learned by the Byzantine-robust federated learning methods.

\subsubsection{Robust Valuation}\label{subsec:robust-valuation}
Robust valuation refers to a fair data valuation method that can resist revenue competitive attacks from data owners. Research work \cite{han2022replication,DBLP:conf/nips/XuWFL21,DBLP:conf/aistats/WangJ23} investigates the robustness against the data replication attack. \citet{DBLP:conf/ec/AgarwalDS19} propose an exponential down-weighting payment division algorithm based on similarity to incentivize the collection of useful and unique data. They adjust the compensation calculated based on Shapley value for data sellers based on the similarity of their data to others in the dataset. The more similar the data of sellers is to others, the less they are compensated. The aim is to discourage sellers from replicating their data merely to increase their compensation, as the compensation does not significantly increase when similar data is detected. \citet{han2022replication} study the problem of replication robust payoff allocation in submodular cooperative games under the replication redundancy assumption based on semivalues including Shapley value, Banzhaf value, and leave-one-out. The submodularity property requires that marginal contributions decrease monotonically with coalition size, and the replication redundancy assumption requires that redundant resources do not contribute additional value to coalitions which already contain another replica or the original resource. They define replication robustness as the inability of players to gain more from replicating resources than the original. With this setting, they prove Banzhaf value and leave-one-out are replication robust, and define Robust Shapley value by adjusting the weighting scheme of the original Shapley value. \citet{DBLP:conf/aistats/WangJ23} consider a brand new scenario for data valuation, where the utility functions are perturbed by noise. They study the robustness of data valuation to noisy model performance scores and show that the Banzhaf value achieves the maximal robustness among all semivalues. Instead of relying on semivalues, \citet{DBLP:conf/nips/XuWFL21} propose a volume-based data valuation with a theoretical guarantee of the replication robustness following the intuition that copying the same data points does not increase the diversity of data. The data diversity is measured by the volume of a data matrix, which is defined as the determinant of its left Gram matrix. \citet{DBLP:conf/nips/IVSHAP} study robust valuation using the Shapley value in the presence of confounding data. They mitigate the influence of confounders by conducting allocations based on a confounder-free model that is trained with instrumental variables.

\subsection{Ensuring Profitability}\label{subsec:profitability} 
To price data products, some data sellers may survey bids from data buyers who are interested in purchasing the data product as a pricing guideline \cite{hartline2013mechanism}. However, this pricing mechanism opens up a new arbitrage opportunity for dishonest buyers. Dishonest buyers may attempt to affect the valuation and pricing of sellers through strategic actions. For example, these buyers can bid strategically to make the seller undervalue the product being sold and thus lose a significant share of earnings. This particular attack launched by strategic data buyers is similar to Sybil attack \cite{DBLP:conf/iptps/Douceur02}, first brought into the context of data markets by \citet{DBLP:conf/sigmod/Fernandez22}.

\begin{definition}(Sybil Attack \cite{DBLP:conf/iptps/Douceur02}) 
A Sybil attack is when a malicious user creates multiple pseudonymous identities and uses their combined effect to affect a system. In the data market, malicious buyers may send artificially low bids through multiple pseudonymous identities to steer prices away from the profitability of data sellers.
\end{definition}

While the Sybil attack can be easily mitigated by limiting the number of bids from a single buyer, there have been instances of malicious users creating or controlling multiple fake identities. To address the challenge, \citet{DBLP:conf/icbc2/AvyuktRK21} propose a novel decentralized crypto-economic system to ensure the credibility of reviews against the possibility of data owners launching Sybil attacks (taking on multiple fake identities) to rate their products when online marketplaces use ratings by buyers to help potential consumers identify good quality products. \citet{DBLP:conf/sigmod/Fernandez22} study the action of the strategic buyers in an online pricing market where prices of products are based on the bids of buyers. They propose a robust pricing algorithm to protect against the impact of artificially low bids by introducing a bid window and limiting the bid time.

\subsection{Ensuring Traceability}\label{subsec:traceability}
In untrusted data markets, malicious data brokers may trade personal data without the knowledge or consent of the data owner; malicious buyers may abuse or illegally distribute data products. Hence, traceability is maintained for claiming data ownership and verifying trade legality. We introduce technology for traceability in Section \ref{subsec:tools-and-tech-trace} and detail how to achieve data ownership verification and data transaction traceability in Section \ref{subsec:trace}. 

\subsubsection{Tools and Techniques}\label{subsec:tools-and-tech-trace}
The most prominent tools and technologies designed to enable traceability are watermarking and blockchain.

\partitle{Watermarking}
Watermarking is a way of tracking digital content like images, audio, or video by embedding a covert mark to it. A watermark can be a simple mark that is not visually perceptible \cite{DBLP:conf/ccs/Zheng24}, or it can carry a number of bits of information \cite{cox2002digital}. Watermarking provides ownership verification and copyright enforcement to prevent unauthorized duplication and distribution of data products in data markets.

A range of watermarking techniques are proposed to verify data ownership in various scenarios \cite{DBLP:conf/icde/Sion04,DBLP:conf/edbt/VlachosLRY08,DBLP:conf/icde/BertinoOYD05}. \citet{DBLP:journals/ijon/WanWZLYS22} present a comprehensive survey on robust image watermarking. \citet{DBLP:conf/icde/Sion04} develop a watermarking technique for categorical data that can cope with data loss, while \citet{DBLP:conf/edbt/VlachosLRY08} develop a watermarking technique for shapes that preserves geodesic distance relationships. To protect individual privacy and data ownership for outsourced medical data, \citet{DBLP:conf/icde/BertinoOYD05} perform a two-step transformation on medical data: binning and hierarchical watermarking. Specifically, the medical data is first binned to meet the k-anonymity specification, and then the binned data is watermarked to enable provable ownership. These approaches are promising for data ownership protection in dataset-based markets.

A few works \cite{DBLP:conf/uss/AdiBCPK18, DBLP:conf/nips/FanNC19,DBLP:conf/kdd/TangDLYH20, DBLP:conf/sp/ChenWPS0JM0S22} focus on watermarking deep learning models. Recent research extends to modifying data samples to mark the models trained with the dataset \cite{DBLP:conf/icml/SablayrollesDSJ20,DBLP:conf/www/Sun0SN022}. \citet{DBLP:conf/icml/SablayrollesDSJ20} makes imperceptible changes to image samples such that classifiers trained on those changed image data are tagged with an identifiable mark. \citet{DBLP:conf/www/Sun0SN022} tag a mark on an audio classification dataset by embedding a pattern in the magnitude of the time-frequency representation such that deep learning models trained with the watermarked dataset would be incorrectly classified as the watermarked class. These approaches are promising for data ownership verification in model-based markets.

\partitle{Blockchain}
Blockchain is a secure, shared, and distributed ledger maintained by all entities in a peer-to-peer network that provides distributed data storage. Its transactions are considered immutable because they cannot be retroactively modified by any entity without altering subsequent transactions once inserted. Due to its inherent immutability, it has been applied to provide data audit trails by recording data operations in the form of transaction records \cite{DBLP:journals/comsur/SalmanZEJS19,DBLP:conf/icbct/SharmaLR20}. 

Smart contracts \cite{szabo1994smart} are computing protocols stored on a blockchain. They automatically enforce contracts in the execution of transactions when predetermined conditions are met without the involvement of a third party. Actions in a smart contract can include sending notifications, issuing payments to involved entities, etc. After the transaction is completed, the blockchain is updated to keep the transaction intact. To realize automatic traceability, \citet{DBLP:conf/IEEEares/NeisseSF17} introduce smart contracts into the traceability system, which enables automatic tracking of data changes and allows managers to customize data access rules. 

\subsubsection{Achieving Traceability}\label{subsec:trace}
A series of research have been proposed to ensure the verification of data ownership and the traceability of data transactions.

\partitle{Data ownership verification}
Frameworks have been developed to manage data ownership \cite{DBLP:conf/icde/HurKSK17,DBLP:journals/pvldb/WangLLLTSZMYSCX22}. \citet{DBLP:conf/icde/HurKSK17} propose a secure data deduplication framework with dynamic ownership management in the cloud storage. To enable data privacy and data integrity against poison attacks, the proposed framework allows data owners to use randomized convergent encryption \cite{DBLP:conf/eurocrypt/BellareKR13}, which encrypts an input file with its plaintext and a randomized generated encryption key. The server can judge whether the files are identical given the ciphertext and perform deduplication. To enable backward and forward secrecy for dynamic ownership management, the proposed framework groups data owners who have identical files into the same ownership group. Then it introduces a re-encryption technique that re-encrypts the encrypted file with an updated ownership group key when an ownership change occurs and passes the key to valid data owners. To keep track of data ownership in the big data marketplace, \citet{DBLP:journals/jiat/SahooH21} design a system, where all actors register to generate login credentials for authentication purposes. A data owner embeds its signature into its own data and uploads the watermarked data using IPFS. They use an access control mechanism that enables trading data through a secure channel. In order to transfer data, they propose to modify the state variables corresponding to the ownership information in the smart contract and delegate the decryption right of the encrypted IPFS-hash through the proxy re-encryption technique. Further, \citet{DBLP:journals/jnca/GuptaDKJ22} extend to track the data ownership spanning across multiple marketplaces designed for trading Internet-of-Things sensors’ data. They add the information of buyers in the watermark and store the ownership records using Ethereum. To implement ownership verification and traceability in general-purpose data marketplaces, \citet{DBLP:conf/kdd/AlvarAY023} propose to store the data of data owners as non-fungible tokens with a watermark containing information about the owner and buyer. In addition, \citet{DBLP:journals/pvldb/WangLLLTSZMYSCX22} design an ownership-preserving database (OPDB) and build an encrypted database framework, Operon. OPDB follows the OPDB paradigm, uses the trusted execution environment to protect sensitive data, and implements a behavior control list to preserve data ownership.

\partitle{Data transaction traceability}
Watermarking, smart contracts, and blockchain and smart contracts are combined to achieve data transaction traceability \cite{DBLP:journals/pvldb/HynesDYCS18,DBLP:journals/isci/ZhaoYLHD19,DBLP:conf/infocom/HuangZ0Y20,DBLP:journals/tifs/DaiDCCZ020,DBLP:journals/jiat/SahooH21,DBLP:journals/tdsc/KoutsosPCTH22,DBLP:journals/tc/LiuHNLS22,DBLP:conf/gis/NguyenGNS19,DBLP:conf/globecom/AlsharifN20,DBLP:conf/coinco/ColmanCC19,DBLP:conf/infocom/JungLHQCHHS17,DBLP:conf/iccS/NasonovVB18,DBLP:conf/ictc/YooK20,DBLP:conf/blockchain2/BajoudahDM19,DBLP:conf/cvcbt/OzyilmazDY18,DBLP:conf/icbct/LawrenzSR19,DBLP:journals/iot/AbbasiPSC23,DBLP:journals/tmc/AnXLXZL23}. The practice is to watermark data products, use smart contracts to automatically perform transaction verification, and record transaction details in the blockchain.

\citet{DBLP:journals/pvldb/HynesDYCS18} propose a privacy-preserving data marketplace, Sterling. In Sterling, a data provider uploads encrypted data, publishes a smart contract that privately stores a data decryption key, and sets constraints such as payment and differential privacy. To purchase desired data, a data consumer creates a smart contract that satisfies the constraints of the selected providers and then gets a data decryption key if the contract passes automatic verification. The verification and operation of the contract are done in TEE to protect privacy.

\citet{DBLP:journals/tifs/DaiDCCZ020} present a similar secure data trading ecosystem, SDTE. In SDTE, a secure data trading platform (SDTP) based on Ethereum is implemented for contract deployment as a data broker between data sellers and data buyers. A seller deploys a data analysis contract on SDTP, finds sellers of interest, and selects trusted nodes of SDTP. Sellers then send their encrypted data to the trusted nodes. To resist data leakage from nodes, the encrypted data is decrypted first in an enclave created by Intel's Software Guard Extensions and then used to perform data analysis on the trusted nodes. The analysis result is encrypted and sent only to the buyer. To prevent payment fraud, a data trading manager contract (DTMC) is introduced to record dataset specifications and bid prices. Also, pre-depositing some Ether is required for buyers. To alleviate exposing the raw data of sellers, buyers are forced to pay huge fees for contracts that require complete or partial raw data. Therefore, SDTP defends against the possible attacks of malicious entities.

\citet{DBLP:conf/infocom/ChenWJXY22} introduce a non-repudiation scheme for IoT data trading. The proposed scheme consists of a trading scheme and an arbitration scheme. In the trading scheme, a two-phase data exchange is adopted instead of only encryption. The data owner first sends unusable partial data to the data buyer, then sends the remaining part of the data after the buyer successfully verifies the partial data, and finally receives money pre-deposited in the contract by the buyer. If a dispute occurs, the arbitration scheme first attempts to resolve it on-chain. The contract uses stored proofs to verify the data transaction. If the on-chain arbitration is unsuccessful, an off-chain arbitration will be activated.

\citet{DBLP:conf/ccs/LiLLX23} propose a verifiable transaction protocol to ensure the data acquirer has faithfully aggregated local models according to pre-agreed weights in federated learning-based data marketplaces and guarantees fair compensation to data providers based on their contributions. The data acquirer commits to selected data providers and the aggregation weights on a blockchain. After receiving the local models, the data acquirer computes the global model and generates a zero-knowledge proof of model aggregation. This proof, along with the aggregation weights and the new global model, is then published for verification. A smart contract on a blockchain manages the reward distribution based on the aggregation weights.

In traditional blockchain systems, each node must process every data transaction, yielding low efficiency. To improve efficiency, \citet{DBLP:journals/jsac/WangCYLWYW22} adopt a layered sharding blockchain, where the nodes in the blockchain are divided into multiple shards, with shards processing transactions in parallel. They generate the sharding structure based on data transaction history. Nodes with the highest number of transactions are selected as core nodes in shards. Other nodes are assigned to these shards if their transaction frequencies with the core node are greater than a threshold.

\nop{
There are also some studies on specific data types. For geosocial data, \citet{DBLP:conf/gis/NguyenGNS19} adopt searchable encryption, digital commitments, and blockchain to protect the location privacy of data owners while incorporating accountability and spam-resilience mechanisms. For medical data, \citet{DBLP:conf/globecom/AlsharifN20} propose a medical data marketplace where medical data can be traded without a central trusted party. In their model, sellers can enforce an access control policy on the encrypted records and buyers can verify the correctness of the encrypted records using their proposed zk-SNARK protocol. Additionally, a zero-knowledge contingent payment protocol is developed to enable a trustless fair exchange of the access key and the record access fee. For wearable data, \citet{DBLP:conf/coinco/ColmanCC19} propose a trusted marketplace based on blockchain, where a contract manager records the contract received from the data broker into smart contracts on the blockchain.
}

While the above efforts attempt to achieve data traceability by using blockchain to immutably record the data transactions that occur on their designed platforms, they do little to prevent unauthorized reselling activities, which is always one of the active requests of the original data owners. To address the limitation, \citet{DBLP:conf/infocom/JungLHQCHHS17} propose AccountTrade, a set of accountable protocols, for trading big data. In AccountTrade, a data seller uploads its dataset to a data broker for sale and the data broker further performs copy detection on the uploaded dataset. They propose a uniqueness index defined by Jaccard Index as a rigorous similarity comparison mechanism to check whether similar datasets have been uploaded before. If there were similar datasets before, then the seller is considered to be dishonestly reselling the dataset. AccountTrade prevents dishonest sellers from reselling data but requires a centralized data broker. Thus, \citet{DBLP:conf/infocom/HuangZ0Y20} propose a distributed protocol based on a smart contract to detect unauthorized data trading where no centralized data broker is needed. In the authorized trading process, three parties, including the data producer, the data reseller, and the data consumer, use their respective public keys to sign the smart contract that contains the data item information, selling price, and revenue sharing ratio between the reseller and the producer. Rogue resellers who resell data without authorization will fail the verification process of the consumer. Once the real producer receives the verification information broadcast by the consumer, she will prove herself and leave the resellers out. However, unauthorized resale activities on platforms that do not deploy these mechanisms cannot be detected. Differently, \citet{DBLP:journals/tmc/HuangZYY23} allow data resale and use smart contracts to ensure data authenticity and legality in data resale. When consumers receive data from a reseller, they verify its authenticity by generating a verification packet from the data via hashing and broadcasting the packet to the network. Data producers then check and confirm the legality of the transaction. If the resale is unauthorized, the producer can generate a new smart contract to trade directly with the consumer.

In addition to transaction details, some systems \cite{DBLP:journals/iot/AbbasiPSC23,DBLP:journals/tmc/AnXLXZL23} allow data buyers to rate the data products they purchased and record it on the blockchain to assess the reliability of data sellers.

\subsection{Discussion on Actual Data Marketplaces}
Actual data marketplaces carefully design secure data exchange technologies. Take Dawex as an example. Dawex experts conduct rigorous security audits regularly. Dawex employs encryption protocols to store and transmit data, as well as secure key management algorithms and protocols. During data transmission and reception, digital signature mechanisms are implemented to verify authenticity and maintain integrity. Furthermore, Dawex enhances security with robust authentication and access control measures, including multi-factor authentication and authorization protocols. Dawex supports data diagnostics for data buyers to evaluate the quality of data products before achieving a transaction.

\section{Government Policies and Industry Status}\label{sec:status}

In this section, we introduce guidelines, policies, laws, regulations, and industry developments of data markets. We present guidelines in Section \ref{subsec:guides-policies}, regulations in Section \ref{subsec:laws-regulations}, and the industry status in Section \ref{subsec:industry-status}.

\subsection{Guidelines}\label{subsec:guides-policies}
In this section, we present guidelines and policies related to data markets in representative countries and regions in alphabetical order.

\partitle{China}
In April 2020, the Communist Party of China Central Committee and the State Council unveiled \emph{a guideline on improving the market-based allocation of production factors}~\cite{productionfactors}. For the first time, data is recognized as the fifth production factor along with the existing four production factors including land, labor, capital, and technology. According to the guideline, China will make efforts to nurture new industries, businesses, and models of the digital economy and support the utilization of data in fields including agriculture, industry, transport, education, and urban management. Later, \emph{the fourteenth five-year plan} for the national economic and social development of China and the outline of long-term goals for 2035 was released in March 2021. The plan makes important arrangements for the development of big data, taking big data as an important raw material of the digital economy \cite{Bigdatawhitepaper2021}.

\partitle{European Union}
The European Union (EU) plans to establish a unified data market so that unused data can be circulated within the EU. The EU regulation on the free flow of non-personal data has been applied since May 2019, which aims at removing barriers to the free exchange of non-personal data in Europe \cite{EUregulationfreeflow}. In February 2020, the European Commission released \emph{the European Digital Strategy}, a plan for digital transformation \cite{european-data-strategy}. The goal is to make enterprises and people benefit in various aspects through the proper use of digital technologies and promote the EU to take a leadership role in establishing a data-enabling society. 

\partitle{International Organizations}
\emph{ISO/IEC 27701:2019} \cite{isoiec27701} is an international standard presented by the International Organization for Standardization (ISO) and the International Electrotechnical Commission (IEC) in August 2019. It is an extension of ISO/IEC 27001 and ISO/IEC 27002. ISO/IEC 27701:2019 specifies requirements and guidelines on the protection of personal data to establish, implement, maintain, and continually improve a Privacy Information Management System (PIMS) for the protection of personal data.

Additionally, \emph{Payment Card Industry Data Security Standard (PCI DSS)} \cite{pcidss} is presented by the founding members of the Payment Card Industry Security Standards Committee (Visa, MasterCard, American Express, Discover Financial Services, JCB, etc.) in order to protect sensitive cardholder data and prevent fraud. PCI DSS puts forward 12 security requirements from the aspects of information security management systems, network security, physical security, data encryption, etc.

\partitle{Japan}
In May 2010, the Headquarters of Japan's Advanced Information and Communication Network Society Promotion Strategy presented \emph{the New Strategy for Information and Communication Technology} \cite{xmetaera-japan}. In July 2012, the Ministry of General Affairs issued a new comprehensive strategy of \emph{``Active ICT Japan''} \cite{Active_ICT_Japan}, which focuses on the development of big data strategy through technological innovation so as to achieve nationally oriented e-government and strengthen mutual assistance between regions.

\partitle{South Korea}
In 2012, the National Science and Technology Commission of South Korea formulated a strategic plan for the future development environment of big data. In the same year, the Ministry of Future Creation Science adopted \emph{the Fifth National Informatization Basic Plan (2013-2017)}.
On August 17, 2023, the Personal Information Protection Commission (PIPC) announced its National My Data Innovation Promotion Strategy.

\partitle{The United Kingdom}
The Department of Digital, Culture, Media and Sports (DCMS) of the United Kingdom released \emph{the National Data Strategy} \cite{National-data-strategy} in September 2020 to support the use of data and help the economy recover from the epidemic. The strategy studies how the UK can use its existing advantages to promote the use of data by enterprises, government, and civil society, and establishes a framework for processing and investing data to promote economic development.

\partitle{The United States}
In May 2014, the Executive Office of the President of the United States issued a report about big data, \emph{``Big Data: Seize Opportunities, Preserving Values''} \cite{whitehousebigdata}. The report presents approaches to protecting open data and privacy, discusses public and private management of data, and gives some recommendations on preserving data values.

In December 2019, the Office of Administration and Budget (OMB) of the White House issued \emph{the Federal Data Strategy and 2020 Action Plan} \cite{federal-data-strategy}, which describes the data vision of the federal government in the next decade and the key actions to be implemented in 2020 from the perspective of government data governance. The core goal is to ``Leveraging Data as a Strategic Asset''.

\partitle{Summary}
Big data has attracted extensive attention from major countries and regions around the world where various big data development strategies have been launched at the national level. These guidelines aim to facilitate the exchange, utilization, and monetization of data, promoting economic growth through the effective management of data.

\subsection{Regulations}\label{subsec:laws-regulations}
Data transactions bring new opportunities for public and private organizations but also come with responsibilities and restrictions. Although specific regulations on data markets are limited, a series of data regulations have been promulgated to enforce organizations to return data ownership to individuals and protect their privacy. In this section, we introduce laws and regulations on privacy protection and data sharing in representative countries and regions.

\subsubsection{Lists} 
We list laws and regulations in representative countries and regions as follows.

\partitle{Brazil}
Inspired by the EU's General Data Protection Regulation (GDPR), Brazil's General Data Protection Law (in Brazilian, \emph{Lei Geral de Proteção de Dados Pessoais, LGPD} \cite{LGPD}) was passed in August 2018 and came into force in August 2020. 
LGPD applies to any enterprise and organization that needs to process the personal data of Brazilian people, no matter where they are located.

\partitle{Canada}
\emph{The Personal Information Protection and Electronic Documents Act (PIPEDA)} \cite{PIPEDA} is a Canadian federal law passed in 2000. PIPEDA applies to the collection, use, and disclosure of personal information by governments and private organizations. Organizations must obtain permission from individuals before collecting or using their personal information. If they do not comply with PIPEDA, they could be fined up to \$100, 000.

\partitle{China}
\emph{The Data Security Law (DSL)} and \emph{The Personal Information Protection Law (PIPL)} \cite{PIPL} was promulgated in 2021. 
The PIPL is the first comprehensive, principle-based personal information protection law in China. It is also inspired by GDPR and other international data protection laws to a certain extent \cite{fieldfisher-pipl}.

\partitle{European Union}
\emph{The General Data Protection Regulation (GDPR)} \cite{GDPR} is a legal framework to protect personal information and harmonize data privacy laws across Europe. GDPR was passed in April 2016 and came into force in May 2018.
\nop{
\emph{The EU Whistleblower Protection Directive} was adopted in October 2019 and entered into force in December 2019. It aims to provide a comprehensive EU approach to whistleblower protection. By December 2021, all Member States must meet the minimum standards stipulated in the EU Whistleblower Protection Directive \cite{whistleblower}, which will affect tens of thousands of organizations.  The relevant elements of the EU Whistleblower Protection Directive are set out and the clarity sought by organizations in this regard is provided  \cite{stappers2021eu}.
}
\emph{The Data Governance Act (DGA)} \cite{DGA} entered into force in June 2022 and became applicable since September 2023. It aims to facilitate fair and trustworthy data sharing. \emph{The Data Act (DA)} \cite{DA} is a regulation that seeks to enhance and encourage the exchange and utilization of data within the European Economic Area. It was officially published in the Official Journal of the European Union in July 2023.

\nop{
\partitle{German}
\emph{The German Supply Chain Due Diligence Law} was promulgated by the German Federal Parliament in June 2021 and entered into force in January 2023 \cite{lexology-lksg}. The bill aims to protect human rights and mitigate environmental risks through supply chain risk management. It protects the rights of people who produce goods for the German market and enables companies in Germany to better fulfill their global responsibilities. 
}

\partitle{Japan}
In May 2003, Japan enacted \emph{the Act on the Protection of Personal Information (APPI)} \cite{APPI}. Comprehensive amendments to APPI were approved on September 3, 2015, and came into full effect on May 30, 2017. On June 5, 2020, Japan's Diet passed a bill to revise APPI known as the ``2020 Amendments''. The bill was promulgated on June 12, 2020, and came into force on April 1, 2022 \cite{japan-data-protection}. The revised version of APPI is close to the EU data protection law.

\partitle{South Africa}
In November 2013, the parliament of the Republic of South Africa promulgated \emph{the Protection of Personal Information Act (POPIA)} \cite{POPIA}, which is a comprehensive personal data and privacy protection law, and came into force in July 2020. Similar to GDPR, POPIA protects the privacy of individuals to determine how private and public organizations use personal data.

\partitle{Thailand}
In Thailand, \emph{the Personal Data Protection Act (PDPA)} \cite{PDPA} is the first law to manage and protect data, which came into effect in June 2022. PDPA puts many small and medium-sized enterprises under pressure for data compliance. For example, illegal collection and disclosure of personal sensitive data can result in fine of $5$ million baht. 

\partitle{The United States}
\emph{The Health Insurance Portability and Accounting Act (HIPAA)} \cite{HIPAA} provides data confidentiality and security provisions to protect patient information, which plays a normative role in a variety of medical and health industries~\cite{fu2022hint}. The act also promotes the adoption of electronic health records to improve the efficiency and quality of the United States health care system through information sharing.

Recently, a growing number of US states are considering and passing data privacy legislation. For example, \emph{The California Consumer Privacy Act (CCPA)} \cite{CCPA} is the first comprehensive privacy law in the United States, which is probably the most influential state-level data privacy law in the United States. It was officially promulgated in June 2018, revised several times in the following two years, and finally implemented in July 2020. CCPA enables consumers to have great control over personal information. \citet{BAIK2020101431} takes CCPA as an example and analyzes the different privacy frameworks advocated by different stakeholders. Although the act only protects the personal information of consumers in California on the surface, enterprises outside California will also be affected due to the size, population, and economic importance of California. Therefore, CCPA has a significant impact on the collection of personal information in data-intensive industries \cite{scope-and-impact}. However, some organizations still do not have CCPA-related content on their platform websites. \citet{https://doi.org/10.48550/arxiv.2205.09897} provides an empirical evaluation of the implementation of CCPA to promote the implementation of CCPA in the organization.
In March 2021, Virginia passed a new privacy law, \emph{the Consumer Data Protection Act (VCDPA)} \cite{VCDPA}, which entered into force in January 2023.
In July 2021, Colorado passed a comprehensive privacy act, \emph{the Colorado Privacy Act (CPA)} \cite{CPA}, which entered into force in July 2023. 

\subsubsection{Discussion}
We outline definitions and measures in various regulations that govern data privacy and facilitate data sharing, highlighting their common threads and unique aspects. We first introduce how regulations determine the definition and scope of personal data, distinguish between general and sensitive personal data, and summarize key principles of personal data processing along with the rights of data subjects. Next, we discuss how regulations define the sharing and sale of personal data. The responsibilities and constraints of data intermediation services, such as data marketplaces, in facilitating data sharing and sales are explored. Given that fair compensation is a critical issue, we present mechanisms specified in regulations to ensure equitable compensation for data subjects and businesses. Finally, we list techniques for privacy preservation mentioned in the regulations.

\partitle{Personal data}
At the heart of regulations, personal data is broadly recognized as information relating to an identified or identifiable individual. For instance, GDPR defines personal data as ``any information relating to an identified or identifiable natural person'', and CCPA describes personal information as ``information that identifies, relates to, describes, is reasonably capable of being associated with, or could reasonably be linked, directly or indirectly, with a particular consumer or household''. While the core principles are similar, there are nuances in the scope of personal data. CCPA expands the scope of personal information beyond an individual to a group of people who ``cohabitate with one another at the same residential address and share use of common devices or services''. The definition of personal data in POPIA includes not only ``an identifiable, living, natural person'' but also ``an identifiable, existing juristic person''. However, POPIA does not apply to the processing of personal data in the context of ``purely personal or household activities''. 

Furthermore, several categories of data are frequently excluded from the definition of personal data and are not subject to protection. 

\begin{enumerate}
    \item \emph{Anonymous data} refers to data that cannot be associated with or identifiable to an individual. Regulations, including GDPR, LGPD, and PIPL, exclude anonymous data from personal data. 

    \item \emph{De-identified data} means ``data that cannot reasonably be used to infer information about, or otherwise be linked to, an identified or identifiable individual, or a device linked to such an individual'' (CCPA). Compared with anonymous data, de-identified data still carries the risk of re-identification. Regulations including CCPA, POPIA, CPA, and VCDPA, exclude de-identified data from personal data. 

    \item \emph{Publicly available information} is ``information that is lawfully made available from federal, state, or local government records, or information that a business has a reasonable basis to believe is lawfully made available to the general public by the consumer or from widely distributed media; or information made available by a person to whom the consumer has disclosed the information if the consumer has not restricted the information to a specific audience'' (CCPA). Regulations, including PIPEDA, CCPA, POPIA, CPA, and VCDPA, exclude publicly available information from personal data. 

    \item \emph{Aggregate information} refers to information combined from multiple individuals, from which individual identities have been removed and cannot be linked to any individual. CCPA excludes aggregate information from personal data.
\end{enumerate}

\partitle{Sensitive personal data}
Sensitive personal data is a subcategory of personal data that encompasses specific data types with enhanced protection. Generally, sensitive personal data includes racial or ethnic origin, political opinions, religious or philosophical beliefs, trade union membership, genetic data, biometric data for uniquely identifying an individual, health-related data, sex life or sexual orientation, and children data. In addition, CCPA considers social security numbers, driver’s license numbers, state identification card numbers, passport numbers, account log-in information, financial account, debit and credit card numbers, precise geolocation, and contents of mail, email, and text messages as sensitive personal information. In POPIA, sensitive personal information also includes criminal behavior. 

\partitle{Principles of personal data processing}
Key principles of personal data processing as outlined in privacy regulations encompass: (1) lawfulness, fairness, and transparency, (2) purpose limitation, (3) data minimization, (4) accuracy, (5) storage limitation, (6) integrity and confidentiality, (7) accountability, and (8) security.

\partitle{Rights of data subjects}
Fundamental rights granted to data subjects under various privacy regulations include the right to access, the right to rectification, the right to deletion (right to be forgotten), the right to restrict processing, the right to data portability, and the right to object (GDPR). More relevant to data markets, CCPA and VCDPA provide data subjects with ``the right to know what personal information is sold or shared and to whom'', as well as ``the right to opt out of sale or sharing of personal data''. Regulations such as CCPA and VCDPA state that businesses must not discriminate against data subjects for exercising their rights. Discriminatory practices, for example, can be charged different prices or rates and provide different levels or qualities of goods or services.

\partitle{Sharing and sale of personal data}
DGA defines data sharing as ``the provision of data by a data subject or a data holder to a data user for the purpose of the joint or individual use of such data, based on voluntary agreements or Union or national law, directly or through an intermediary, for example under open or commercial licenses subject to a fee or free of charge''. The sale of personal data is defined as ``the exchange of personal data for monetary or other valuable consideration'' by CCPA, CPA, and VCDPA. By allowing data subjects to seamlessly transmit their personal data from one business to another in ``a structured, commonly used, and machine-readable format'', the right to data portability makes it feasible and convenient for data subjects to sell their personal data in data markets.

\partitle{Data intermediation services}
Under DGA, a data intermediation service is defined as ``a service which aims to establish commercial relationships for the purposes of data sharing between an undetermined number of data subjects and data holders on the one hand and data users on the other, through technical, legal or other means''. Data intermediation services are expected to support, promote, and facilitate data sharing. Data marketplaces that ``make data available to others'' are identified as an example of data intermediation services.

DGA explicitly outlines services that are not considered data intermediation services. These include (1) services that obtain, aggregate, enrich, or transform data from data holders but do not establish a commercial relationship between data holders and data users; (2) services that focus on copyright-protected content; (3) services that are used exclusively by one data holder or multiple legal persons in a closed group; and (4) public sector data sharing services that are not intended to establish a commercial relationship.

DGA imposes a set of rules on data intermediation service providers to ensure that they operate in a trustworthy, fair, and neutral manner. (1) Commercial relationship establishment: Providers should aim to establish a commercial relationship between data holders and data users; providers can charge for their services and can use data from data holders to improve their services, but cannot directly use data that they broker for financial profit. (2) Support for the rights of data subjects: Providers should support data subjects in exercising their rights to personal data and act in the best interest of the data subjects. (3) Neutrality, fairness, transparency, and non-discriminatory: Providers should keep neutral in the data exchange between data holders or data subjects and data users; providers should provide fair, transparent, and non-discriminatory services. (4) Penalties: Providers should have procedures to impose penalties for fraudulent or abusive practices. (5) Privacy preservation: Providers should ``take necessary measures to ensure an appropriate level of security for the storage, processing and transmission of non-personal data'' and ``further ensure the highest level of security for the storage and transmission of competitively sensitive information''. (6) History: Providers should keep a log record of data intermediation activities. (7) Compliance: Providers should comply with competition law.

\partitle{Fair compensation for data subjects}
CCPA and VCDPA permit businesses to provide financial incentives for collecting, selling, or retaining personal data. Businesses may also offer different prices, rates, levels, or qualities of goods or services, provided that these variations are reasonably related to the value of personal data to the business. Businesses are required to notify data subjects about financial incentives and can only engage them in such programs with their clear consent, which can be revoked at any time. Importantly, any financial incentive practice that is ``unjust, unreasonable, coercive, or usurious in nature'' is prohibited. Ensuring fairness in compensation allocation among data subjects is a key focus of academic research.

However, under the Data Act, gatekeepers are prohibited from soliciting or commercially incentivizing a data subject in any manner to make data available to one of its services, including by providing monetary or any other compensation. Also, gatekeepers are not allowed to receive data that a data subject has requested from another business. This prohibition aims to prevent gatekeepers from their market dominance to collect and control large amounts of personal data, leading to unfair competition.

\partitle{Compensation for businesses}
Regulations, including GDPR, APPI, and the Data Act, recognize that businesses incur costs when supporting the rights of data subjects. To ensure fairness, these regulations allow for the recovery of costs associated with data requests and management. They differ in the specific provisions on compensation for businesses. APPI allows charging data subjects not only for the disclosure of data but also for the notification about the purpose of data use. The charges shall be reasonable to reflect ``the actual costs''. GDPR specifies that data subjects have the right to obtain ``a copy of the personal data undergoing processing'', but businesses may charge ``a reasonable fee based on administrative costs'' for further copies. The Data Act establishes a comprehensive framework for non-discriminatory and reasonable compensation. When determining compensation, factors considered can include (1) ``costs incurred in making the data available'' (e.g., formatting, electronic dissemination, or storage), (2) ``investments in the collection and production of data'', and (3) ``the volume, format and nature of the data''. In addition, businesses are required to provide data at necessary costs incurred or free of charge in special situations (e.g., to support not-for-profit organizations, or public emergencies).

\partitle{Techniques for privacy preservation}
Many techniques for privacy preservation have been proposed, including anonymization, pseudonymization, differential privacy, generalization, suppression, randomization, the use of synthetic data, and smart contracts. These techniques are not only utilized by academia to construct privacy-preserving data markets, but are also recommended in regulations such as GDPR and DGA.


\subsection{Industry Status}\label{subsec:industry-status}
In this section, we introduce two perspectives on the current status of the industry: data marketplaces in Section \ref{subsubsec:practical-data-market} and data black market in Section \ref{subsubsec:data-black-market}.

\subsubsection{Data Marketplaces}\label{subsubsec:practical-data-market}
Data marketplaces are spreading across countries and regions, processing and trading data using their own developed technologies while complying with regulations. The European Commission began its research on the European data market in 2013, and for the first time tried to provide facts and data on the scale and trend of the European data economy by developing a monitoring tool for the European data market. Its results for the period 2017-2020 are reported in \cite{european-data-market-study}. They continue to study the new European data market for the period 2021-2023 and report results in \cite{new-european-data-market-study}. In China, 25 data marketplaces have been established by local government departments and private enterprises. 
As of January 15, 2024, Global Big Data Exchange (Guiyang) has accumulated 777 data vendors, 79 data intermediaries, and 1523 listed products. The total transaction volume of digital data has reached 2.514 billion yuan \cite{news-guiyang}.
The opportunities and forecasts of the big data market in Japan are studied in \cite{japan-big-data-market}. The rise of social media and the large growth of information volume have promoted the big data market in Japan while privacy issues and low adoption rates in some industries are some of the biggest obstacles to its development. Table \ref{tab:data-marketplaces} shows actual data marketplaces established so far.


\begin{table}[htbp]
  \centering
  \caption{Data marketplaces.}\label{tab:data-marketplaces}
    \adjustbox{width=\linewidth}{
    \begin{tabular}{|l|l|l|l|l|l|l|p{2cm}|p{2cm}|p{2cm}|p{2cm}|p{2cm}|p{2cm}|}
    \hline
    \multicolumn{1}{|c|}{\multirow{2}[4]{*}{\textbf{Market}}} & \multicolumn{1}{c|}{\multirow{2}[4]{*}{\textbf{Country}}} & \multicolumn{1}{c|}{\multirow{2}[4]{*}{\textbf{Date Est.}}} & \multicolumn{1}{c}{\textbf{User}} & \multicolumn{1}{c}{} &       & \multicolumn{4}{c|}{\textbf{Data Product}} & \multicolumn{2}{c|}{\textbf{Transaction}} & \multicolumn{1}{c|}{\multirow{2}[4]{*}{\textbf{Tech Stack}}} \\
\cline{4-12}          &       &       & \multicolumn{1}{c|}{\textbf{Buyer}} & \multicolumn{1}{c|}{\textbf{Seller}} & \multicolumn{1}{c|}{\textbf{Broker}} & \multicolumn{1}{c|}{\textbf{Type}} & \multicolumn{1}{c|}{\textbf{Time Frame}} & \multicolumn{1}{c|}{\textbf{Origin}} & \multicolumn{1}{c|}{\textbf{Storage}} & \multicolumn{1}{c|}{\textbf{Pricing Model}} & \multicolumn{1}{c|}{\textbf{Exchange Method}} &  \\
    \hline
    \href{https://market.aliyun.com/}{Ali Data Market} & China & Sep-09 & \checkmark &       & \checkmark & any data, query, model &       &       &       & subscription &       &  \\
    \hline
    \href{http://apistore.baidu.com/}{Baidu Data} & China & Unknown & \checkmark & \checkmark & \checkmark & any data, query, model &       &       &       & subscription &       &  \\
    \hline
    \href{}{Beibu Gulf Big Data Trading Center} & China & Aug-20 & \checkmark &       &       & any data &       &       &       &       & API   &  \\
    \hline
    \href{https://www.bjidex.com}{Beijing International Big Data Exchange} & China & Mar-21 & \checkmark &       &       & any data & static & organizations &       &       &       &  \\
    \hline
    \href{}{Central China Data Exchange} & China & Jul-16 & \checkmark & \checkmark &       &       &       &       &       &       &       &  \\
    \hline
    \href{http://www.datatang.com/}{Data Tang} & China & Jun-11 & \checkmark &       & \checkmark & any data, model & static &       &       & negotiation &       & artificial intelligence \\
    \hline
    \href{http://www.chinadatatrading.com/}{East Lake Trading Center for Big Data} & China & Jul-15 & \checkmark &       &       & any data, query & static &       &       & negotiation & API   & artificial intelligence \\
    \hline
    \href{https://www.gzdex.com.cn/}{Global Big Data Exchange (Guiyang)} & China & Dec-14 & \checkmark &       &       & any data, model & static &       &       & subscription, negotiation, free &       &  \\
    \hline
    \href{https://wx.jcloud.com/}{JD Wan Xiang} & China & Unknown & \checkmark & \checkmark &       & any data, query, model & static &       &       & subscription & API   &  \\
    \hline
    \href{http://www.bigdatahd.com/}{Jiangsu Big Data Exchange} & China & Nov-15 & \checkmark &       &       & any data, query &       &       &       &       &       &  \\
    \hline
    \href{http://www.markwaymall.com/}{Markway Mall} & China & 2001  & \checkmark &       &       & any data, model & static &       &       & free, negotiation & download & artificial intelligence \\
    \hline
    \href{https://www.datadmz.com/}{Northern Big Data Trading Center} & China & Nov-21 &       &       &       &       &       &       &       &       &       &  \\
    \hline
    \href{http://www.qddata.com.cn/}{Qingdao Big Data Exchange} & China & Apr-17 & \checkmark &       & \checkmark & any data, query & static &       &       & subscription, negotiation &       &  \\
    \hline
    \href{https://www.sddep.com/}{Shandong Data Trading Company} & China & Dec-19 & \checkmark &       &       & any data, query & static &       &       & free, negotiation & API   &  \\
    \hline
    \href{}{Shanxi Data Transaction Service Platform} & China & Jul-20 & \checkmark &       &
    \checkmark   & any data & static &   &       &  & API  & artificial intelligence \\
    \hline
    \href{https://www.chinadep.com/}{Shanghai Data Exchange Corp} & China & Apr-15 & \checkmark &       &       & any data & static &       &       &       &       &  \\
    \hline
    \href{http://www.szdex.com}{Shenzhen Data Exchange} & China & Nov-22 & \checkmark &       &       & any data, query, model & static & organizations &       & negotiation &       &  \\
    \hline
    \href{http://www.zjdex.com/}{Zhejiang Big Data Exchange} & China & Mar-16 &       &       &       &       &       &       &       &       &       &  \\
    \hline
    \href{https://www.dawex.com/en}{Dawex} & France & 2015  & \checkmark & \checkmark & \checkmark & any data & static &       & cloud storage &       & API   & two-factor authentication, asymmetric encryption protocol \\
    \hline
    \href{https://blogs.sap.com/2021/12/13/sap-data-warehouse-cloud-data-marketplace-an-overview/}{SAP} & Germany &       & \checkmark & \checkmark &       & any data &       &       & cloud storage &       &       & SAP Data Warehouse Cloud \\
    \hline
    \href{https://www.statista.com/}{Statista} & Germany & 2007  & \checkmark &       &       & any data &       &       &       & subscription &       & cloud storage, ETL, AutoML \\
    \hline
    \href{}{Data plaza} & Japan & Apr-23 &       &       &       &       &       &       &       &       &       &  \\
    \hline
    \href{}{EverySense Japan} & Japan &       &       &       &       &       &       &       &       &       &       &  \\
    \hline
    \href{}{SCRY.INFO} & Japan & Oct-17 &       &       &       &       &       &       &       &       &       &  \\
    \hline
    \href{https://www.qlik.com/}{Qilk} & Sweden & 2020  &       &       &       &       &       &       &       &       &       &  \\
    \hline
    \href{https://streamr.network/}{Streamer} & Switzerland & 2017  & \checkmark & \checkmark &       & any data & real-time &       & decentralized storage & subscription, free & API   & Ethereum, xDai, Data Union \\
    \hline
    \href{https://aws.amazon.com/data-exchange/}{AWS Data Exchange} & the US & 2019  & \checkmark & \checkmark &       & any data &       &       &       & subscription, negotiation &       &  \\
    \hline
    \href{https://www.bdex.com/}{BDEX} & the US & 2014  & \checkmark &       & \checkmark &       &       &       &       & subscription &       &  \\
    \hline
    \href{https://www.bloomberg.com/professional/product/market-data/}{Bloomberg} & the US & 1981  & \checkmark & \checkmark & \checkmark & financial data &       &       &       &       &       &  \\
    \hline
    \href{https://www.factual.com/}{Factual} & the US & 2007  &       &       &       &       &       &       &       &       &       &  \\
    \hline
    \href{https://www.informatica.com/}{Informatica} & the US &       & \checkmark & \checkmark &       & any data &       &       & cloud storage &       &       & cloud storage \\
    \hline
    \href{https://www.narrative.io/}{Narrative} & the US & 2016  & \checkmark & \checkmark &       & any data &       &       &       & pay-as-you-go, committed-use discounts &       &  \\
    \hline
    \href{https://data.nasdaq.com/}{Nasdaq Data Link} & the US & 1971  & \checkmark & \checkmark &       & financial data & real-time, delayed, time-series & central banks, governments, organizations &       & subscription, free & API   & authentication \\
    \hline
    \href{https://www.oracle.com/}{Oracle} & the US & 1997  & \checkmark &       &       & any data & static &       &       &       &       &  \\
    \hline
    \href{https://www.safegraph.com/}{SafeGraph} & the US & 2016  & \checkmark &       &       & geospatial data &       &       &       &       & bulk download, Snowflake, request a sample &  \\
    \hline
    \href{https://www.snowflake.com/data-marketplace/}{Snowflake} & the US & 2012  & \checkmark &       &       & any data & static &       & cloud storage &       &       & cloud storage \\
    \hline
    \href{https://www.xignite.com/}{Xignite} & the US & 2000  & \checkmark &       & \checkmark & financial data & real-time, delayed, historical, reference & >250 data vendors & cloud storage & negotiation & API   & Amazon Web Services \\
    \hline
    \end{tabular}
    }
\end{table}%

\partitle{Geospatial data}
SafeGraph \cite{safegraph} was founded in 2016. SafeGraph aims to help answer various questions by providing access to large data sets that analyze the past to predict the future. Its first product is a geospatial data platform, which aims to obtain accurate records for urban planners, retailers, academic researchers, marketers, and investors, and use them as key information to predict future related decision-making. In April 2017, SafeGraph announced that it had obtained A round of financing of \$35M. Auren Hoffman, CEO of SafeGraph, said that the financing would be used to ``build and maintain data sets to help accelerate machine learning and artificial intelligence''. 
Spectus \cite{spectus} is a Platform-as-a-Service solution that provides a secure and privacy-focused environment for building custom location data-based solutions. Spectus uses differential privacy technologies to fully anonymize sensitive data and offers tools, data, and infrastructure to expedite the development of human mobility analytics and solutions. With a focus on patent-pending privacy solutions, curated data, and fast insights, Spectus enables organizations to build and commercialize solutions while maintaining control of their data. Spectus also provides a platform that enables data owners of core mobility location data and adjacent assets to securely monetize their properties.

\partitle{Financial Data}
Bloomberg \cite{bloombergchina} was founded in 1981. Its products include Bloomberg terminal and trade and order management, which provide data, news, and analysis tools for financial and business professionals. Bloomberg helps customers acquire, integrate, distribute, and manage cross-agency data and information and provides solutions for enterprises with the help of technology \cite{bloombergchina}. In 2012, Bloomberg group's global operating revenue reached \$7.6B, making Bloomberg the world's largest financial information service provider.

Xignite \cite{Xignite} was founded in 2000 and was listed in Forbes' top 50 global financial technologies. Xignite provides data distribution solutions for fintech companies and financial service providers. It also provides real-time and reference financial data to emerging companies and established enterprises through cloud-based API.

Narrative \cite{Narrative} is a data flow platform founded in 2016 and headquartered in New York. According to the report in April 2017, Narrative has simplified the purchase and sale of online data. It strictly audits the data on the platform and establishes a transparent connection between buyers and sellers \cite{techcrunch-narrative}. According to the report in July 2021, Narrative proposes Data Shops to achieve data commercialization, where all enterprises and all types of data can participate in transactions, including financial services and marketing \cite{prnewswire-narrative}.

\partitle{Social Media Data}
X's \cite{Twitter} data business profits gave birth to the social data analysis industry. In 2012, X had \$47.50M revenue from selling data to some fast-growing companies \cite{techweb-twitter}. These companies analyze the data provided by X to gain insight into consumer habits and psychology, and perspective on news events and development trends. On social networking sites, the expression of user experience, views, and feelings has become a business ecosystem. As a part of the data industry, emerging social data companies are committed to tracking the development trend of social events.
DataSift \cite{DataSift} provides analysis services for massive social data, offering real-time or historical social data to brand companies, traditional enterprises, financial markets, news organizations, and more. While DataSift primarily focuses on social data, it is about to expand its business scope to include other types of data.


\partitle{Real-time data}
Streamr was founded in 2014  and headquartered in Zug, Switzerland \cite{golden-streamr}. It serves as a global data marketplace for selling real-time data and develops a decentralized Streamr network. Users can publish and subscribe to real-time data streams through Streamr. Typical application scenarios include intelligent driving, intelligent transportation, and smart cities. 
In May 2018, the New York Blockchain Consensus Conference announced that the blockchain data platform Streamr would cooperate with Nokia and the California software company OSIsoft. 

\partitle{Clould storage}
Snowflake \cite{Snowflake} is a cloud-based company headquartered in Potzmann, Montana. Snowflake Marketplace mainly provides storage, transmission, security, and other technical services for data products. As of July 2022, Snowflake Marketplace has provided access to more than 1300 real-time and readily searchable data sets from more than 260 third-party data suppliers and data service providers for those who need data-driven decisions.

Amazon Web Services (AWS) \cite{AWS}, a subsidiary of Amazon, mainly provides on-demand cloud computing platforms and APIs for individuals, companies, and governments. AWS launched AWS Data Exchange in 2019, including more than 1000 licensable data products from more than 80 data providers \cite{amazon-aws}. This includes a variety of free and paid products, such as financial services, health care/life sciences, geospatial, weather, and mapping.

\partitle{Data Innovation}
Dawex \cite{dawex} is a data exchange platform that enables organizations to buy, sell, and share data securely and efficiently. The platform offers a range of data exchange services, including data monetization, data licensing, and data sourcing. Dawex also provides a suite of tools for managing data exchange operations, including data quality control, legal and contractual agreements, and analytics. The platform supports a wide range of data types, including structured, unstructured, and semi-structured data, and offers features for anonymizing and protecting sensitive data.

\partitle{Data Compliance}
OneTrust \cite{OneTrust} is a data compliance enterprise founded in 2016, which has developed rapidly. As a data privacy management service platform, OneTrust can help enterprises understand and comply with many data privacy regulations, such as GDPR \cite{GDPR} and CCPA \cite{CCPA}. In the six years since its establishment, OneTrust has continuously acquired other data privacy software companies, achieving growth in capacity and revenue. In the round C financing in April 2021, OneTrust raised \$920M and its valuation reached \$5.3B.

TrustARC \cite{TrustARC} is a California-based company that provides technology-driven privacy compliance and risk management solutions. Its predecessor was TRUSTe, a non-profit organization founded in 1997. In July 2019, TrustARC completed the financing of \$70M, further improving the guiding position in the privacy market. Similar to OneTrust, TrustArc provides enterprises with compliance assessments according to different laws and regulations.

BigID \cite{BigID}, founded in 2016, is an enterprise privacy management company based in New York, and was shortlisted in the top 100 Forbes cloud computing in 2020. By the end of 2020, BigID had completed round D of financing, with a total financing amount of \$216 million. Its post-financing valuation exceeds \$1B. BigID provides big data solutions and services for enterprises through Software-as-a-Service (SaaS) tools and uses machine learning to find sensitive data to reduce the risk of breach.

\partitle{Summary}
Data marketplaces in various countries have witnessed remarkable development. Successful applications of data markets are primarily concentrated in developed countries across Europe and the United States. The data products on sale cover a variety of fields, mainly including finance, health care, entertainment media, transportation, public sector, telecommunications, and energy. Active data marketplaces mainly serve as a platform to connect buyers and sellers while adjusting to compliance with data regulations but are not focused on developing pricing services.

\subsubsection{Data Black Markets}\label{subsubsec:data-black-market}
In this section, we introduce the data black market and its relevant news coverage.

The data black market has attracted extensive attention and is also known as the stolen data market in literature. \citet{kennedyrevisiting} point out that personal information such as passwords or other authentication information is often sold on the black markets. By analyzing the forum operated by several data thieves, \citet{doi:10.1080/14786011003634415} find that personal data can be obtained through a small part of its value. With the increasing number of attacks and public concerns, malicious hackers and network attacks are receiving more and more attention \cite{ablon2015hacker}. 
\citet{holt2013exploring} conduct many studies on the stolen data market, including the organizational structure and relationships of participants, the black market economy \cite{hutchings2015crime}, the revenues and profits obtained by participants \cite{holt2016exploring}, and methods of intervening and destructing the online stolen data market \citet{hutchings2017online}.
Among them, \citet{hutchings2015crime} study the online black market economy and the economy related to stolen data and find that it is possible to commercialize stolen data by providing multiple products and services, such as software and hardware that steal data. 

The surge of online data and customer demand is the reason for the black market of online data \cite{marketingweek}. The data quality of online data suppliers varies widely. Personal data is usually obtained through malicious software, phishing, or trust fraud on social networking sites and accounts. This sensitive information will be sold in some forums or the dark web. Cryptocurrency (usually Bitcoin) is adopted as a payment method in transactions. Medical data is at serious risk of being stolen due to its economic effect and special sensitivity~\cite{cn-healthcare,fu2022reinforced,fu2022antibody}. As early as 2015, the amount of stolen medical data in the United States accounted for 67\% of the total amount of stolen data \cite{cn-healthcare}. In 2019, more than 4 billion data were hijacked due to leakage, an increase of 54\% over the previous year. According to the British Guardian in October 2022, many websites and companies have been attacked by hackers recently. Among them, Medibank, the largest medical insurance company in Australia, leaked 3.9 million customers' personal information due to hacker attacks, which may cause a loss of 25 million to 35 million dollars. 
The cost of the first visit to the Australian data black market is about 500 dollars, but there is no unified pricing standard \cite{ikanchai}.  
The four stages of the data black market mechanism are analyzed and simple countermeasures are given in \cite{sanctionscanner}.

In China, the scale of data black market transactions has exceeded $¥150$ billion by January 2022. 
Excessive requests for permission to collect data from various applications are a major factor causing data breaches. Other ways of data breaches include insider leakage and various technical means to steal individual privacy. Personal information is illegally traded for precise marketing or fraud. To alleviate the problem, the State Internet Information Office and other institutions issue the Provisions on the Scope of Necessary Personal Information for Common Types of Mobile Internet Applications \cite{personal-data-by-mobile-apps}. The ``Fourteenth Five Year'' Digital Economy Development Plan, officially released in January 2022, proposes to severely crack down on data black market transactions and create a safe and orderly market environment \cite{zhongguoqingnianwang}. 
In January 2022, nine departments, including the National Development and Reform Commission, jointly issued Several Opinions on Promoting the Standardized, Healthy, and Sustainable Development of the Platform Economy, which prohibits illegal activities of platform enterprises, such as collecting personal information beyond their scope and using personal information beyond their authority.

It is worth noting that the data black market is thriving while the legitimate formal data market is still in its infancy. Reasons are summarized as follows: (1) Due to the absence of guidelines, standards, laws, and regulations, sellers and buyers prefer to conduct private transactions to evade the potential legal risks and sanctions; (2) There are indeed some data that cannot be traded legally; (3) There are huge profits in the data black market, which follows the operation law of the market economy; (4) Even if law enforcement authorities ban one black market or service provider, another will take its place.

\section{Conclusion and Future Work}\label{sec:conclusion}

Data markets come from practical demands and have been tackled in multiple disciplines. We introduce the framework of data markets including differences with other products, key entities, and main procedures. We summarize important desiderata for designing well-functioning data markets. Based on this, articles concentrating on data productization, data pricing, data transaction as well as issues on privacy, security, and trust are thoroughly investigated to present an overview of state-of-art data markets. Also, guidelines, policies, laws, regulations, and actual data marketplaces are carefully investigated to gain a clear understanding of current government and industry status. Although there is a rich body of literature addressing a series of issues in data markets, there are still many questions that are far from practical applications and remain unexplored. In this section, we discuss some interesting and promising challenges for possible future work. By no means our list is exhaustive. Instead, we hope that our discussion can intrigue more extensive interest and research effort into this fast-growing area.

\subsection{Data Discovery}
One of the primary challenges in data markets is data discovery. With the increasing volume of data being generated every day, it becomes crucial to have efficient methods for buyers to find data products that cater to their specific needs. Traditional search methods may not be effective in such scenarios. Instead, more advanced techniques such as vector database indexing can be employed. This would enable similarity searches over datasets, ensuring buyers get the most relevant results for their queries.

\subsection{Participant Selection}
In data markets, participant selection can be viewed through the lens of cooperative game theory, particularly focusing on ``Core Stability''. A core-stable coalition implies that no group within the data market—be it sellers or buyers—would find it beneficial to operate outside the existing structure. By ensuring that the chosen combination of participants offers the maximum utility, the market safeguards against potential fragmentations. This not only emphasizes the quality and credibility of the participants but also reinforces the market's resilience. 


\subsection{Multiple Data Brokers}
A data broker aggregates data resources from multiple data owners and sells data products to data buyers while maximizing its profit. Existing work assumes that there is only one data market in which only one data broker works. However, this scenario is not realistic as there are multiple data marketplaces with one or more data brokers in each marketplace. In this situation, data brokers must engage in competition to maximize surplus but may also choose to cooperate to share information or otherwise.

\subsection{Data Ownership}
Data ownership becomes a crucial issue when data holds value, particularly in data marketplaces. The concept of data ownership often involves having complete control over data and the power to grant access to others. However, answering the question of ``who owns data?'' has proven to be difficult. The practice of undisclosed data reselling is prevalent, where the same data product is sold across various data marketplaces and to multiple users. The absence of unified data traceability standards and technologies adds to the complexity of determining data ownership. In the realm of data markets, the challenge of managing data ownership is to establish the authority of a data product before, during, and even after a transaction.

\subsection{Data Externality}
Externality indicates the indirect effect (compared to the direct data utility) from data to a data buyer when other buyers get the data. Once other buyers get the same data, the utility of the data is diluted (negative externality) or dilated (positive externality) to some extent. For example, among competitive companies, if the competitors get the same data of potential data buyers, the chance of one company to sell products to these data buyers declines, bringing about a negative externality,  while in collaborating settings a positive externality may arise. 

Although externality is not exclusive to data and can be found in other goods, it is particularly pronounced in data markets due to the distinctive characteristic, i.e., the replicability of data and the correlation between different datasets \cite{acemoglu2022too}. Besides the linear externality model borrowed from classical economics, a rich class of externality functions including the contributing factors tailored to data are worth further investigation. 
While \citet{DBLP:journals/corr/abs-2302-08012} give insights into the joint externality model in data markets, the relationship between the externality and data pricing has not been approached.
The data externality among data buyers links their utility with each other, where game theory may provide an approach. The data externality among data sellers could also be an interesting research topic. 

\subsection{Data Product Cost}
Measuring the cost of a data product involves considering various factors, such as infrastructure costs for data storage. While the marginal cost of duplicating a data product can be low, there are significant fixed costs associated with developing, deploying, maintaining, and using the data product. A key question in data markets is how to manage and optimize the costs of data products for achieving long-term profitability.

\subsection{The Dynamicity of Data Trading}
The replicability of data also implies that a data seller can repeatedly sell its data (maybe in distinctive versions) across time, which brings about several challenges. From the perspective of data sellers, the privacy cost induced from the sold data in one round of data transaction can be affected by the sales in later transactions, and how to formulate the cost of sellers considering the long-time trading opportunity can be a tough problem. From the perspective of data buyers, it is possible that everyone can finally get the data if unlimited data transactions over time are allowed, which can make the data price extremely low and thwart the price mechanism. Moreover, as freshness can be a significant dimension for the measurement of data quality, the time when buyers get the data makes a difference in their acquired utility, which should be considered in the design of the market mechanism. The appropriate framework catering to the dynamic data trading across time instead of the one-shot one needs to be further investigated.

\subsection{Data Security Insurance}
Data security insurance, also referred to as cyber liability insurance, is a type of insurance policy designed to mitigate financial losses associated with data breaches, cyber-attacks, and other forms of unauthorized access to sensitive information. As data markets rely on digital infrastructure and handle vast amounts of personal and proprietary data, the risks of cyber incidents have become significant. Data security insurance helps entities in data markets cover the costs of these incidents and manage the associated liabilities. Therefore, specific data security insurance for data markets is highly desired to benefit risk management, thus incentivizing the entities in data markets.

\subsection{Interdisciplinary Perspective}
Due to the interdisciplinary nature of data market research, various fields such as economics, marketing, and mechanism design are integrated into the study of data markets, particularly data pricing. As highlighted in our work, different auction formats and game-theoretic models have been adopted in data pricing to satisfy diverse desiderata across distinct scenarios. Further exploration is needed to bridge the theories and methodologies from these disciplines to fully leverage the benefits of cross-disciplinary approaches. Strengthening the communication between theoretical and applied disciplines, as well as fostering dialog across different fields, is crucial for advancing this area. Moreover, studying the hierarchical design and modularization of data markets holds great potential for developing a complex yet flexible data market system. Approaches addressing different aspects, from data pricing to revenue allocation, can be encapsulated into modules, which can enforce the reuse and migration of specific functionalities. 

As data science continues to transform numerous application domains, the construction of data markets must consider a wide range of practical scenarios. To enable domain-specific data markets, it is important to customize general technologies of data pricing, revenue allocation, and other significant aspects in data markets for niche domains. Taking data pricing as an example, user behavior is difficult to model completely and information transparency remains uncertain in practical data markets, which calls for sophisticated adaptations of classical theories. Understanding how data pricing models and their underlying assumptions can be implemented and enforced in practice is vital. Experimental studies on these models should be closely linked to theoretical investigations to ensure their effectiveness in real-world applications.

\bibliographystyle{ACM-Reference-Format}
\bibliography{references/Introduction,references/Framework,references/Desiderata,references/DataSearch,references/DataTransaction,references/DataProductizationPricingRevenueAllocation,references/PrivacySecurityTrust,references/GovernmentIndustry,references/Conclusion}

\end{document}